\documentclass[
superscriptaddress,
preprint,
 amsmath,amssymb,
prfluids,
floatfix,
]{revtex4-2}

\usepackage{graphicx}
\usepackage{dcolumn}
\usepackage{bm}

\usepackage[caption=false]{subfig}
\usepackage{xcolor}

\graphicspath{{figures/}}

\newcommand\bs{\ensuremath\boldsymbol} 	
\newcommand{\vect}[1]{\boldsymbol{#1}} 

\newcommand{\bx}{\ensuremath{\bs{x}}}
\newcommand{\utotal}{\ensuremath{ \widetilde{\vect{u}} }} 	
\newcommand{\umean}{\ensuremath{ \overline{u} }}         	
\newcommand{\ufluc}{\ensuremath{ \vect{u} }} 	
\newcommand{\ufric}{\ensuremath{ u_\tau }}
\newcommand{\ubulk}{\ensuremath{ u_B }}

\newcommand{\ptotal}{\ensuremath{ \widetilde{p} }}  		
\newcommand{\pmean}{\ensuremath{ \overline{p} }} 		
\newcommand{\pfluc}{\ensuremath{ p }}  		
\newcommand{\rtotal}{\ensuremath{ \widetilde{\rho} }}
\newcommand{\rmean}{\ensuremath{ \overline{\rho} }}
\newcommand{\rfluc}{\ensuremath{ \rho }}

\newcommand{\q}{\ensuremath{ \hat{\bs{q}} }} 	
\newcommand{\f}{\ensuremath{ \hat{\bs{f}} }} 	

\newcommand{\kperp}{\ensuremath{ k_{\perp}^2 }}
\newcommand{\D}{\ensuremath{ D_y }}
\newcommand{\DD}{\ensuremath{ D_{yy}}}
\newcommand{\nablatwo}{\ensuremath{ \DD - \kperp }}
\newcommand{\uf}{\ensuremath{ \hat{u} }}
\newcommand{\vf}{\ensuremath{ \hat{v} }}
\newcommand{\wf}{\ensuremath{ \hat{w} }}
\newcommand{\rf}{\ensuremath{ \hat{\rho} }}
\newcommand{\pf}{\ensuremath{ \hat{p} }}
\newcommand{\fuf}{\ensuremath{ \hat{f}_u }}
\newcommand{\fvf}{\ensuremath{ \hat{f}_v }}
\newcommand{\fwf}{\ensuremath{ \hat{f}_w }}
\newcommand{\frf}{\ensuremath{ \hat{f}_\rho }}

\newcommand{\kx}{\ensuremath{k_x}}
\newcommand{\kz}{\ensuremath{k_z}}
\newcommand{\lx}{\ensuremath{\lambda_x}}
\newcommand{\lz}{\ensuremath{\lambda_z}}

\newcommand{\wavespeed}{\ensuremath{ c }}
\newcommand{\wavetriplet}{\ensuremath{\kx, \kz, \omega}}
\newcommand{\bk}{\ensuremath{\bs{k}}}

\newcommand{\resolventOp}{\ensuremath{ \mathcal{H} }}
\newcommand{\linOp}{\ensuremath{ \mathcal{A} }}

\newcommand{\yplus}{\ensuremath{ y^+ }}

\newcommand{\im}{\ensuremath{ \text{i} }}

\newcommand{\pdt}{\ensuremath{ \partial_t }}

\newcommand{\Reynolds}{\ensuremath{ Re }}
\newcommand{\Retau}{\ensuremath{ \Reynolds_\tau }}
\newcommand{\Ri}{\ensuremath{ Ri }}
\newcommand{\Rifric}{\ensuremath{ \Ri_\tau }}
\newcommand{\Ritau}{\ensuremath{ \Ri_\tau }}
\newcommand{\Rigrad}{\ensuremath{ \Ri_g }}

\newcommand{\Ribulk}{\ensuremath{ \Ri_B }}

\newcommand{\Prandtl}{\ensuremath{ Pr }}
\newcommand{\diffusivity}{\ensuremath{ \gamma }}

\newcommand{\unity}{\ensuremath{ \bs{e}_y }}

\newcommand{\Ny}{\ensuremath{ N_y }}

\begin{document}


\title{Resolvent analysis of stratification effects on wall-bounded shear flows}

\author{M. A. Ahmed}
 \email{arslan@caltech.edu}
 \affiliation{Graduate Aerospace Laboratories, California Institute of Technology, Pasadena, CA 91125, USA}
\author{H. J. Bae}
 \email{jbae@caltech.edu}
 \affiliation{Harvard University, Cambridge, MA 02139, USA}
 \affiliation{Graduate Aerospace Laboratories, California Institute of Technology, Pasadena, CA 91125, USA}
\author{A. F. Thompson}%
 \email{andrewt@caltech.edu}
\affiliation{Geophysical and Planetary Sciences, California Institute of
Technology, Pasadena, CA 91125, USA}
\author{B. J. McKeon}
 \email{mckeon@caltech.edu}
 \affiliation{Graduate Aerospace Laboratories, California Institute of Technology, Pasadena, CA 91125, USA}

\date{\today}

\begin{abstract}
The interaction between shear driven turbulence and stratification is a key process in a wide array of geophysical flows with spatio-temporal scales that span many orders of magnitude. A quick numerical model prediction based on external parameters of stratified boundary layers could greatly benefit the understanding of the interaction between velocity and scalar flux at varying scales. For these reasons, here, we use the resolvent framework \cite{McKeon2010} to investigate the effects of an active scalar on incompressible wall-bounded turbulence. We obtain the state of the flow system by applying the linear resolvent operator to the nonlinear terms in the governing Navier-Stokes equations with the Boussinesq approximation. This extends the formulation to include the scalar advection equation with the scalar component acting in the wall-normal direction in the momentum equations \cite{Dawson2018}. We use the mean velocity profiles from a direct numerical simulation (DNS) of a stably-stratified turbulent channel flow at varying friction Richardson number $\Rifric$. The results obtained from the resolvent analysis are compared to the premultiplied energy spectra, auto-correlation coefficient, and the energy budget terms obtained from the DNS. It is shown that despite using only a very limited range of representative scales, the resolvent model is able to reproduce the balance of energy budget terms as well as provide meaningful insight of coherent structures occurring in the flow. Computation of the leading resolvent models, despite considering a limited range of scales, reproduces the balance of energy budget terms, provides meaningful predictions of coherent structures in the flow, and is more cost-effective than performing full-scale simulations. This quick model can provide further understanding of stratified flows with only information about the mean profile and prior knowledge of energetic scales of motion in the neutrally-buoyant boundary layers. 
\end{abstract} 

\maketitle

\section{Introduction}\label{sec:introduction} 

Stable boundary layers can be generated by the advection of warm air over a colder surface. Stably-stratified atmospheric boundary layers are observed during clear nights as a result of radiative cooling of the ground surface \cite{Nieuwstadt1984,Stull2000}. Oceans, unlike the lower atmosphere, are heated from above and are usually stably stratified \cite{Wunsch2004,Thorpe2005}. In both the atmosphere and oceans, stratification has a significant effect on turbulence production, propagation, and decay. The interaction between shear-driven turbulence and stratification is a key process in a wide array of relevant geophysical flows for which the spatio-temporal scales span many orders of magnitudes.

Classical understanding of stably-stratified boundary layers is well described in a number of textbooks \cite{Panofsky1984,Sorbjan1989,Stull1988,Wyngaard2010} and reviews \cite{Garratt1994,Ivey2008,Mahrt2014}.  However, fundamental features of the stably-stratified turbulent boundary layer still remain elusive from a modeling standpoint. The strong intermittency observed in stable boundary layers causes the upper portion of the boundary layer to decouple from the near-wall region due to the inhibition in vertical mixing \cite{Stull1988,Mahrt1999,Williams2017}. Strong stable stratification also significantly changes the flow structures prevalent in a boundary layer with additional features becoming prominent such as large-scale intermittency, gravity waves and Kelvin-Helmholtz instabilities \cite{Mahrt1999}, and the near parallel downstream tilting of flow structures \cite{Chauhan2013,Salesky2018,Salesky2020}.  

One way to study the stably-stratified turbulent boundary layer is through on-site experiments. Researchers in the past decades have conducted field experiments in the stably-stratified atmospheric boundary layer to study turbulent energy budgets \cite{Wyngaard1971}, heat and momentum transfer \cite{Kondo1978}, regime characterization \cite{Mahrt1998, Mahrt1999}, flow structures \cite{Chauhan2013}, and the complexities of atmospheric stable boundary layers \cite{Fernando2010}. Measurements of turbulence quantities in the ocean near the bottom boundary are difficult to measure and as such the literature is sparse. Smedman \emph{et al.}~\cite{Smedman1994}, using data from a marine coastal experiment over the Baltic sea, found that the near-wall turbulence was virtually independent of forcing from large-scale structures embedded in the flow. Experiments performed in the northern bay of San Francisco \cite{Stacey1999} found that active turbulence is confined near the wall.  Additionally, tidal channel experiments \cite{Lu2000} demonstrated that the production of turbulent kinetic energy is generally greatest near the bottom boundary while the buoyancy flux is weakest in this region. Still, real-world atmospheric and oceanic boundary layers are complicated by non-turbulent motions occurring simultaneously on a variety of scales, the possible importance of radiative flux divergence of the air within the boundary layer, surface condensation, and variable cloudiness \cite{Large1994,Garratt1994,Mahrt2014}. In order to isolate instances where the secondary effects are minimized, restrictions on nonstationarity or conditions on the minimum allowed value of turbulence energy may be applied to the data collected. Nonetheless, certain assumptions that are applied for analyses of these real-world stratified boundary layers are not always valid. As such, researchers supplement their work with laboratory experiments as well as simulations.

Laboratory experiments of stratified wall-bounded flows show that buoyancy effects play an important role in the transfer of heat and momentum in both the inner and outer layers of the boundary layer \cite{Arya1975, Britter1974, Piat1981, Komori1983, Fukui1983}. In general, the experiments show that with increasing stratification, the turbulence shear production rate is strongly affected by buoyancy and greatly reduced far from the wall. One measure of stratification strength is the local gradient Richardson number, $\Rigrad$. Since shear originates at the wall, the local gradient Richardson number, which is inversely proportional to the shear, is generally smaller in the near-wall region as the shear term overpowers the buoyancy term.  The stabilizing effect of stratification has a greater impact farther from the wall. Indeed, works listed here demonstrated that velocity fluctuations become weaker from the wall and in some cases, turbulence intensity is reduced as the buoyancy frequency in the system is increased. Linear inviscid stability analysis \cite{Miles1961} showed that there exists a critical value for the gradient Richardson number, $\Rigrad \geq 0.25$, that serves as a sufficient condition for stability. Additionally, the experiments of Komori \emph{et al.}~\cite{Komori1983} show that the correlation coefficients associated with the Reynolds shear stress approach zero at values of $\Rigrad \simeq 0.2 - 0.3$. 

There have been many large-eddy simulations (LES) \cite{Garg2000,Armenio2002,Basu2006,Stoll2008} and direct numerical simulations (DNS) \cite{Iida2002,Nieuwstadt2005,Brethouwer2007,Flores2011,Garcia2011} of density stratified channel flows. The results support the experimental observations: strengthening the stratification leads to the reduction (or even suppression) of turbulent velocity fluctuations further from the wall. Garg \emph{et al.}~\cite{Garg2000} showed in their work that the mean velocity profiles of the stratified channel were similar in the near-wall region but differed in the logarithmic region.  The difference is characterized by a reduction in the value of both the slope of the log-law of the mean velocity and the gradient of the mean velocity profile. It should be noted that the authors used the friction Richardson number to categorize the stratification strengths investigated in their simulations and concluded that the friction Richardson number is superior to the local gradient Richardson number in characterizing flow regimes as it is a global flow property. 

Performing experiments (both on-site and in laboratories) of stratified wall-bounded turbulence can be challenging for reasons such as topography or secondary effects and simulations suffer from computational constraints. Moreover, laboratory experiments and simulations can attain only a limited range of Reynolds and Richardson numbers that are often orders of magnitude smaller than real-world geophysical phenomena. A quick numerical model prediction of key features of stratified boundary layers could greatly benefit the understanding of the interaction between velocity and scalar flux at varying scales. For these reasons, in this paper, we aim to explore the interaction between velocity and scalar fluctuations using the resolvent model \cite{McKeon2010}. 

The resolvent model provides an optimal basis, in an energy sense, that allows an in-depth comparison of the underlying mechanisms in the flow. Moreover, the model is computationally efficient with only a singular value decomposition of the largest singular value required to obtain the leading order model. Resolvent analysis has been widely applied to a range of flow configurations to identify dominant flow structures and the underlying forcing, e.g. Ref.~\cite{McKeon2010,Yeh18, Towne18,Bae2020,McMullen2020, Nogueira2021}, and has been reviewed in detail in Ref. \cite{McKeon2017} and Ref. \cite{Jovanovic2021}. We use the model to provide analysis of the flow using only mean quantities, which are easy to obtain even in field experiments, along with knowledge from the energetics of the unstratified case, which is better documented than the stably-stratified case. The predictions from the resolvent model are then compared to the flow statistics from a DNS of a stably-stratified turbulent channel flow. The Reynolds number under consideration in the current study is considerably lower than those observed in geophysical flows, which is dictated by the available DNS data for comparison, rather than by the resolvent model. Resolvent analysis of unstratified wall-bounded flows shows that the results of the model are still relevant for moderate Reynolds numbers \cite{Moarref2013} with the resolvent modes in the logarithmic layer showing self-similar behavior. We expect the capability of the model in stably-stratified regimes to extend to higher Reynolds numbers as well.

The paper is organized as follows. In \S\ref{sec:modelling_and_analysis}, we introduce the resolvent framework with the inclusion of the scalar advection-diffusion equation and discuss the relevant energy norm, boundary conditions, and computational methods.  In \S\ref{sec:lowrank}, we examine the sensitivity of the low-rank properties of the resolvent operator to the stable stratification strength and compare these properties with the most energetic scales in each flow.  In \S\ref{sec:mode_shapes}, we analyze the characteristics of the forcing and response modes of both velocity and scalar. We compare the mode shapes with correlations obtained from DNS data. In \S\ref{sec:energy_balance}, we study the turbulent kinetic energy budget in the resolvent formulation and compare the results with the energy budget obtained from the DNS data. Finally, our conclusions on the application of the resolvent framework to a stably-stratified boundary layer are given in \S\ref{sec:conclusions}.

\section{Modeling active scalar dynamics in the Navier-Stokes equations}
\label{sec:modelling_and_analysis}

\subsection{Navier-Stokes equation with active scalar}
\label{sec:NSE_Boussinesq}

We consider a density-stratified turbulent channel flow where the density acts in the direction of gravitational acceleration. We use a Cartesian co-ordinate system $\bx=(x,y,z)$ such that the force of gravity acts in the $-y$ direction, with $x$, $y$ and $z$ being the streamwise, wall-normal and spanwise directions, respectively. The governing equations are given by the non-dimensional Navier-Stokes equation under the Boussinesq approximation, 
\begin{subequations}
\label{eqn:NSE_BL}
\begin{align}
\frac{ \partial \utotal }{ \partial t } + (\utotal \cdot
\nabla)\utotal & =  -\nabla \ptotal + \frac{\nabla^2 \utotal}{\Retau}
- \Rifric\rtotal\unity,\\
\frac{ \partial \rtotal }{ \partial t } + (\utotal \cdot
\nabla)\rtotal & =  \frac{\nabla^2 \rtotal}{\Retau\Prandtl},\\
\nabla \cdot \utotal & = 0.
\end{align}
\end{subequations}
Here, $\utotal = (\tilde{u},\tilde{v},\tilde{w})$ is the instantaneous velocity vector in the reference system $(x,y,z)$, $t$ is time, $\ptotal$ is the kinematic pressure field that remains after removing the part that is in hydrostatic balance with the mean density field, $\rtotal$ is the density deviation from the reference density $\rho_0$ ($\rtotal \ll \rho_0$), and $\unity$ is the unit vector acting in the $y$-direction. The velocity and length scales are non-dimensionalized using the friction velocity $u_\tau$ and channel half-height $\delta$, respectively, and the density is non-dimensionalized using $\Delta\rho$, the difference in density between the two channel walls. We define the walls to be located at $y=0$ and $y=2$. The non-dimensional quantities are given by the Reynolds, Prandtl and Richardson numbers, defined as \refstepcounter{equation}
$$
\Retau = \frac{u_{\tau}\delta}{\nu}, \qquad
\Prandtl = \frac{\nu}{\diffusivity}, \qquad 
\Rifric = \frac{g \Delta \rho \delta}{\rho_0\ufric^2},
\eqno{(\theequation{\mathit{a},\mathit{b},\mathit{c}})}\label{nondims_rey_pra}
$$
where $\nu$ is the kinematic viscosity, $\diffusivity$ is the molecular diffusivity of density, and $g$ is the acceleration due to gravity.

\subsection{Resolvent framework with an active scalar}\label{sec:NSE_Resolvent}

The total fields $\utotal$, $\ptotal$ and $\rtotal$ can be split into mean and fluctuating parts as
\begin{subequations}
\begin{align}\label{eqn:means}
\utotal(\bx,t) & = {\boldsymbol{\umean}}(y) + \ufluc(\bx,t),\\
\ptotal(\bx,t) & = \pmean(y) + \pfluc(\bx,t),\\
\rtotal(\bx,t) & = \rmean(y) + \rfluc(\bx,t),
\end{align}
\end{subequations}
where the mean is taken in the homogeneous directions, $x$ and $z$, and time. Note that ${\boldsymbol{\umean}} = (\bar{u},\bar{v},\bar{w})$ and $\bar{v}=\bar{w}=0$. We substitute the decomposed variables into Eq. (\ref{eqn:NSE_BL}) to obtain the fluctuation equations 
\begin{subequations}
\begin{align}
\pdt\ufluc + (\umean \cdot \nabla)\ufluc + (\ufluc \cdot \nabla)\umean
&= -\nabla \pfluc + \frac{\nabla^2 \ufluc}{\Retau} -
\Rifric\rfluc\unity + \bs{f}_{\bs{u}} \\ 
\pdt\rfluc + (\umean \cdot \nabla)\rfluc + (\ufluc \cdot \nabla)\rmean
&= \frac{\nabla^2 \rfluc}{\Retau\Prandtl} + f_\rfluc,\\ \nabla \cdot
\ufluc &= 0,
\end{align}
\end{subequations}
where $\bs{f}_{\bs{u}} = - \ufluc\cdot\nabla\ufluc$ and $f_\rfluc = -\ufluc\cdot\nabla\rfluc$ are the nonlinear terms.

Taking the Fourier transform of the fluctuation equations above in homogeneous directions and time, the variables can be expressed as
\begin{equation}
\begin{bmatrix}
\boldsymbol{u}(x,y,z,t)\\p(x,y,z,t)\\\rho(x,y,z,t)
\end{bmatrix}
= \iiint^{\infty}_{-\infty}
\begin{bmatrix}
\hat{\boldsymbol{u}}(y;\wavetriplet)\\
\hat{p}(y;\wavetriplet)\\
\hat{\rho}(y;\wavetriplet)
\end{bmatrix}
e^{\im(\kx x + \kz z -\omega t)}
d\kx d\kz d\omega,
\end{equation}
for $\bk = (\wavetriplet) \neq (0,0,0)$, where $(\hat{\cdot})$ denotes the Fourier transformed variables.  Here, the streamwise and spanwise wavenumbers are $\kx$ and $\kz$, respectively, and $\omega$ is the temporal frequency defined as $\omega = \wavespeed\kx$, where $c$ is the wavespeed. The streamwise and spanwise wavelengths are defined as $\lx = 2\pi/\kx$ and $\lz = 2\pi/\kz$, respectively. Critical-layers can be identified when the wavespeed $\wavespeed$ is equivalent to the mean velocity, i.e. $y_c$ is the critical layer location for wavespeed $\wavespeed=\umean(y_c)$. Assuming the mean velocity and density profiles are known, the fluctuations equations are expressed compactly in a  linear equation as
\begin{equation}\label{eqn:linear_compact}
-\im\omega\q - \linOp\q = \f,
\end{equation}
where we define $\q = [ \uf\;\vf\;\wf\;\pf\;\rf ]^T$ as the state vector and $\f = [ \fuf\;\fvf\;\fwf\;0\;\frf ]^T$ as the forcing vector. The linear operator is given by
\begin{equation}\label{eqn:linearOp}
\linOp = \begin{pmatrix}
A & -\partial\umean/\partial y & 0 & -\im\kx & 0 \\
0 & A & 0 & -\D & -\Rifric \\
0 & 0 & A & -\im\kz & 0 \\
-\im\kx & -\D & -\im\kz & 0 & 0 \\
0 & -\partial\rmean/\partial y & 0 & 0 &  A_\rfluc
\end{pmatrix},
\end{equation}
where
\begin{subequations}
\begin{align}
A &=  -\im\kx\umean +\frac{\hat\Delta}{\Retau},\\
A_\rfluc &=  -\im\kx\umean +\frac{\hat\Delta}{\Retau Pr},
\end{align}
\end{subequations}
$\D$ is the wall-normal derivative operator and $\hat\Delta \equiv \nablatwo$ is the Laplacian with $\kperp = \kx^2 + \kz^2$. The block matrix $\linOp$ describes the linear dynamics of the system. Equation (\ref{eqn:linear_compact}) can be rearranged to yield
\begin{equation} \label{eqn:resolvent}
\q\; =\; \resolventOp(\bk)\; \f,
\end{equation}
where $\resolventOp(\bk) = (-\im\omega I - \linOp)^{-1}$ is the resolvent of the linear operator and $I$ is the identity matrix. A related analysis has been performed in Ref. \cite{madhusudanan2020coherent}.

From Eq. \eqref{eqn:resolvent}, we wish to find a decomposition of the resolvent operator that enables us to identify high gain input and output modes with respect to the linear operator. For resolvent analysis, this is given by the Schmidt decomposition. However, this decomposition must be accompanied by a choice of inner product and the corresponding norm. The natural and physically meaningful norm is given by the non-dimensionalized energy norm, which is the sum of kinetic and potential energies \cite{Lorenz1955,Turner1979}  
\begin{equation}\label{eqn:norm}
\frac{1}{2}\|\boldsymbol{q}\|^2_E = \frac{1}{2}
(\boldsymbol{q},\boldsymbol{q})_E= 
\frac{1}{2} \int_{0}^2\left( u^*u + v^*v + w^*w  +
\Rifric(\rho^*\rho) \right)dy,
\end{equation}
where $(\cdot)^*$ denotes the conjugate transpose.

We perform the Schmidt decomposition of the resolvent operator $\resolventOp$ to generate a basis based on the most highly amplified forcing and response directions such that
\begin{equation}\label{eqn:svd}
\resolventOp(\bk) = \sum_{j=1}^{\infty} \sigma_j(\bk)
\bs{\hat{\psi}}_j(y;\bk) \bs{\hat{\phi}}^{*}_j(y;\bk),
\end{equation}
where the right and left Schmidt bases (or singular vectors in the discrete case) are given by $\bs{\hat{\phi}}_j$ and $\bs{\hat{\psi}}_j$ along with their corresponding gains $\sigma_j$. The singular values are in descending order such that $\sigma_1 \geq \sigma_2 \geq \cdots \geq 0$.  The forcing and resolvent modes are orthonormal such that
\begin{equation}
(\bs{\hat{\phi}}_j, \bs{\hat{\phi}}_k)_E =
(\bs{\hat{\psi}}_j, \bs{\hat{\psi}}_k)_E = \delta_{jk},
\end{equation}
where $\delta_{jk}$ denotes the Kronecker delta. The basis pair defined above is used to decompose the nonlinear forcing and response field at a specified wavenumber triplet as 
\begin{subequations}
\begin{align}
\label{eqn:bases}
\f(y;\bk) & = \sum_{j=1}^{\infty} \bs{\hat{\phi}}_j(y;\bk)
\chi_j(\bk),\\
\q(y;\bk) & = \sum_{j=1}^{\infty} \chi_j(\bk) \sigma_j(\bk)
\bs{\hat{\psi}}_j(y;\bk).
\end{align}
\end{subequations}
Here, $\chi_j$ is a projection variable that is obtained by projecting the nonlinear forcing onto the forcing modes, and subsequently use to weight the response modes.  Note that the largest energy is obtained when the forcing is aligned with the leading singular vector, i.e.\ when $\chi_j=\delta_{j1}$.

\subsection{Computational approach}\label{sec:computational_approach}

\subsubsection{Mean velocity and density profiles} \label{sec:mean_velocity}

\begin{table} 
\caption{Comparison of our DNS and the results of Garc\'ia-Villalba \& del \'Alamo~\cite{Garcia2011} denoted under columns titled GV11, both at $\Retau=180$. $\Reynolds_B$ is the bulk Reynolds number defined as $\ubulk\delta/\nu$ where the bulk velocity is $\ubulk = \int_0^{2}\umean dy/2$.  $\Ribulk$ is the bulk Richardson number which is defined as $\Ribulk = \Rifric(\ufric/\ubulk)^2/2$.  $Nu$ is the Nusselt number defined as $Nu = 2\delta q_w/(\diffusivity \Delta\rfluc)$. For laminar flow $Nu=1$.}
\begin{center}
\def~{\hphantom{0}}
\begin{ruledtabular}
\begin{tabular}{rcccccc}
& \multicolumn{2}{c}{$Re_B$} & \multicolumn{2}{c}{$\Ri_B$} & \multicolumn{2}{c}{$Nu$}\\
\colrule
$\Rifric$  & \textit{GV11} & DNS & \textit{GV11} & DNS & \textit{GV11} & DNS \\
\colrule
0   & \textit{2820} & 2823 & \textit{0.000} & 0.000 & \textit{6.03} & 6.08 \\[2pt]
10  & \textit{   -} & 2970 & \textit{    -} & 0.018 & \textit{   -} & 4.78 \\[2pt]
18  & \textit{3043} & 3060 & \textit{0.031} & 0.031 & \textit{4.02} & 4.15 \\[2pt]
60  & \textit{3436} & 3473 & \textit{0.082} & 0.081 & \textit{2.80} & 2.82 \\[2pt]
100 & \textit{   -} & 3850 & \textit{    -} & 0.109 & \textit{   -} & 2.37 \\
\end{tabular} 
\end{ruledtabular}
\label{tab:dns}
\end{center}
\end{table}

\begin{figure} 
\centering
\subfloat[][]{\includegraphics[width=0.42\textwidth]{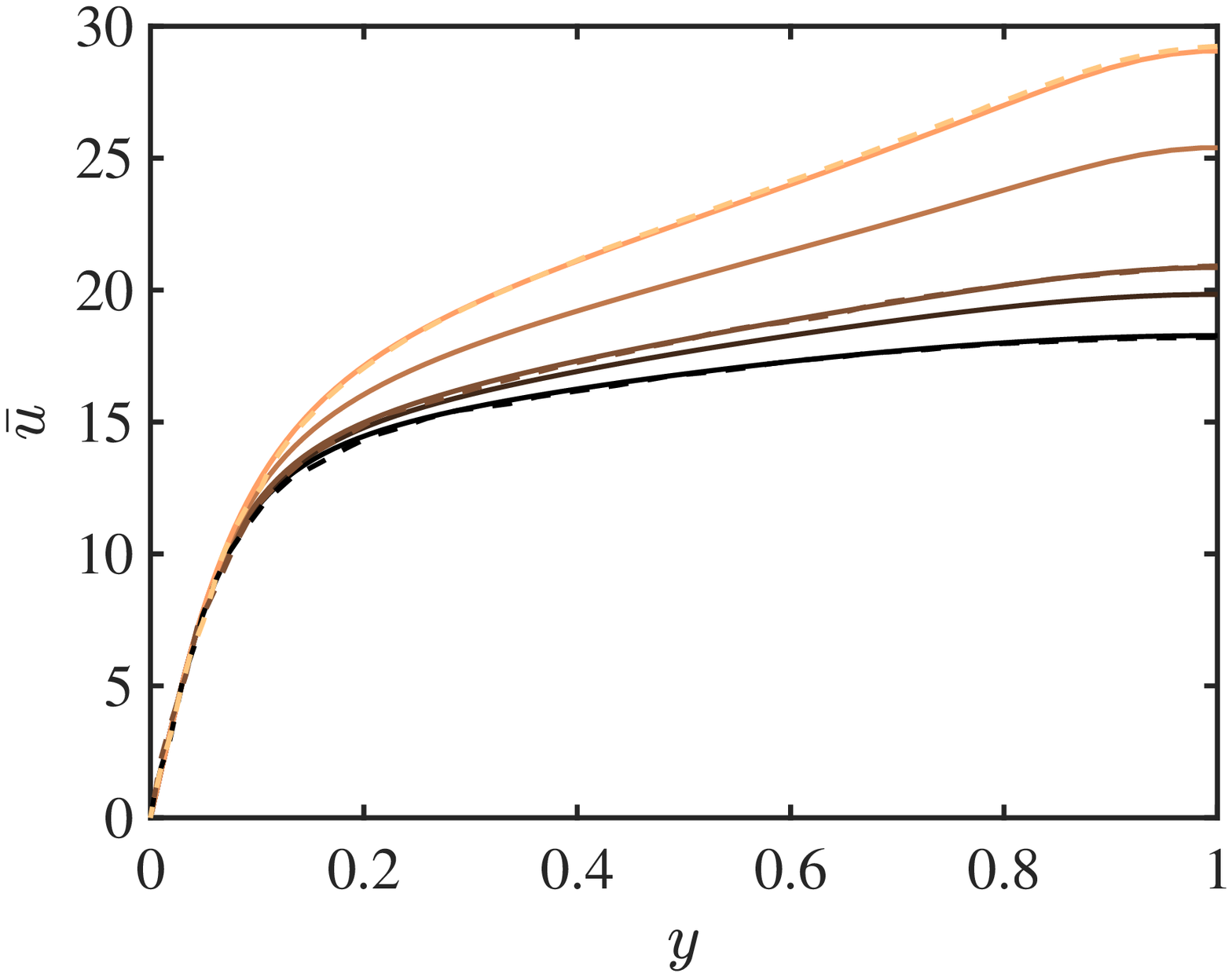}}
\hspace{0.1cm}
\subfloat[][]{\includegraphics[width=0.42\textwidth]{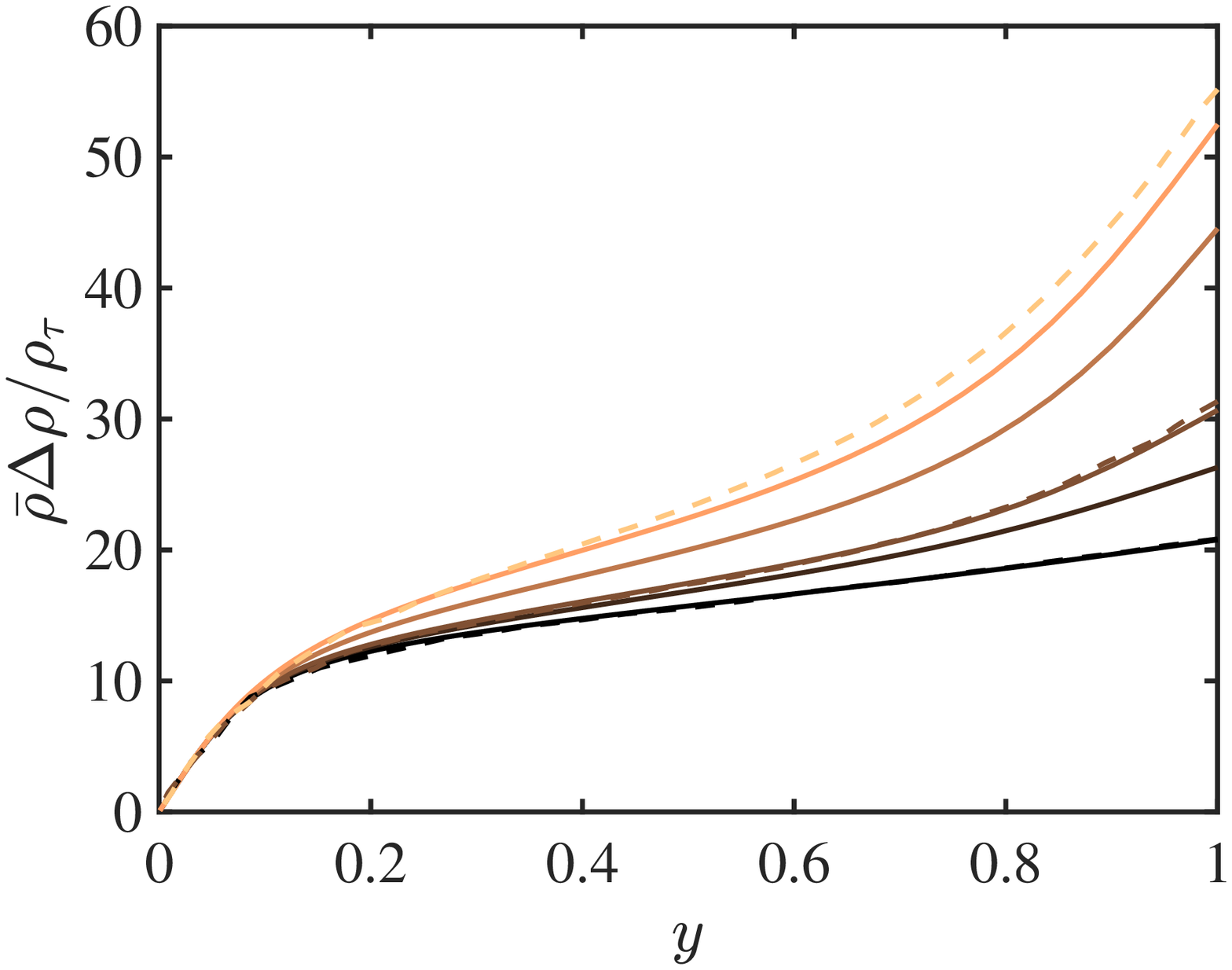}}\\
\subfloat[][]{\includegraphics[width=0.42\textwidth]{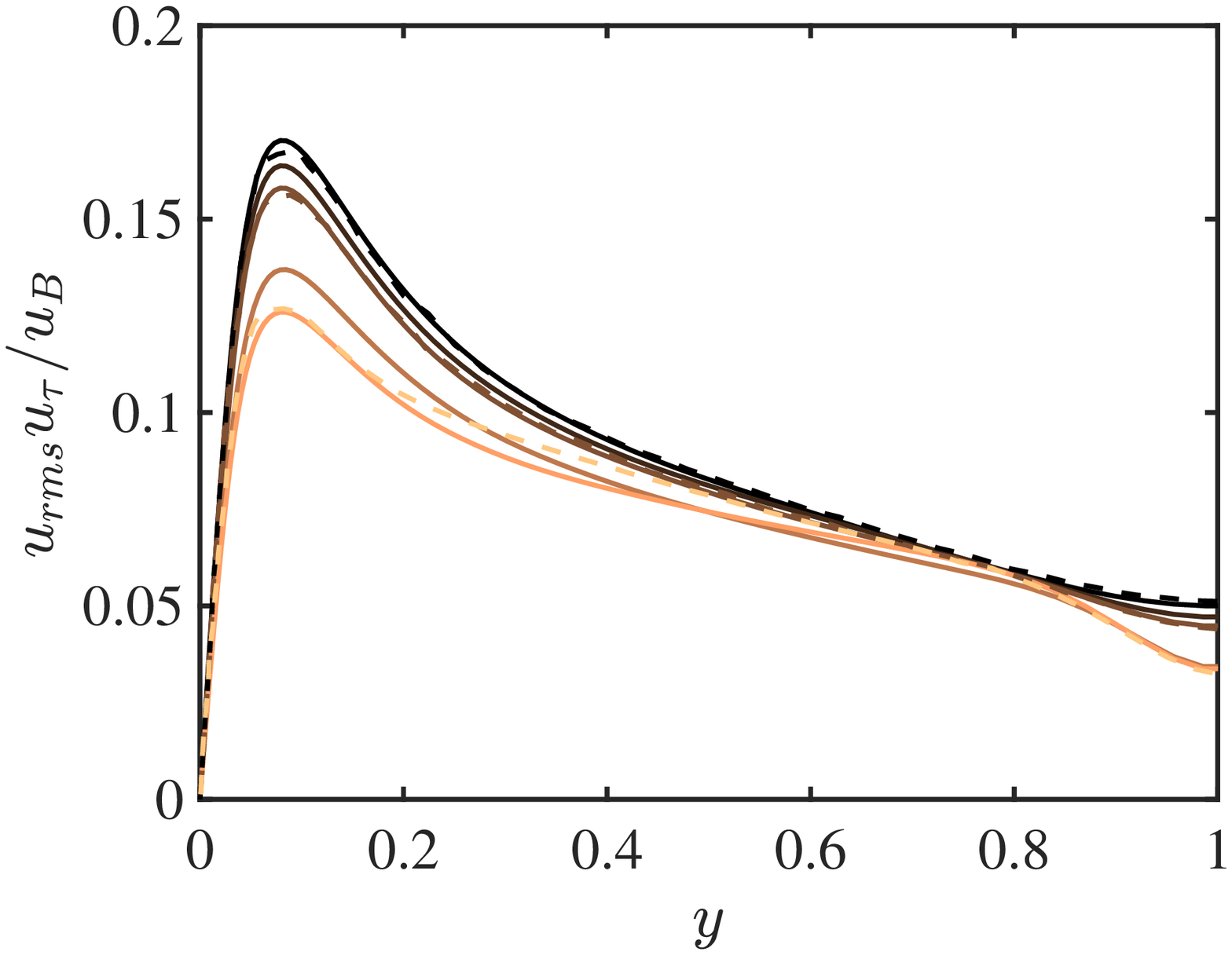}}
\hspace{0.1cm}
\subfloat[][]{\includegraphics[width=0.42\textwidth]{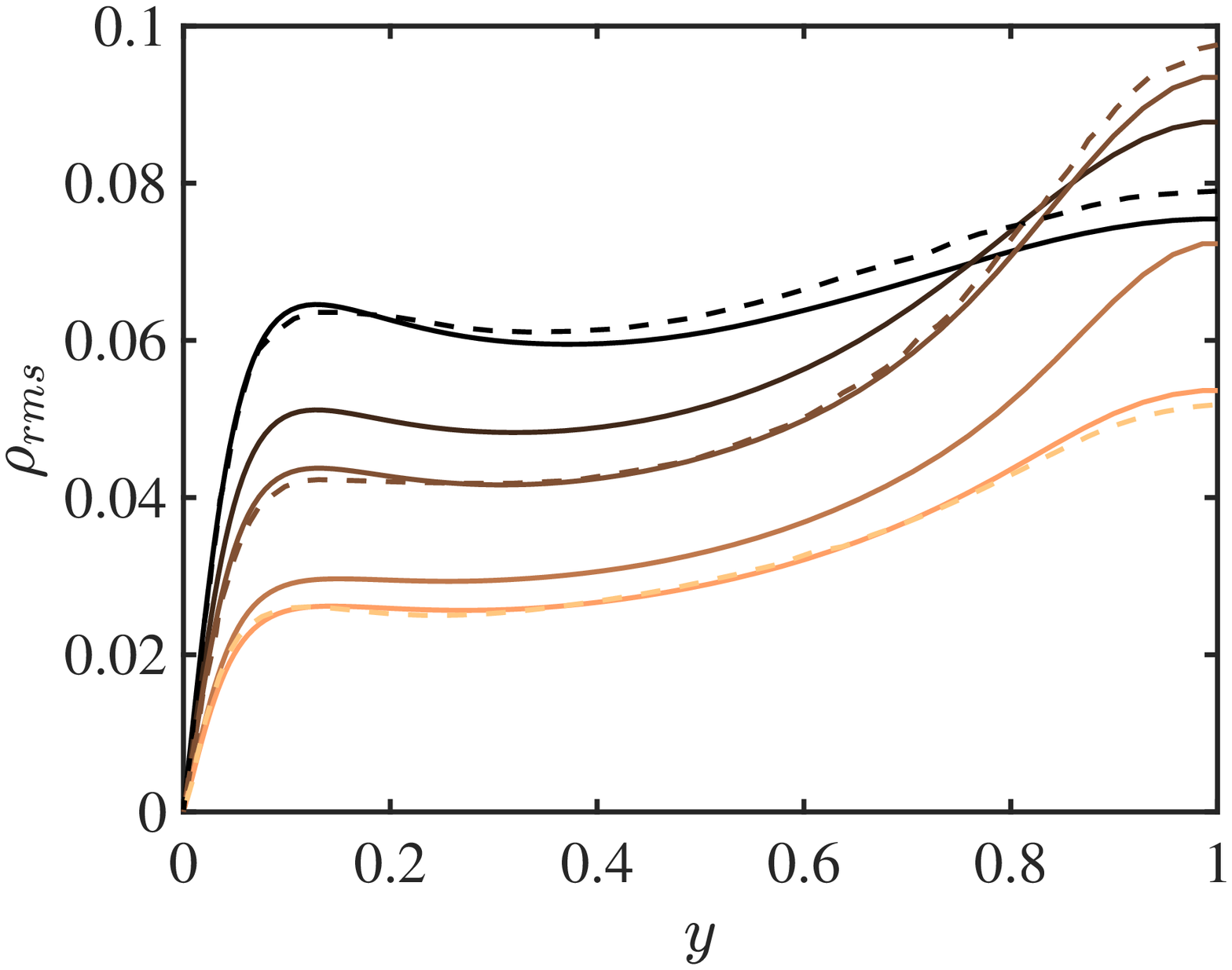}}
\caption{Mean (a) streamwise velocity and (b) density profiles and root-mean-square (r.m.s.) (c) streamwise velocity and (d) density profiles from the current DNS for $\Ri_\tau = 0,10,18,60,100$ (solid lines darker to lighter), compared to the mean profiles of Ref. \cite{Garcia2011} for $\Ri_\tau = 0, 18, 120$ (dashed lines darker to lighter). The friction density is defined as $\rho_\tau = q_w / u_\tau$, where $q_w$ is the density flux at the wall.}
\label{fig:mean_profiles}
\end{figure}

Mean velocity and density profiles are required to close the resolvent model. We obtain the one-dimensional mean velocity and density profiles from a DNS of a stratified turbulent channel at $Re_\tau=180$ for a wide range of $\Rifric$. The simulations are performed by discretizing the incompressible Navier-Stokes equations with a staggered, second-order accurate, central finite-difference method in space \cite{Orlandi2000}, and an explicit third-order accurate Runge-Kutta method for time advancement \cite{Wray1990}. The system of equations is solved via an operator splitting approach \cite{Chorin1968}. The code has been verified for neutrally-buoyant cases in Ref. \cite{Bae2019,Lozano-Duran2019}.

Periodic boundary conditions are imposed in the streamwise and spanwise directions, the no-slip and no-penetration condition with $\tilde{\rho}=0$ is applied at the bottom boundary, and a no-slip and no-penetration condition with $\tilde{\rho}=1$ is applied at the top boundary. The streamwise, wall-normal, and spanwise domain sizes are $4\pi$, $2$, and $2\pi$ respectively. The grid spacings in the streamwise and spanwise directions are uniform with $\Delta x^+=8.8$ and $\Delta z^+=4.4$; non-uniform meshes are used in the wall-normal direction, with the grid stretched toward the wall according to a hyperbolic tangent distribution with $\min(\Delta y^+ ) = 0.31$ and $\max(\Delta y^+ ) = 5.19$, where the superscript $+$ indicates length scales in wall units normalized by $\nu/u_\tau$ rather than $\delta$. A constant pressure gradient is applied to drive the flow. The simulation was run over $100$ eddy-turnover times, defined as $\delta/u_\tau$, after transients. 

The work of Garc\'ia-Villalba \& del \'Alamo~\cite{Garcia2011} at $\Retau = 180$ is used to validate the results. The comparison of a few key quantities is shown in Table \ref{tab:dns}, which indicate a good agreement for all Richardson numbers. The mean and root-mean-squared streamwise velocity and density profiles are shown in Fig. \ref{fig:mean_profiles} for all current cases and select cases from Ref. \cite{Garcia2011}. The profiles show good agreement among all statistics. 

\subsubsection{Resolvent mode computation}

The Schmidt decomposition of the resolvent operator outlined in \S\ref{sec:NSE_Resolvent} is numerically implemented as the singular value decomposition (SVD). We solve the discrete equations using a spectral collocation method with the number of points in the wall-normal direction given by $\Ny$, thus limiting the number of singular values to $5\Ny$ because the state vector $\q\in\mathbb{C}^{5\Ny\times 1}$. In this study, after conducting a grid convergence study examining the singular values, we selected a wall-normal grid resolution of $N_y=400$. Thus, the computational cost of the resolvent mode computation is at most $O(N_y^3)$ (less if randomized algorithms are employed \cite{Moarref2013,Ribeiro20}), often only requiring a leading order singular value decomposition (see \S\ref{sec:lowrank}  for more information), and can be performed in seconds on a personal computer.

The discretized linear operator is constructed using Chebyshev differentiation matrices and is shifted to integrate between $y\in[0, 2]$ rather than $y\in[-1, 1]$. The mean velocity and density profiles obtained from DNS as well as their wall-normal derivatives are interpolated to the Chebyshev grid points to form the resolvent operator as in Eq. \eqref{eqn:linearOp}. The no-slip and no-penetration boundary conditions for the fluctuating velocities and density, i.e. $u,v,w,\rho=0$, are applied at the walls.   

In the case of a turbulent channel, due to the symmetry in the geometry, the resolvent modes appear in pairs that can be linearly combined to produce symmetric and antisymmetric modes. Depending on the support of these modes, the singular values may be identical or similar in magnitude. For the results in the following sections, only results in the bottom half-channel will be shown, but the corresponding upper half-channel results are analogous in all cases.  

\section{Results} \label{sec:results}

In this section, we explore how the resolvent analysis provides insight to changes in flow characteristics with increasing stratification from only a limited range of representative scales. We compare (i) the resolvent energy spectra, obtained from the ratio of the energy in the leading resolvent response mode to the total response, $(\sigma_1^2+\sigma_2^2)/\sum_j\sigma_j^2$, to the premultiplied energy spectra of the DNS, (ii) the structure identified by the leading resolvent mode to the correlation computed from DNS, and (iii) the energy budgets of the resolvent modes to that of the DNS. 

In order for full representation of the system, a wide range of scales as well as information of all other subsequent modes in addition to the leading resolvent modes are necessary \cite{McKeon2010,McKeon2017}. However, the goal here is to provide a quick model for characterising the flow. The simplest and quickest model can be provided via a rank-one approximation, where only the leading resolvent mode is computed. Thus, our focus will be on the representation given by the leading resolvent mode for a limited number of scales. 

\subsection{Resolvent energy spectra}\label{sec:lowrank}

The resolvent norm is the principal singular value, $\sqrt{\sigma_1^2+\sigma_2^2}$ in this case, of the resolvent operator $\resolventOp$, and quantifies the system’s sensitivity to temporal forcing \cite{Symon2018}.  The energetic contribution from broadband forcing is quantified as the square of the resolvent norm. The resolvent operator $\resolventOp$ can be described as low-rank if the majority of its response to broadband forcing in the wall-normal direction is captured by the first few response modes. Theoretically, there are an infinite number of singular values and corresponding modes because the wall-normal coordinate is continuous. However, not all of the singular vectors are energetically significant. As described in \S\ref{sec:NSE_Resolvent}, a self-sustaining representation of the flow will correspond to a weighted assembly of forcing modes rather than a broadband forcing \cite{Nogueira2021}; however, past studies have showed that broadband forcing is successful in identifying the important component of the flow, e.g. Ref. \cite{McKeon2010,Bae2020}. McKeon \& Sharma \cite{McKeon2010} demonstrated that the characteristics of the leading response modes for a range of wavenumber-frequency combinations agree with experimental observations in pipe flow and with scaling concepts in wall-bounded turbulence. Moarref \emph{et al.} \cite{Moarref2013} showed that the first two resolvent modes account for more than 80\% of the total response in a channel. Bae \emph{et al.} \cite{Bae2020} investigated the low-rank nature of a compressible turbulent boundary layer and highlighted the similarities in the region where the low-rank approximation is valid for the incompressible regime. 

Assuming the resolvent operator is low-rank ($\sigma_1 \simeq \sigma_2 \gg \sigma_3$) allows us to approximate the operator as 
\begin{equation}
\resolventOp(\bk)\; \approx\; \sigma_1\; \bs{\hat{\psi}}_1\;
\bs{\hat{\phi}}^{*}_1 + \sigma_2\; \bs{\hat{\psi}}_2\;
\bs{\hat{\phi}}^{*}_2,
\end{equation}
for each $\bk$ since most of the energy in the system is modelled by the principal singular value. The low-rank behavior of $\resolventOp$ is typically representative of there being a dynamically significant physical, spatio-temporal structure at the scale dictated by $\bk$.  

\begin{figure}	
\centering
\includegraphics[width=0.8\textwidth]{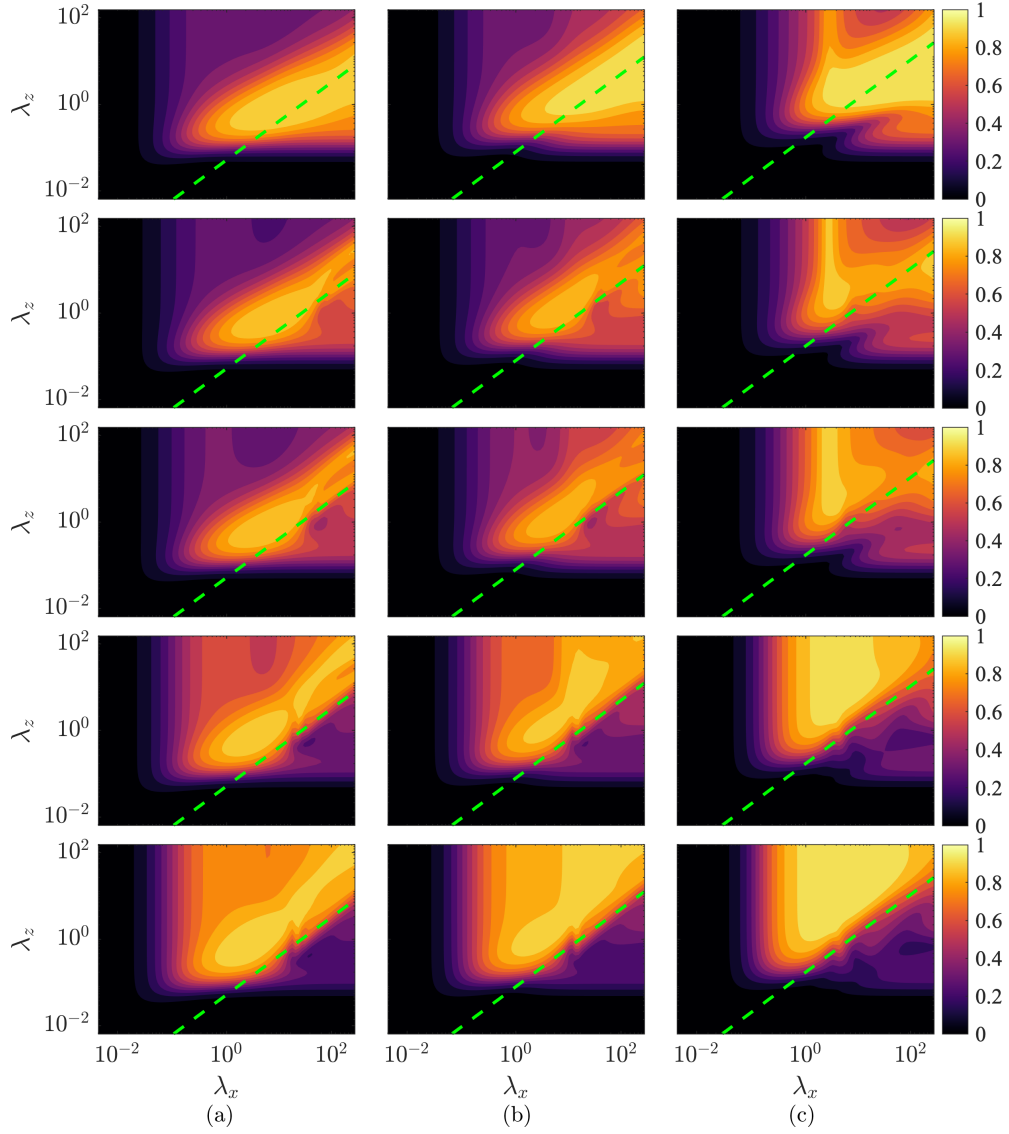}
\caption{Contour plots depicting the energy contained in the leading response mode relative to the total response, $(\sigma_1^2+\sigma_2^2) / \Sigma_j \sigma_j^2$, for different streamwise and spanwise wavelengths at (a) $\wavespeed =\umean(\yplus=15)$, (b) $\wavespeed =\umean(\yplus=30)$ and (c) $\wavespeed =\umean(\yplus=100)$ for $\Ritau =$ 0, 10, 18, 60, 100 (top to bottom). Green dashed lines are (a) $\lambda_x = 15\lambda_z$, (b) $\lambda_x = 10\lambda_z$ and (c) $\lambda_x = 5\lambda_z$. }
\label{fig:lowrank}
\end{figure}
\begin{figure}	
\centering
\includegraphics[width=0.8\textwidth]{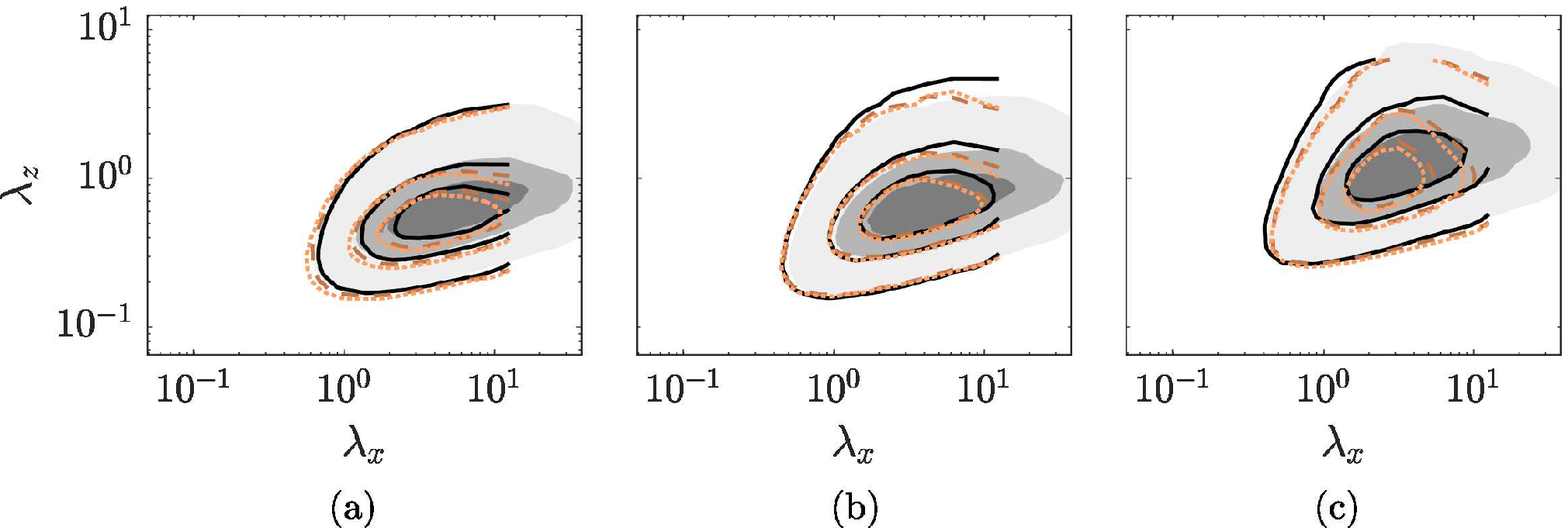}
\caption{Contour plots depicting the premultiplied streamwise kinetic energy spectra as functions of the streamwise and spanwise wavelengths obtained from DNS at (a) $y^+=15$, (b) $y^+=30$, and (c) $y^+=100$ for $\Ritau = 0$ (solid line), $\Ritau = 60$ (dashed line) and $\Ritau = 100$ (dotted line). The shaded contours are from the $\Retau = 180$ neutral channel \cite{delAlamo2004}. The levels plotted are $0.1, 0.3, 0.5$ times the maximum value of the corresponding spectrum.}
\label{fig:DNSspec}
\end{figure}
To study the variation in the low-rank behavior for different magnitudes of stratification, we plot the energetic contribution of the principal response mode to the total response in the model for a given $\bk$ quantified by $(\sigma_1^2+\sigma_2^2) / \Sigma_j \sigma_j^2$ for a range of wall-parallel wavelengths (Fig. \ref{fig:lowrank}). The leading response modes account for more than $80\%$ of the total response over a large range of homogeneous wavelengths for the three wavespeeds selected.

The range of wavenumbers for which the resolvent operator is low-rank changes significantly with stratification. In the neutrally-buoyant case ($\Rifric=0$), we see that $\resolventOp$ is low-rank in a range of moderate-to-large streamwise wavelengths. For the neutrally-buoyant case, it is known that the low-rank region coincides with the most energetic wavenumbers from the premultiplied energy spectra of a turbulent channel \cite{Moarref2013}.  As the friction Richardson number first increases, the low-rank behavior shifts to only a small range of streamwise wavelengths. We see a similar phenomenon in the premultiplied streamwise energy spectra from the DNS (Fig. \ref{fig:DNSspec}), where with increasing $\Ritau$, the larger streamwise wavelength content is suppressed. This was also observed in the premultiplied energy spectra of Garc\'ia-Villalba \& del \'Alamo~\cite{Garcia2011} for a wider range of $\Retau$ and $\Ritau$.

However, after $\Ritau = 18$, the low-rank behavior of the principal resolvent modes intensifies along a vertical band $\lx/\delta\geq 1$ until the system becomes low-rank at large spanwise wavelengths with almost no low-rank behavior below the green dashed line in Fig. \ref{fig:lowrank} ($\lx=15\lz$, $10\lz$ and $5\lz$ for $y^+ = 15$, $30$ and $100$, respectively). This seems to indicate a low-rank behavior in structures that are descriptive of quasi-two-dimensional flow where $\lambda_z\gg\lambda_x$. Hopfinger \cite{Hopfinger1987} details the emergence of two-dimensional modes for a variety of flows with strong stratification.  Moreover, Mahrt \cite{Mahrt2014} alludes to  the emergence of two-dimensional modes (often referred to as pancake modes) owing to the conversion of vertical kinetic energy to potential energy in the presence of strong stable stratification. The premultiplied energy spectra for higher $\Ritau$  indicate high energy in the vertical band as well \cite{Garcia2011}.  

\subsection{Mode shapes}\label{sec:mode_shapes}

\begin{table} 
\caption{Representative wavenumber combinations that we will explore in \S\ref{sec:mode_shapes}.}
\centering
\begin{ruledtabular}
\begin{tabular}{lccc}
Mode name                               & $\kx$ 	& $\kz$         & $\wavespeed$          \\ 
\colrule
E1: most energetic mode for $y^+=15$    & $\pi/2$	& $4\pi$       	& $\umean(\yplus=15)$   \\
E2: most energetic mode for $y^+=30$    & $\pi/2$	& $3\pi$       	& $\umean(\yplus=30)$   \\ 
E3: most energetic mode for $y^+=100$   & $\pi/2$	& $2\pi$    	& $\umean(\yplus=100)$  
\end{tabular}
\end{ruledtabular}

\label{tab:combos}
\end{table}

\begin{figure}
\centering
\subfloat[][]{\includegraphics[height=4.5cm]{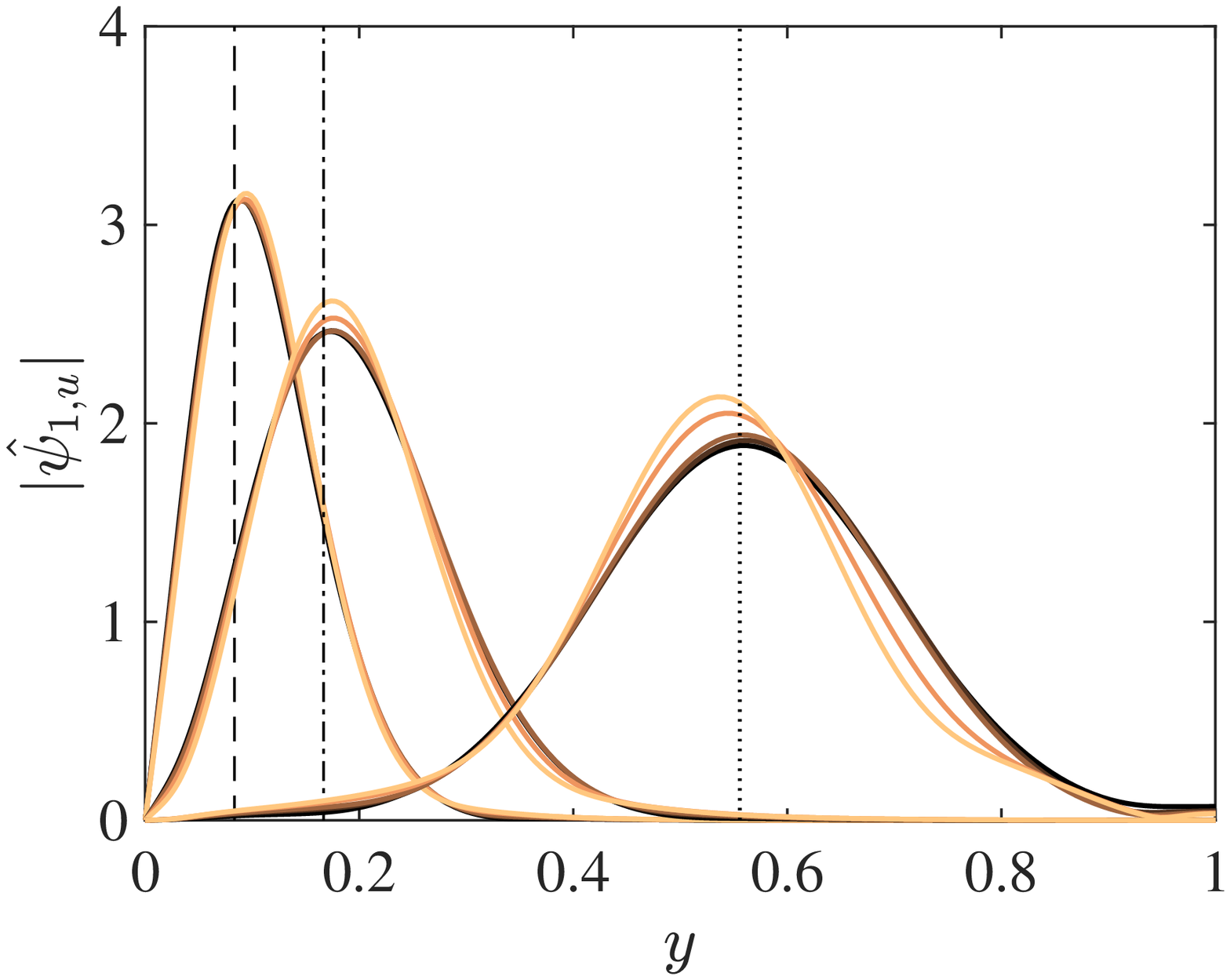}}
\hspace{0.1cm}
\subfloat[][]{\includegraphics[height=4.5cm]{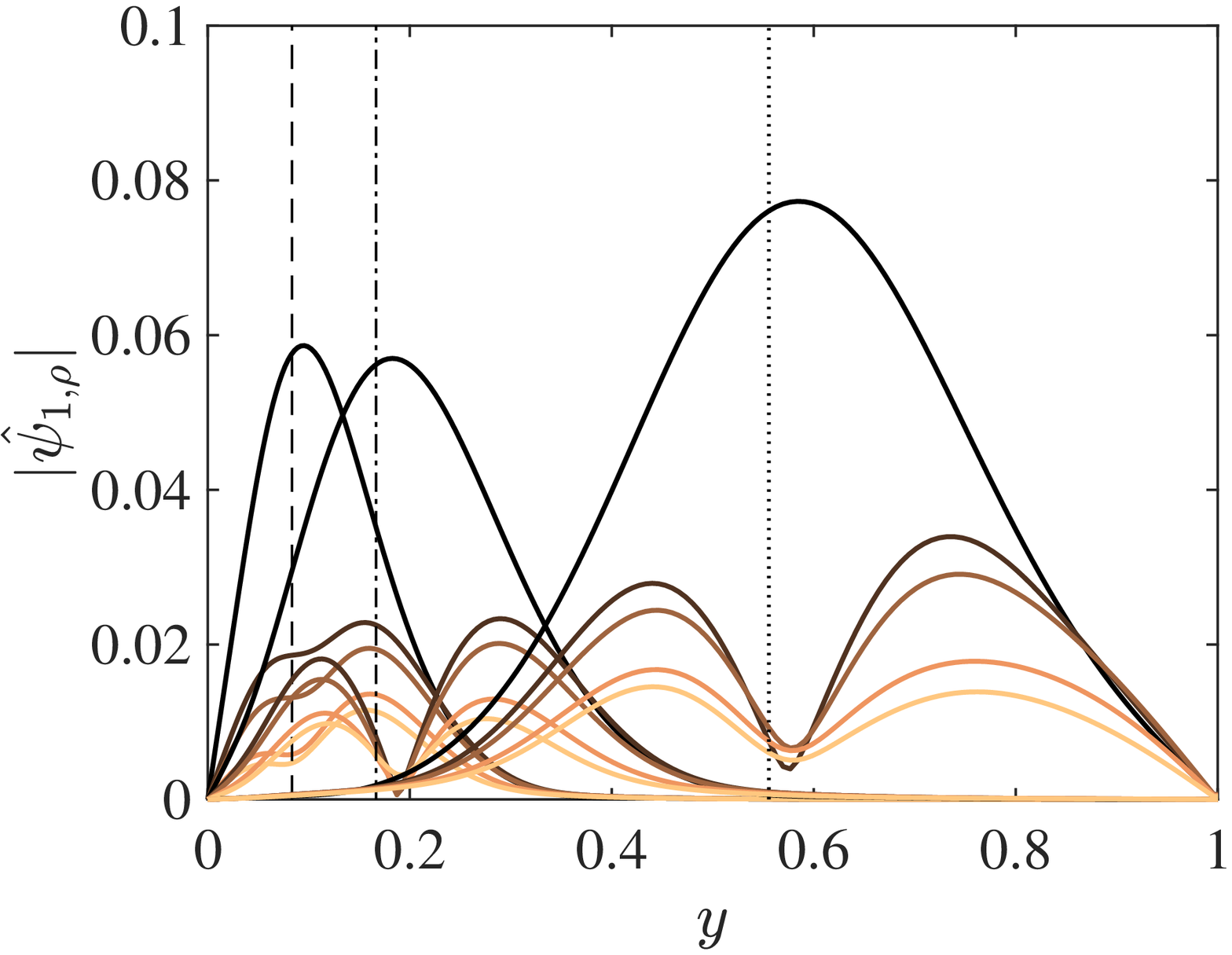}}
\caption{Amplitudes of the leading resolvent response modes for the (a) streamwise velocity and (b) density for $\Ritau = 0,10,18,60,100$ (darker to lighter) at $c = \bar{u}(y^+=15)$ (dashed line), $\bar{u}(y^+=30)$ (dot-dashed line) and $\bar{u}(y^+=100)$ (dotted line) for wave-parameters corresponding to E1, E2 and E3, respectively.  The subscripts $u$ and $\rho$ indicate the corresponding components  of the resolvent response mode.} 
\label{fig:rms_res}
\end{figure}

In order to study the flow structures, we compute the resolvent response modes for a set of wave parameters. The most energetic scales for the various $\Ritau$ under consideration for the different wall-normal heights still coincide with the neutrally-buoyant case (Fig. \ref{fig:DNSspec}), falling in the low-rank region despite the fact that including the scalar advection-diffusion equation in the governing equations changes the wavelengths at which the resolvent operator is low-rank (Fig. \ref{fig:lowrank}). In this section, we study the resolvent response mode shapes for these wavenumber and wavespeed combinations. The list of mode combinations under consideration is listed in Table \ref{tab:combos}. In particular, mode E1 is the most energetic mode for $y^+=15$, E2 for $y^+=30$ and E3 for $y^+=100$. 

The predictive capabilities of the resolvent modes are first shown through the amplitudes of the leading resolvent response modes (Fig. \ref{fig:rms_res}) of the streamwise velocity and density. The resolvent modes compare well to the streamwise and density turbulence intensities in Fig. \ref{fig:DNSspec}(c,d). The streamwise root-mean-square (r.m.s.) quantities and resolvent amplitudes show no variation among different Richardson numbers closer to the wall and increase slightly with $\Ritau$ farther away from the wall. On the other hand, the density r.m.s. and resolvent amplitudes decrease significantly with Richardson number at all wall-normal heights.  Despite only using the leading resolvent mode, the relative magnitude at each corresponding wall-normal height is well captured for the range of Richardson numbers considered here. 

\begin{figure}
\centering
\subfloat[][]{\includegraphics[height=4.5cm]{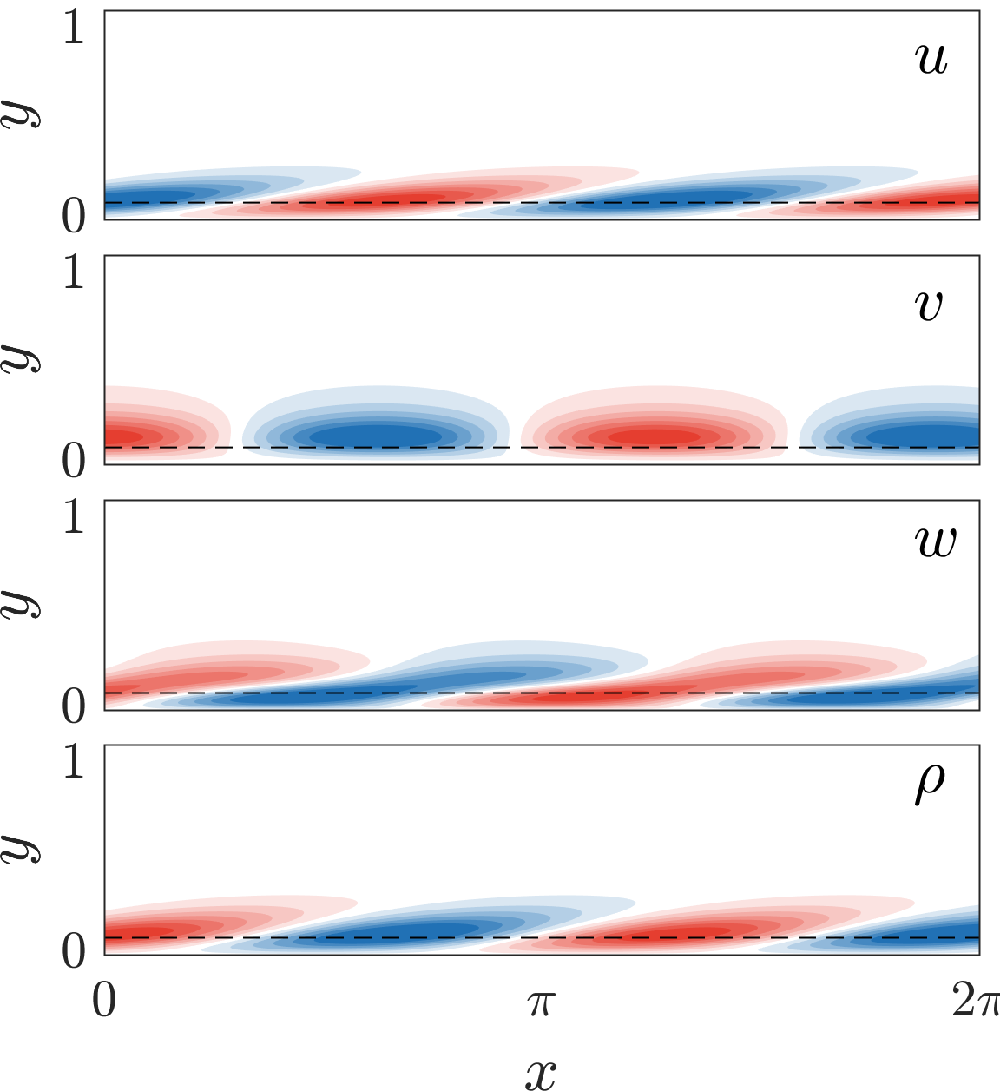}}
\hspace{0.1cm}
\subfloat[][]{\includegraphics[height=4.5cm]{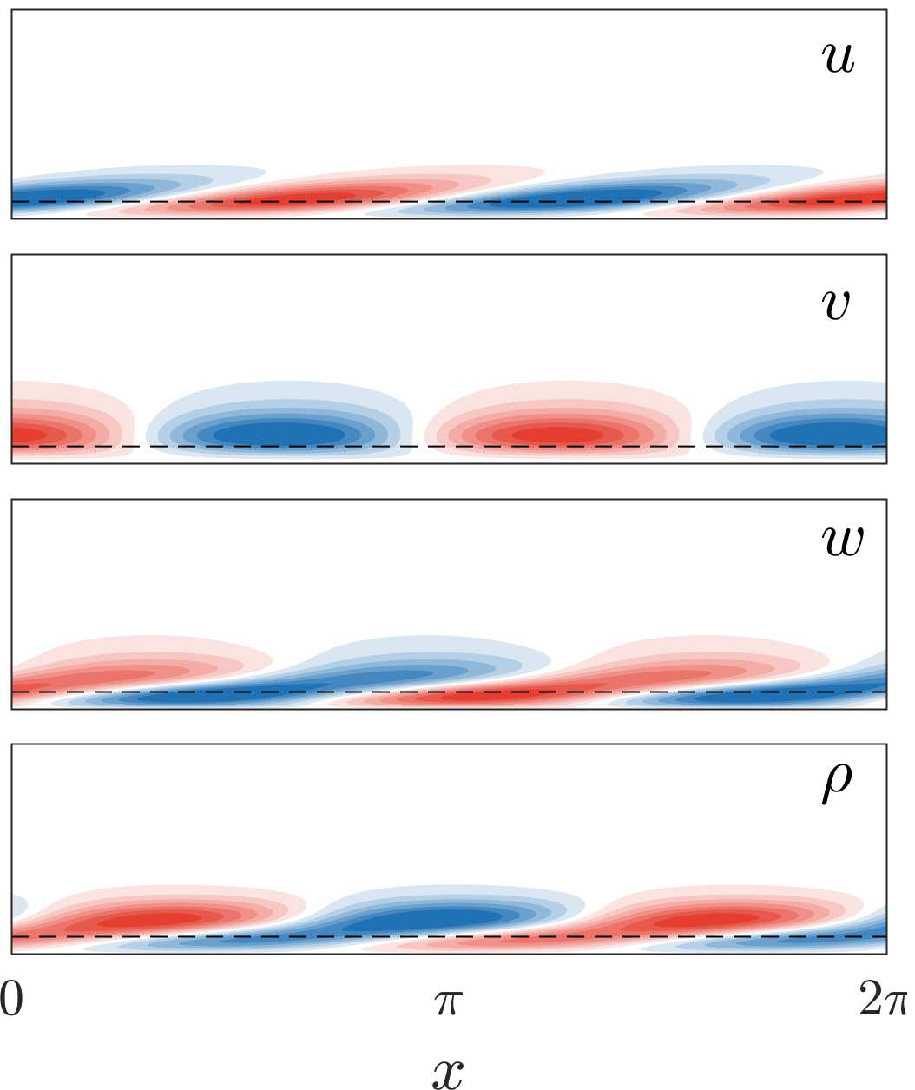}}
\hspace{0.1cm}
\subfloat[][]{\includegraphics[height=4.5cm]{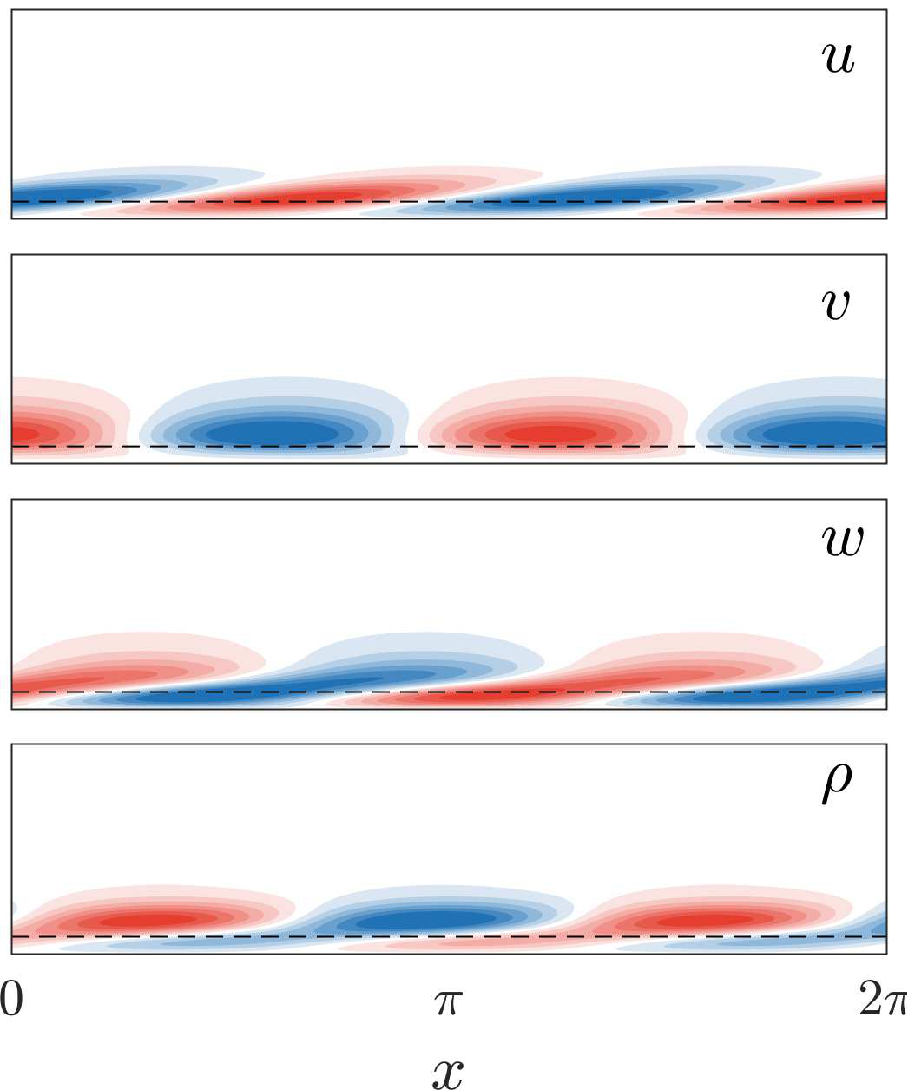}}
\caption{Two-dimensional response mode shapes for $(\kx,\kz) = (\pi/2,4\pi)$ at a critical-layer location of $\yplus=15$ for (a) $\Ritau = 0$, (b) $18$, and (c) $100$.  Red and blue contours represent positive and negative fluctuations, respectively.  The contour levels are scaled by the maximum of each mode component. The dashed black line in each sub-plot is the location of the critical-layer where $\wavespeed = \umean(\yplus=15)$.}
\label{fig:resp_yp15}
\end{figure}

\begin{figure}
\centering
\subfloat[][]{\includegraphics[height=4.5cm]{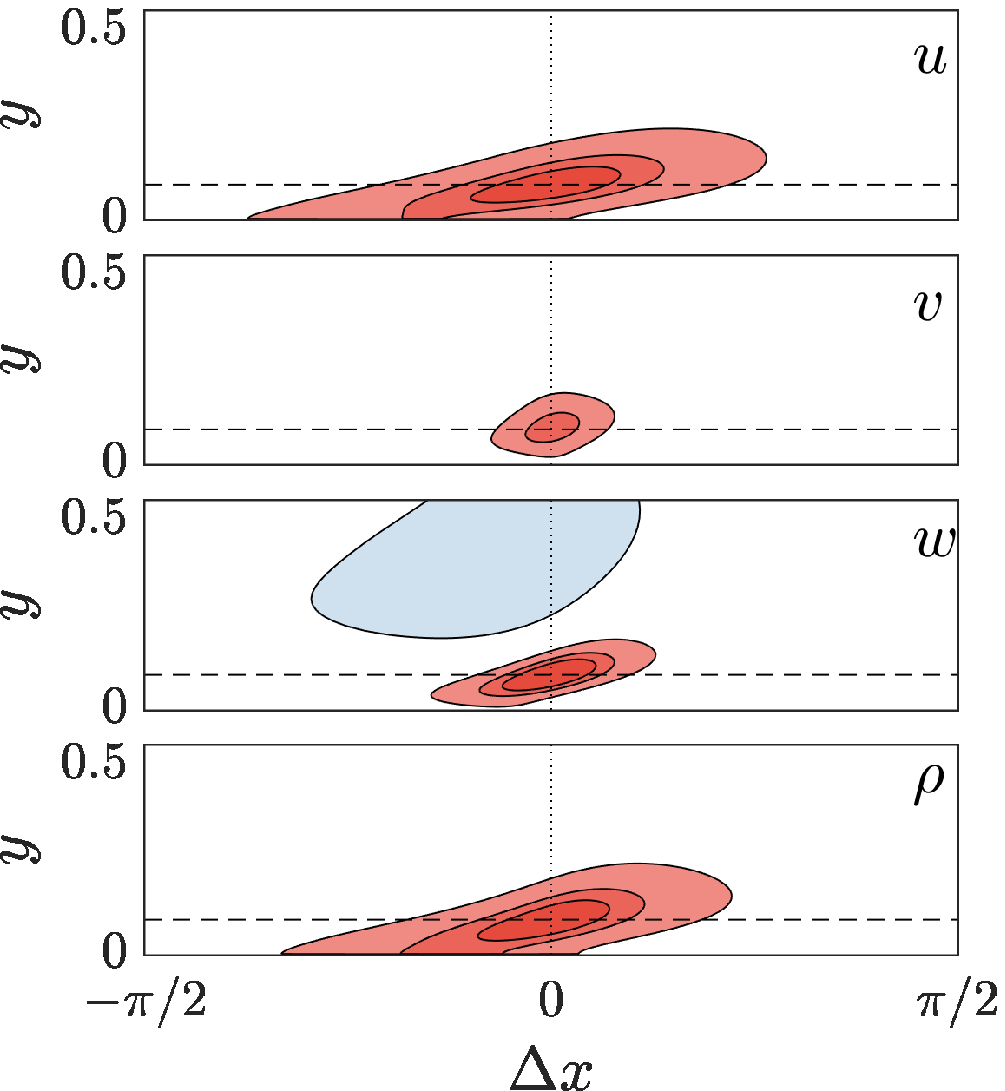}}
\hspace{0.1cm}
\subfloat[][]{\includegraphics[height=4.5cm]{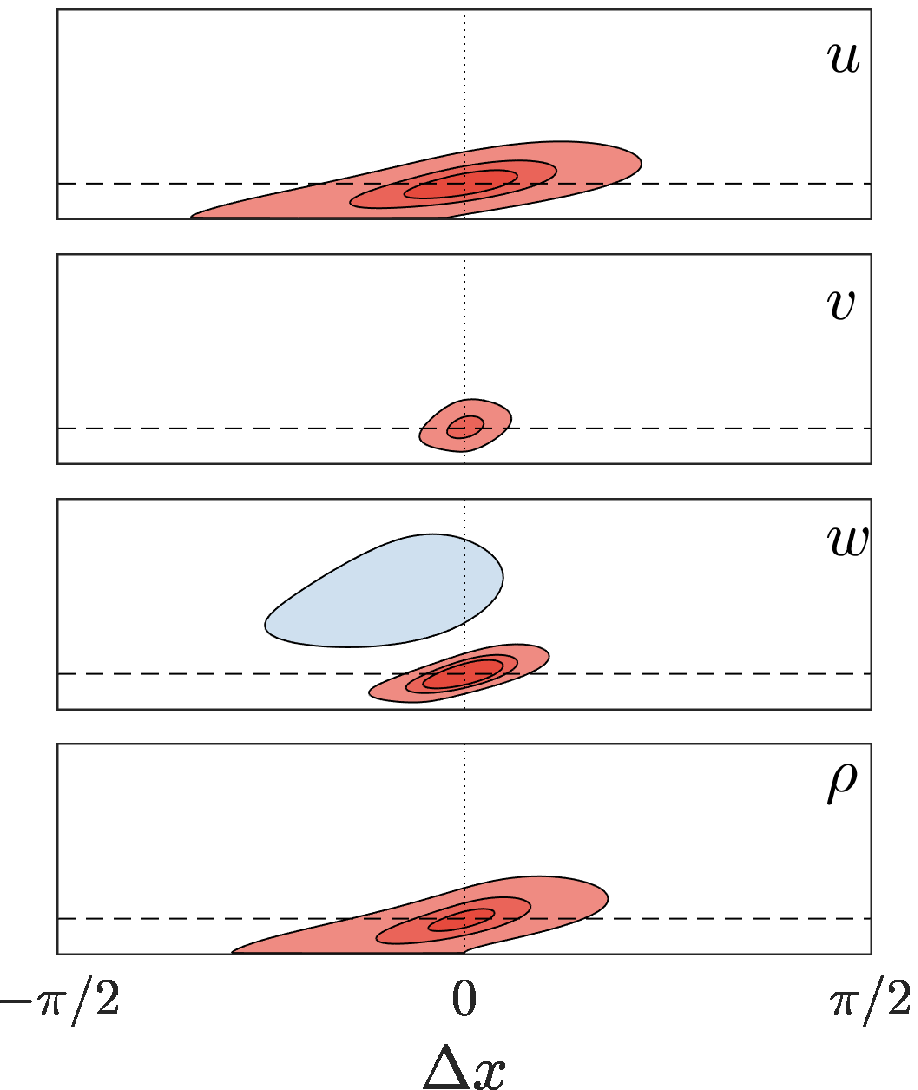}}
\caption{Autocorrelation coefficients $C_{uu}$, $C_{vv}$, $C_{ww}$ and $C_{\rho\rho}$ of the DNS at $\yplus=15$ for (a) $\Ritau = 0$ and (b) $100$.  Red and blue contours represent positive (0.4, 0.6, 0.8) and negative (-0.2) correlation, respectively, with each contour level signifying 0.2 increments. The horizontal dashed line is $\yplus=30$ and the vertical dotted line is $\Delta x = 0$.} 
\label{fig:corr_yp15}
\end{figure}

Additionally, we examine the response mode shapes in two dimensions for the different regions and compare the structures observed in the resolvent modes with the autocorrelation coefficient from the DNS data. We first define the streamwise auto-covariance as 
\begin{equation}
\hat{R}_{qq}(\kx,y,y',\kz) = \langle
\hat{q}(\kx,y,\kz)\hat{q}^*(\kx,y',\kz)
\rangle,
\end{equation}
where $q$ is a generic variable of zero mean and $\langle\cdot\rangle$ is the expected value. The auto-covariance in physical space, $R_{qq}(\Delta x,y,y',\Delta z)$, is obtained as the inverse Fourier transform of $\hat{R}$, where $\Delta x =x-x'$ and $\Delta z= z-z'$ are the distances between the two points in the homogeneous directions. The autocorrelation coefficient,
\begin{equation}
C_{qq}(\boldsymbol{x},\boldsymbol{x}') = 
\frac{R_{qq}(\boldsymbol{x},\boldsymbol{x}')}
{\varsigma_q(\boldsymbol{x})\varsigma_q(\boldsymbol{x}')},
\end{equation}
is obtained by normalising the covariance with the product of the standard deviations, $\varsigma$, at the two points involved in the measurements, which is the normalization adopted by most researchers \cite{Tritton1967,Liu2001,Ganapathisubramani2005,Lee2011,Pirozzoli2011,Sillero2014}.

The two-dimensional structures of mode E1, which coincides with the size of the near the wall structures observed previously in experiments and simulations \cite{Kline1967, Smith1983}, are plotted in Fig. \ref{fig:resp_yp15}.  The autocorrelations of the streamwise, wall-normal and spanwise velocity fields as well as the density field are shown in Fig. \ref{fig:corr_yp15}, for a two-dimensional slice at $\Delta z = 0$.  The reference location $y^{\prime+} = 15$. 

The LES of Armenio \emph{et al.}~\cite{Armenio2002} and the DNS of Garc\'ia-Villalba \& del \'Alamo~\cite{Garcia2011} demonstrated that structures in the near-wall region are largely unaffected by stable stratification. As expected, both the resolvent response modes and the correlations do not change significantly for the range of $\Ritau$ considered. For the velocities, the main difference is a reduction in the autocorrelation coefficient in the stratified case. The largest difference occurs for density properties as the phase in the wall-normal direction along the resolvent response modes are shifted, creating structures that are more detached from the wall. Similarly, the density correlations are wall-attached for $\Ritau=0$ whereas they are more detached for $\Ritau=100$.

\begin{figure}
\centering
\subfloat[][]{\includegraphics[height=4.5cm]{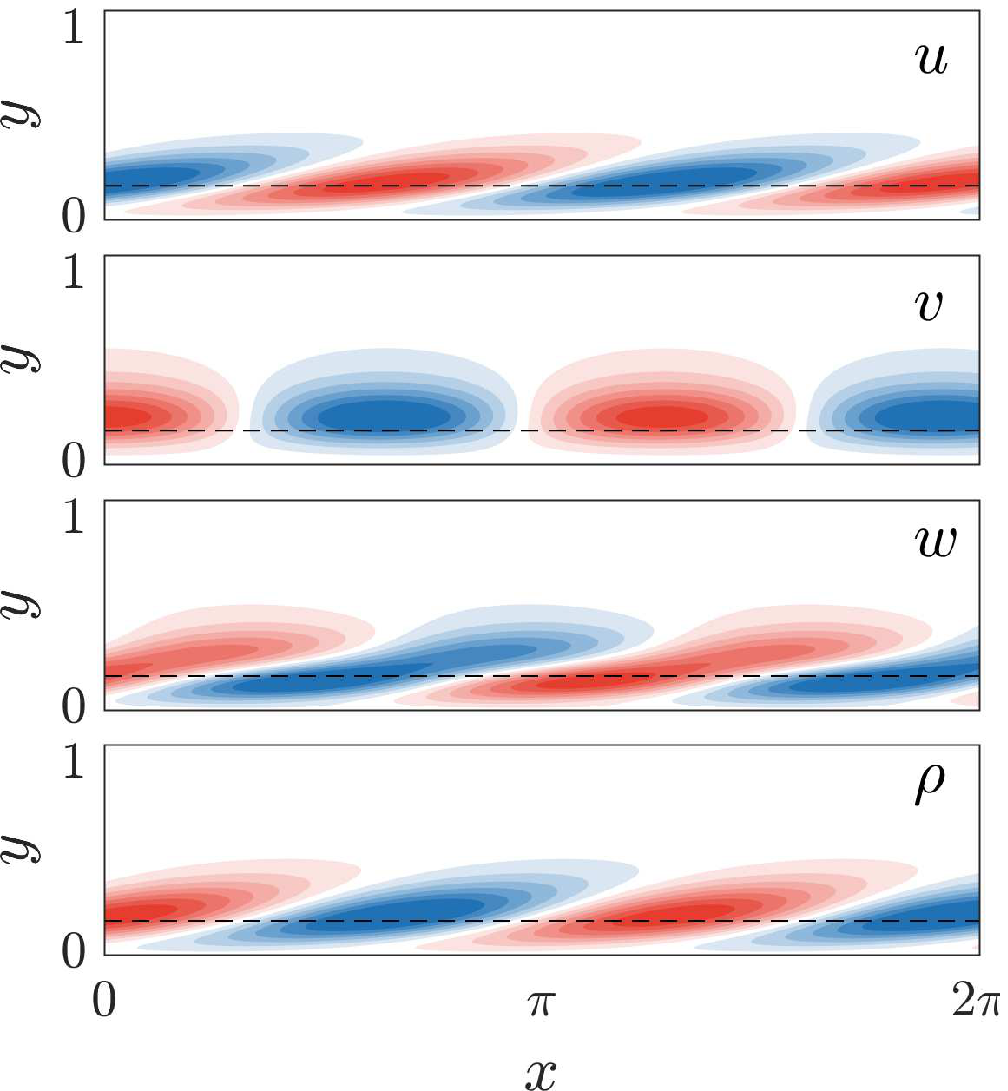}}
\hspace{0.1cm}
\subfloat[][]{\includegraphics[height=4.5cm]{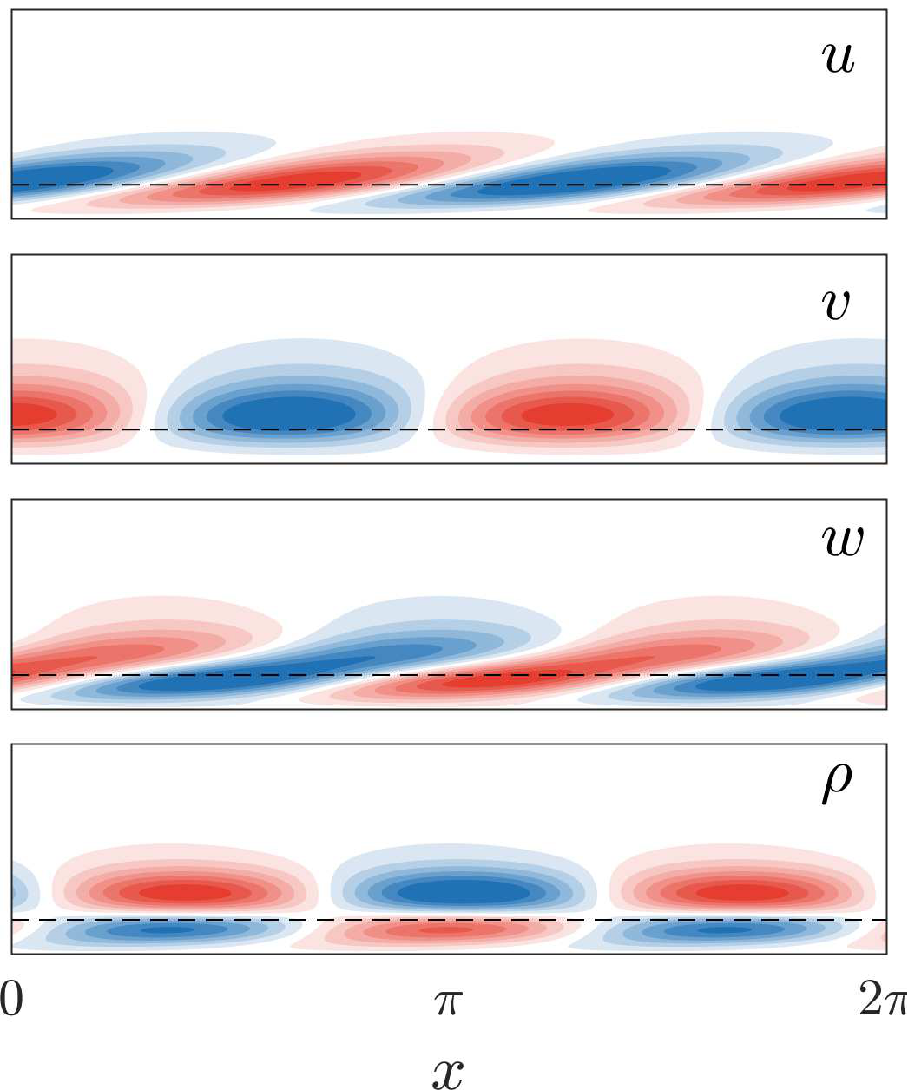}}
\hspace{0.1cm}
\subfloat[][]{\includegraphics[height=4.5cm]{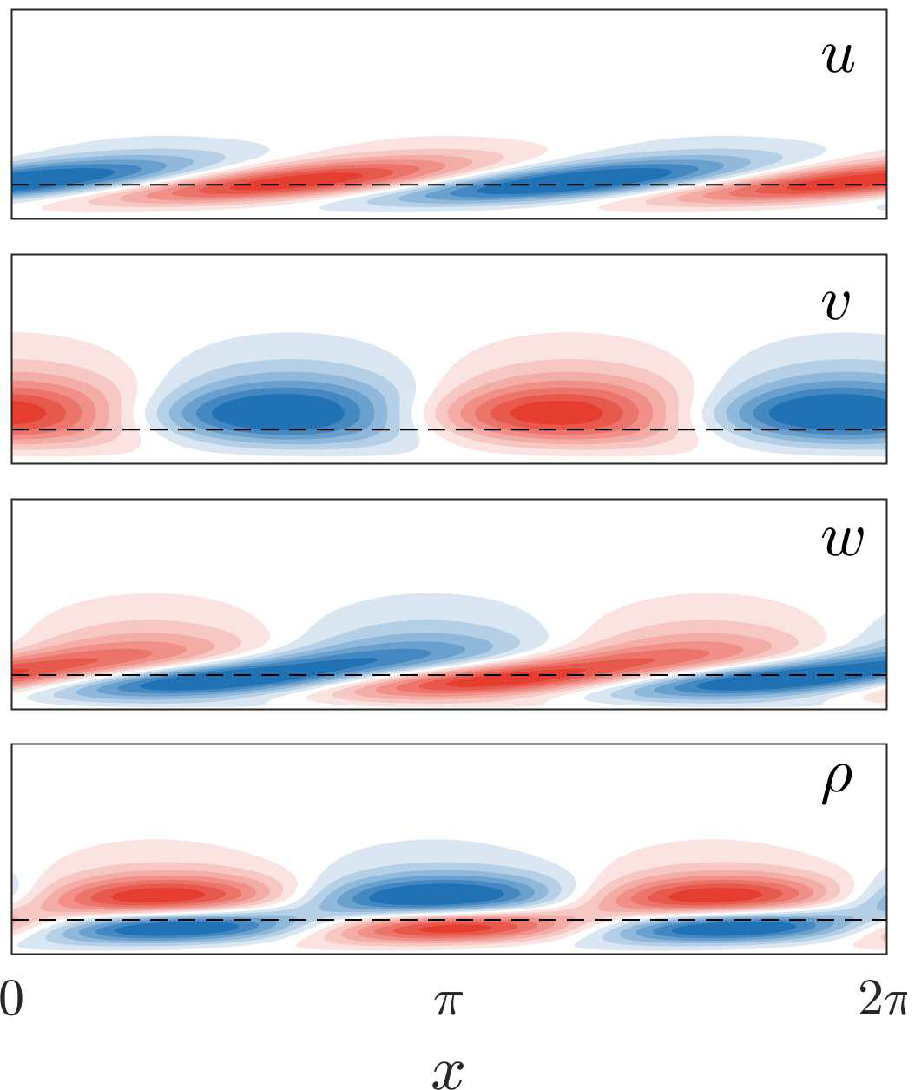}}
\caption{Two-dimensional response mode shapes for $(\kx,\kz) = (\pi/2,3\pi)$ at a critical-layer location of $\yplus=30$ for (a) $\Ritau = 0$, (b) $18$, and (c) $100$.  Red and blue contours represent positive and negative fluctuations, respectively.  The contour levels are scaled by the maximum of each mode component. The dashed black line in each sub-plot is the location of the critical-layer where $\wavespeed = \umean(\yplus=30)$.}
\label{fig:resp_yp30}
\end{figure}

\begin{figure}
\centering
\subfloat[][]{\includegraphics[height=4.5cm]{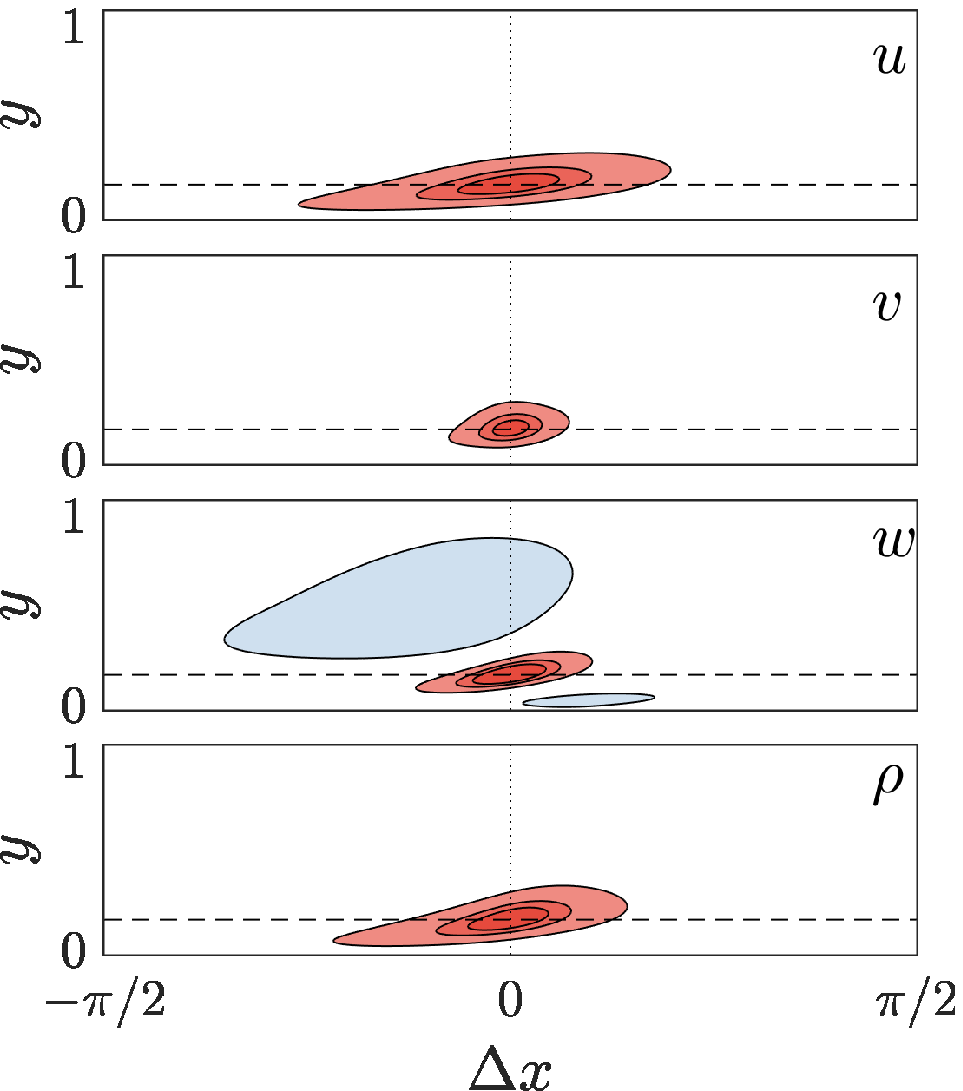}}
\hspace{0.1cm}
\subfloat[][]{\includegraphics[height=4.5cm]{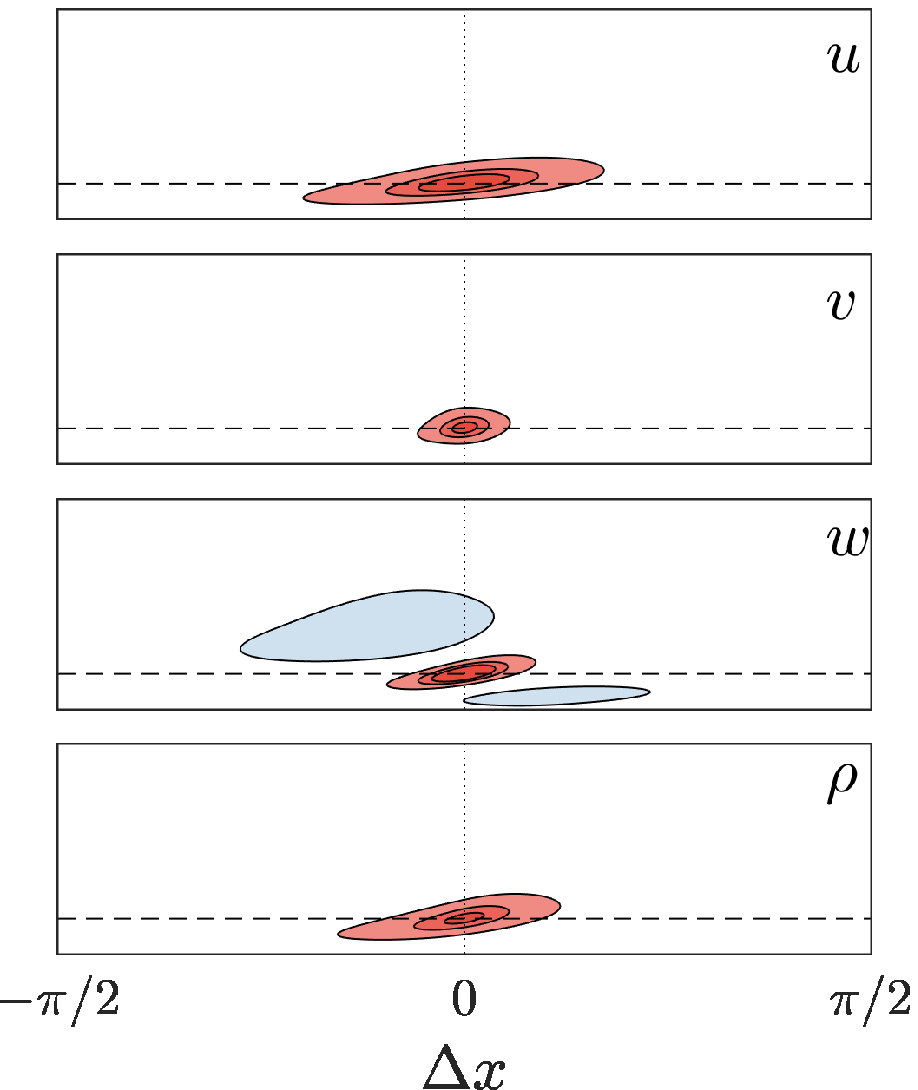}}
\caption{ Autocorrelation coefficients $C_{uu}$, $C_{vv}$, $C_{ww}$ and $C_{\rho\rho}$ of the DNS  at $\yplus=30$ for (a) $\Ritau = 0$ and (b) $100$.  Red and blue contours represent positive (0.4, 0.6, 0.8) and negative (-0.2) correlation, respectively, with each contour level signifying 0.2 increments. The horizontal dashed line is $\yplus=30$ and the vertical dotted line is $\Delta x = 0$.} \label{fig:corr_yp30}
\end{figure}

We plot the resolvent response modes for the wavenumbers and wavespeed corresponding to E2 (Fig. \ref{fig:resp_yp30}) and the correlations for $y^{\prime+}=30$ at $\Delta z = 0$ (Fig. \ref{fig:corr_yp30}). The results are similar to that of E1, since the velocity response modes do not vary across $\Ritau$, but a difference is observed in the density modes as a phase change along $y$. The correlation for the density modes are both wall-detached in the $\Ritau = 0$ and $\Ritau = 100$ case, although the centre of the density correlation for the $\Ritau = 100 $ case lies farther away from the wall.

\begin{figure}
\centering
\subfloat[][]{\includegraphics[height=4.5cm]{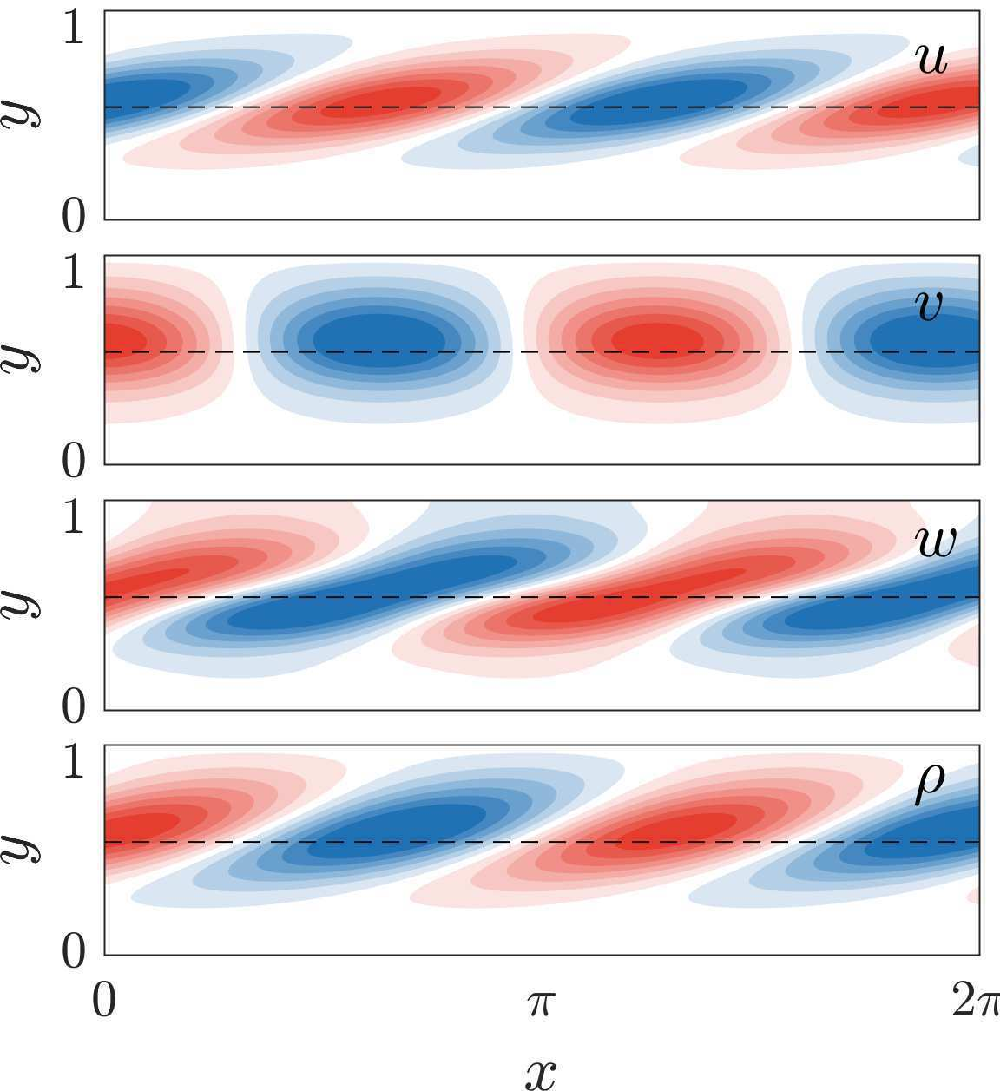}}
\hspace{0.1cm}
\subfloat[][]{\includegraphics[height=4.5cm]{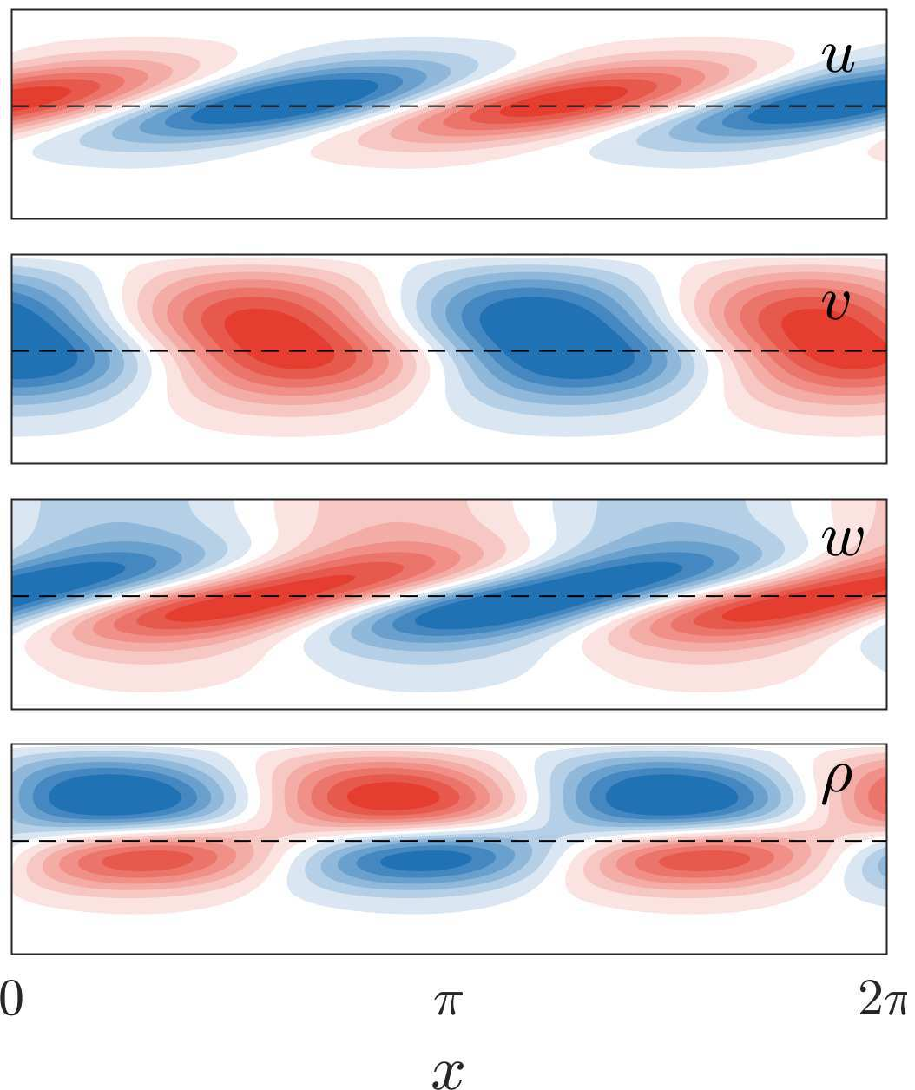}}
\hspace{0.1cm}
\subfloat[][]{\includegraphics[height=4.5cm]{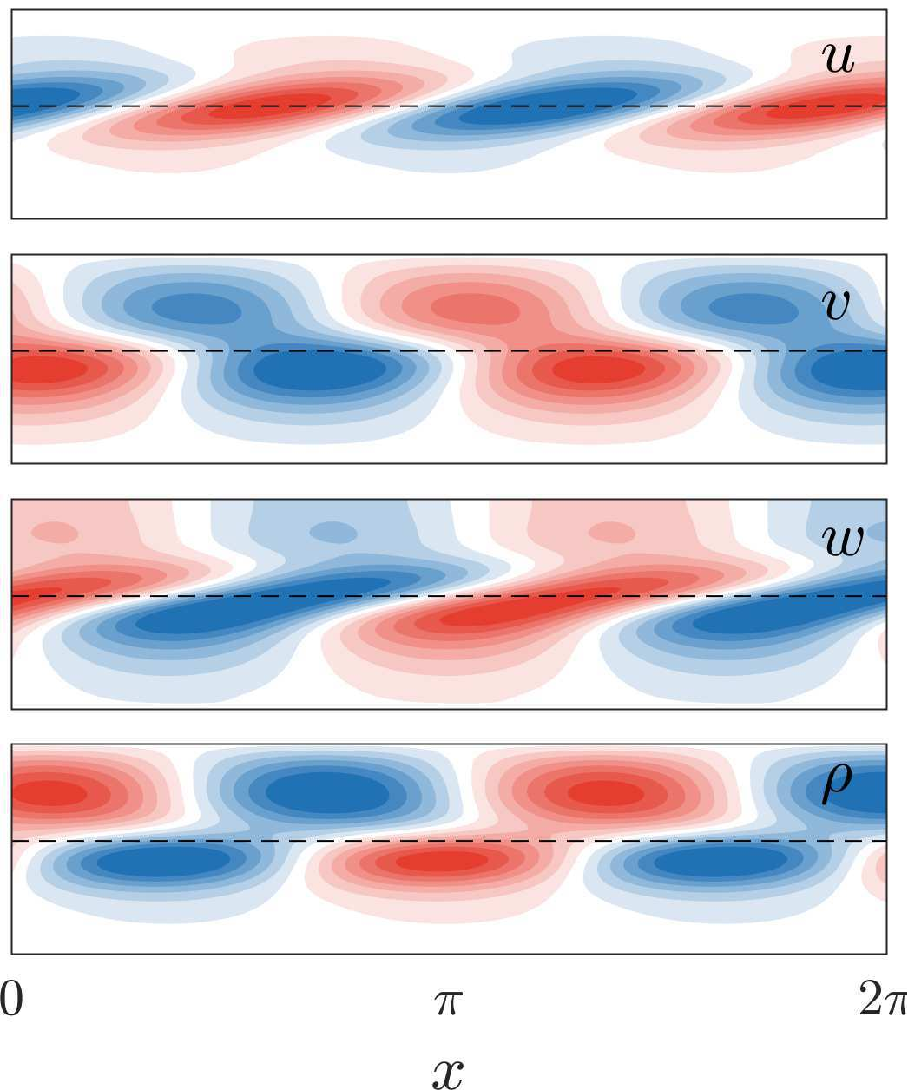}}
\caption{Two-dimensional response mode shapes for $(\kx,\kz) = (\pi/2,2\pi)$ at a critical-layer location of $\yplus=100$ for (a) $\Ritau = 0$, (b) $18$, and (c) $100$.  Red and blue contours represent positive and negative fluctuations, respectively.  The contour levels are scaled by the maximum of each mode component. The dashed black line in each sub-plot is the location of the critical-layer where $\wavespeed = \umean(\yplus=100)$.}
\label{fig:resp_yp100}
\end{figure}

\begin{figure}
\centering
\subfloat[][]{\includegraphics[height=4.5cm]{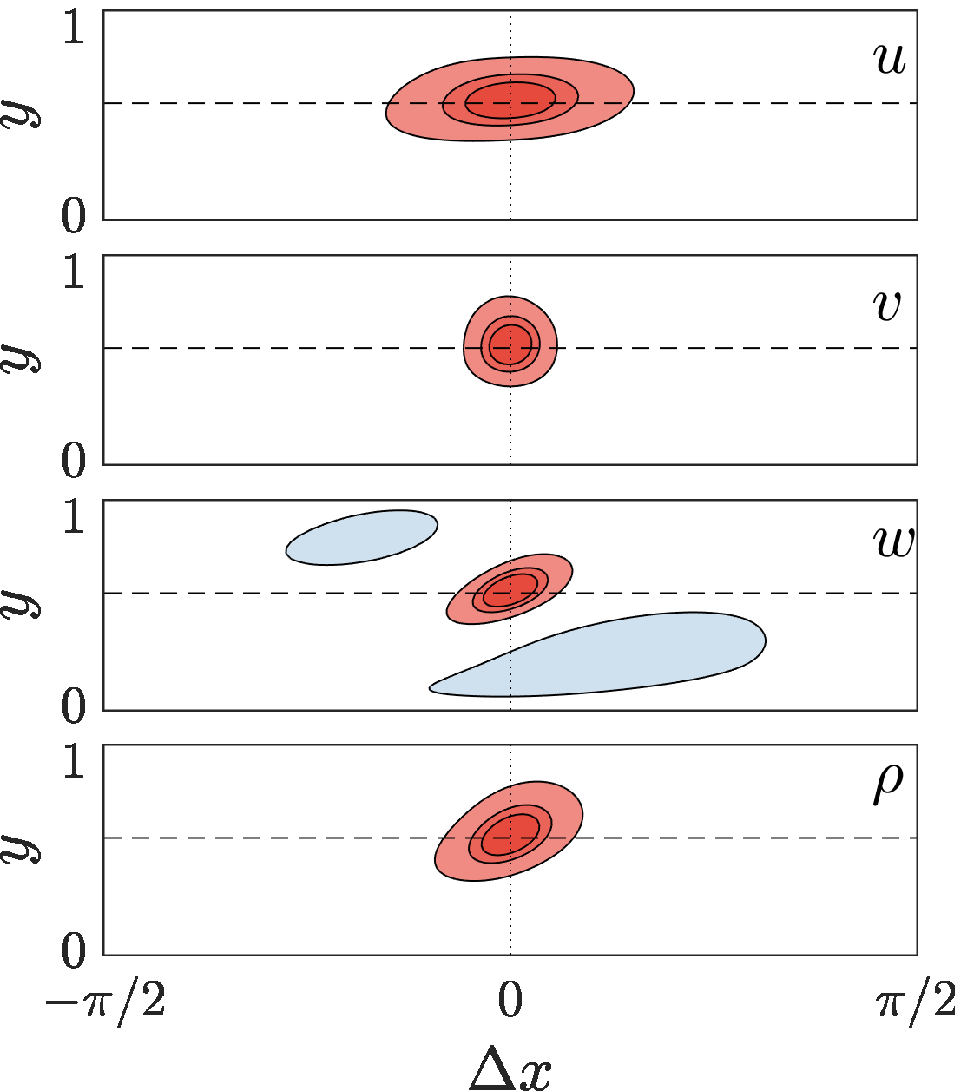}}
\hspace{0.1cm}
\subfloat[][]{\includegraphics[height=4.5cm]{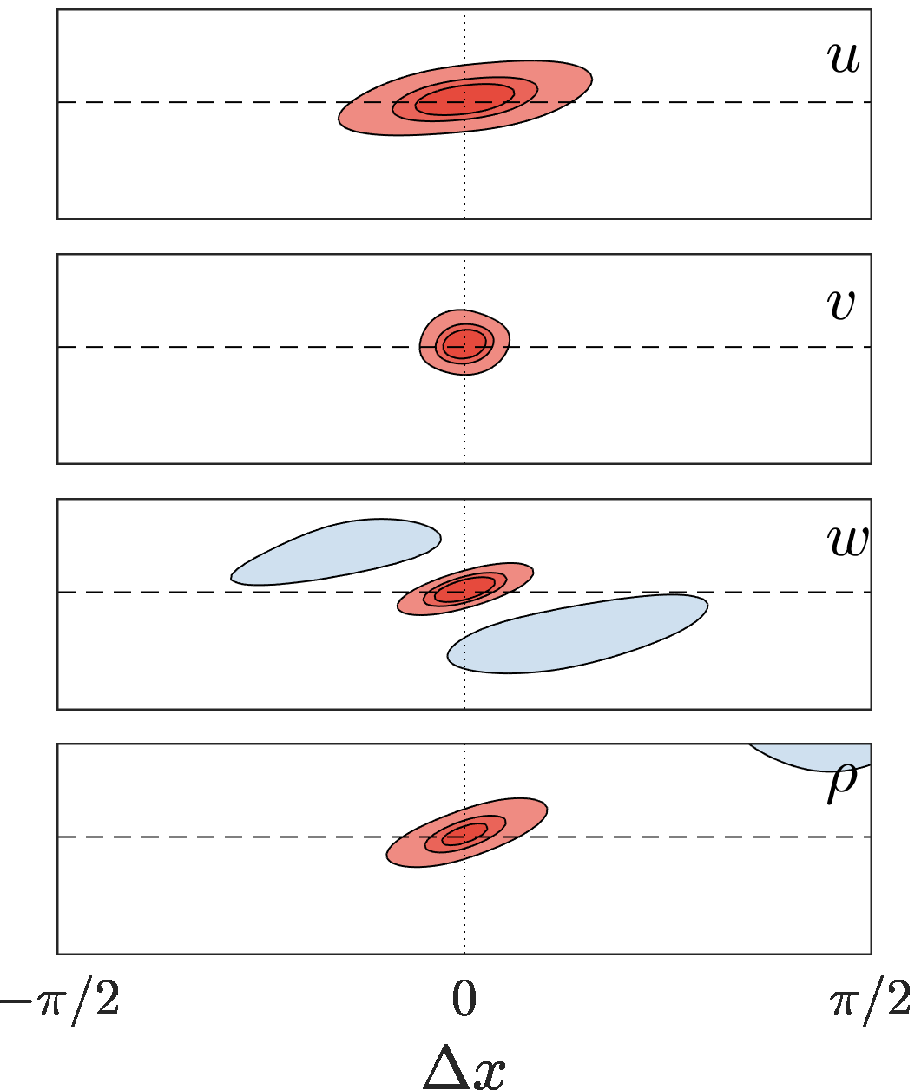}}
\caption{Autocorrelation coefficients $C_{uu}$, $C_{vv}$, $C_{ww}$ and $C_{\rho\rho}$ of the DNS at $\yplus=100$ for (a) $\Ritau = 0$ and (b) $100$.  Red and blue contours represent positive (0.4, 0.6, 0.8) and negative (-0.2) correlation, respectively, with each contour level signifying 0.2 increments. The horizontal dashed line is $\yplus=30$ and the vertical dotted line is $\Delta x = 0$.} 
\label{fig:corr_yp100}
\end{figure}

The biggest difference in the resolvent response modes for the different Richardson numbers can be seen for the wavenumber and wavespeed corresponding to E3.  We plot the resolvent response modes for the wavenumbers and wavespeed corresponding to E3 (Fig. \ref{fig:resp_yp100}) and the correlations for $y^{\prime+}=100$ at $\Delta z = 0$ (Fig. \ref{fig:corr_yp100}). 

Here, all resolvent modes show significant differences in the stratified case compared to the unstratified case. In particular, the backwards tilting of the wall-normal velocity modes, the forward tilting of the density modes, as well as the phase difference across $y$ of the density mode are pronounced. These phenomena occur in the correlations as well. There is noticeable backwards tilting in the $C_{vv}$ term and forwards tilting in the $C_{\rho\rho}$ term for $\Ritau = 100$ compared to the neutrally stratified case. The biggest differences come in the form of the wall-normal and density models because they are coupled through the Richardson number in the stratified Navier-Stokes equations.

\subsection{Energy balance at selected scales}\label{sec:energy_balance}

\begin{figure}
\centering
\includegraphics[width=0.9\textwidth]{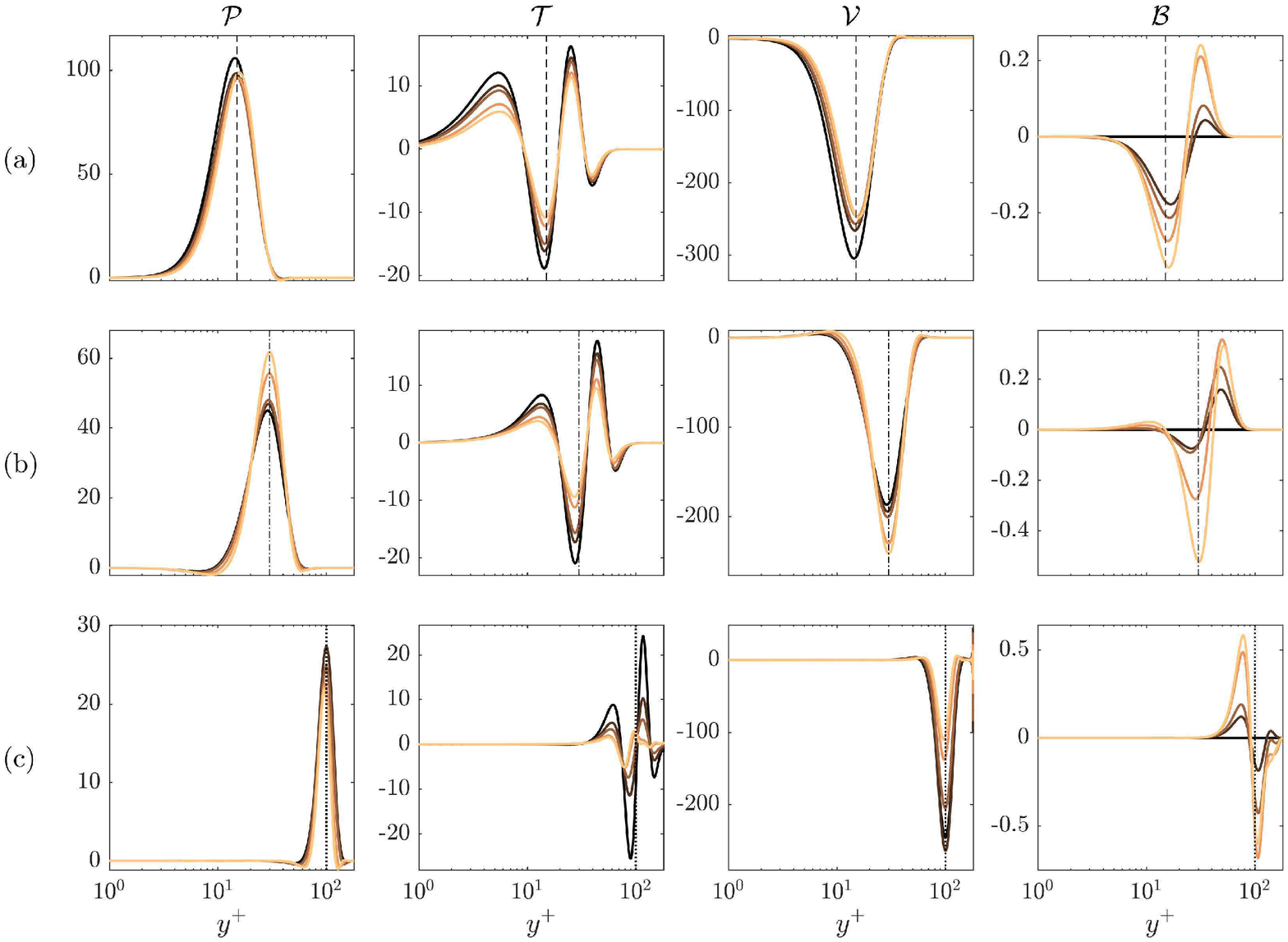}
\caption{Energy budget terms computed from resolvent modes Eq. (\ref{eqn:prod_k}--\ref{eqn:diff_k}) for wavenumbers given by (a) E1 at $\wavespeed=\umean(\yplus=15)$, (b) E2 at $\wavespeed=\umean(\yplus=30)$, and (c) E3 at $\wavespeed=\umean(\yplus=100)$ for $\Ritau = 0, 10,  18, 60, 100$ (darker to lighter).}
\label{fig:energy_budget_res}
\end{figure}

\begin{figure}
\centering
\includegraphics[width=0.9\textwidth]{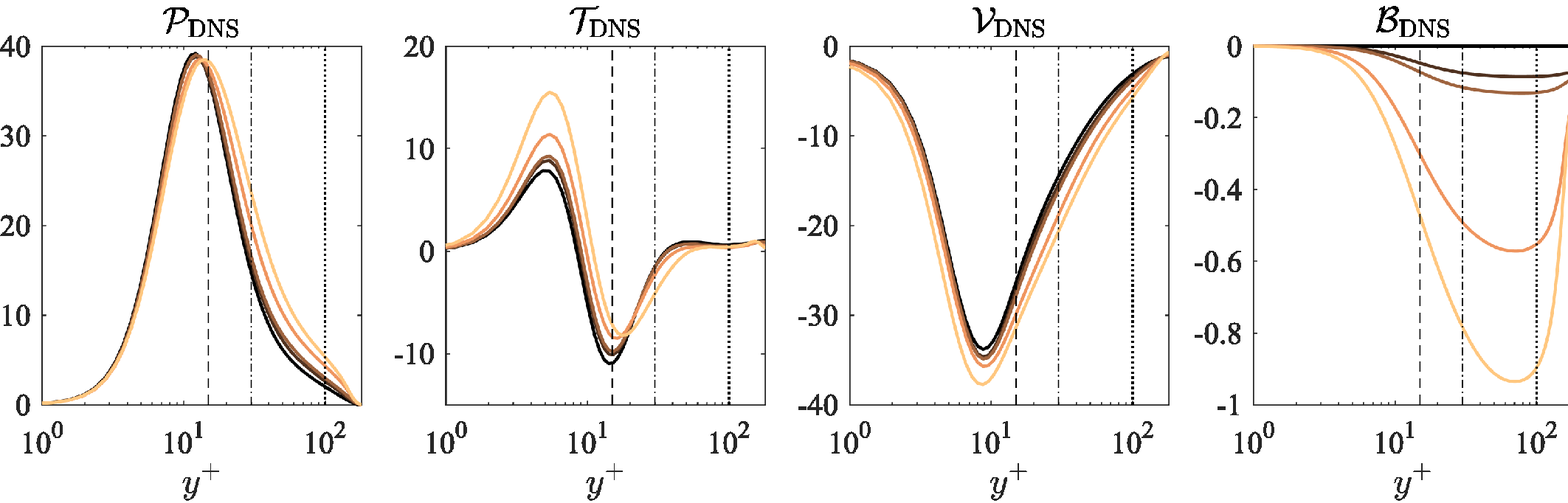}
\caption{(Energy budget terms computed from the DNS for $\Ritau = 0, 10,  18, 60, 100$ (darker to lighter).}
\label{fig:energy_budget_DNS}
\end{figure}

Finally, we study the energy budget terms of the stratified channel. We define the production, transport, buoyancy flux, and viscous dissipation budget terms in the resolvent formulation \cite{Symon21,madhusudanan2020coherent} as 
\begin{subequations}
\begin{align}
\mathcal{P}_{\text{tot}}(y) &=  \mathbb{R}\left[ 
-\frac{\partial \umean}{\partial y}
\sum_j\int_{-\infty}^{\infty} 
\sigma_j^2\chi_j^2 \Big(
\bs{\hat{\psi}}^{*}_{j,u}\bs{\hat{\psi}}_{j,v}  
\Big) d\bk \right],
\label{eqn:prod}\\
\mathcal{T}_{\text{tot}}(y) &=  \mathbb{R}\left[
\sum_j\sum_i\int_{-\infty}^{\infty}  
\sigma_j\chi_j\chi_i \D\Big(
\bs{\hat{\phi}}^{*}_{i,u}\bs{\hat{\psi}}_{j,v}  +
\bs{\hat{\phi}}^{*}_{i,v}\bs{\hat{\psi}}_{j,v} +
\bs{\hat{\phi}}^{*}_{i,w}\bs{\hat{\psi}}_{j,v}  
\Big) d\bk \right],
\label{eqn:transp}\\
\mathcal{B}_{\text{tot}}(y) &= \mathbb{R}\left[ 
-\Rifric
\sum_j\int_{-\infty}^{\infty} 
\sigma_j^2\chi_j^2 \Big(
\bs{\hat{\psi}}^{*}_{j,v}\bs{\hat{\psi}}_{j,\rfluc}  
\Big) d\bk \right],
\label{eqn:buoy}\\
\mathcal{V}_{\text{tot}}(y) &= \mathbb{R}\left[
\frac{1}{\Retau} 
\sum_j\int_{-\infty}^{\infty}
\sigma_j^2\chi_j^2 \Big(
\bs{\hat{\psi}}^{*}_{j,u}\hat\Delta\bs{\hat{\psi}}_{j,u} +
\bs{\hat{\psi}}^{*}_{j,v}\hat\Delta\bs{\hat{\psi}}_{j,v} +
\bs{\hat{\psi}}^{*}_{j,w}\hat\Delta\bs{\hat{\psi}}_{j,w}  
\Big) d\bk \right],
\end{align}
\end{subequations}
where $\chi_j$, $\sigma_j$, $\bs{\hat{\psi}}_j$ and $\bs{\hat{\phi}}_j$ are functions of $\bk$ and the subscript $u,v,w,\rfluc$ indicate the corresponding components of the response or forcing mode. To get a global sense of the energy balance, the equations above are integrated over all wavenumber triplets. Here, we will examine only the principal resolvent mode contribution to the local components of the total budgets for particular $\bk$, defined as  
\begin{subequations}
\begin{align}
\mathcal{P}(y,\bk) &=  \mathbb{R}\left[ 
-\frac{\partial \umean}{\partial y}
\sigma_1^2 \Big(
\bs{\hat{\psi}}^{*}_{1,u}\bs{\hat{\psi}}_{1,v} 
\Big) \right],
\label{eqn:prod_k}\\
\mathcal{T}(y,\bk) &=  \mathbb{R}\left[
\sigma_1\D\Big(
\bs{\hat{\phi}}^{*}_{1,u}\bs{\hat{\psi}}_{1,v} +
\bs{\hat{\phi}}^{*}_{1,v}\bs{\hat{\psi}}_{1,v} +
\bs{\hat{\phi}}^{*}_{1,w}\bs{\hat{\psi}}_{1,v} 
 \Big) \right],
\label{eqn:transp_k}\\
\mathcal{B}(y,\bk) &= \mathbb{R}\left[ 
-\Rifric 
\sigma_1^2\Big(
\bs{\hat{\psi}}^{*}_{1,v}\bs{\hat{\psi}}_{1,\rfluc}  
\Big) \right],
\label{eqn:buoy_k}\\
\mathcal{V}(y,\bk) &= \mathbb{R}\left[
\frac{1}{\Retau}
\sigma_1^2\Big(
\bs{\hat{\psi}}^{*}_{1,u}\hat\Delta\bs{\hat{\psi}}_{1,u} +
\bs{\hat{\psi}}^{*}_{1,v}\hat\Delta\bs{\hat{\psi}}_{1,v} +
\bs{\hat{\psi}}^{*}_{1,w}\hat\Delta\bs{\hat{\psi}}_{1,w}  
\Big) \right].
\label{eqn:diff_k}
\end{align}
\end{subequations}
The results for wavenumber combinations E1, E2 and E3 are shown in Fig. \ref{fig:energy_budget_res}. Since the wavenumber combinations E1, E2 and E3 are the most energetic at each wavespeed, we predict that the local components of the budget term should indicate the overall trend of the total budget term at the corresponding wall-normal height. These quantities are compared to the energy budget computed from the DNS, shown in Fig. \ref{fig:energy_budget_DNS}. 

The trends observed in the energy budget computed from the DNS are also recovered in the resolvent budgets. The production is mostly balanced by viscous dissipation and has larger magnitudes compared to the transport (approximately 10\% of the production term) or buoyancy flux (approximately 0.1-1\%, depending on $\Ritau$ of the production term) terms.  Comparing the quantities at the wall-normal heights of interest, we see that at $y^+=15$, there is little variation in the production and viscous dissipation terms in both DNS and resolvent modes. The difference in relative magnitude over the various values of $\Ritau$ increases farther away from the wall, and at $y^+=100$, the production (and viscous diffusion) of the $\Ritau = 100$ case is double the production (and viscous diffusion) of the neutrally-buoyant case in both the DNS and resolvent. 

Direct comparison of the integrated magnitudes is more difficult for the transport and buoyancy flux terms as they are not uniformly positive or negative. However, this indicates that, locally, buoyancy flux acts as a energy transfer term, much like the turbulent transport, as the term adds energy in one wall-normal location and removes it from another. Because the DNS energy budget is integrated for all spatio-temporal scales, it is impossible to deduce that the buoyance flux term acts as a local energy transfer term from Fig. \ref{fig:energy_budget_DNS}, which shows a net negative energy balance from $\mathcal{B}$ at all wall-normal locations. In contrast, the resolvent buoyancy flux term indicates a non-monotonic distribution of energy in the wall-normal direction. Similar results could be obtained through spatio-temporal deconstruction of the DNS energy budget term as in Ref. \cite{Mizuno2016}, but this would require a time-resolved dataset for a longer time domain. The resolvent turbulent transport stays relatively similar among different $\Ritau$, as does the turbulent transport term from DNS. The buoyancy flux is much more dependent on $\Ritau$, with variations becoming greater farther away from the wall in both the DNS and resolvent results.  

\begin{figure}
\centering
\subfloat[][]{\includegraphics[width=0.42\textwidth]{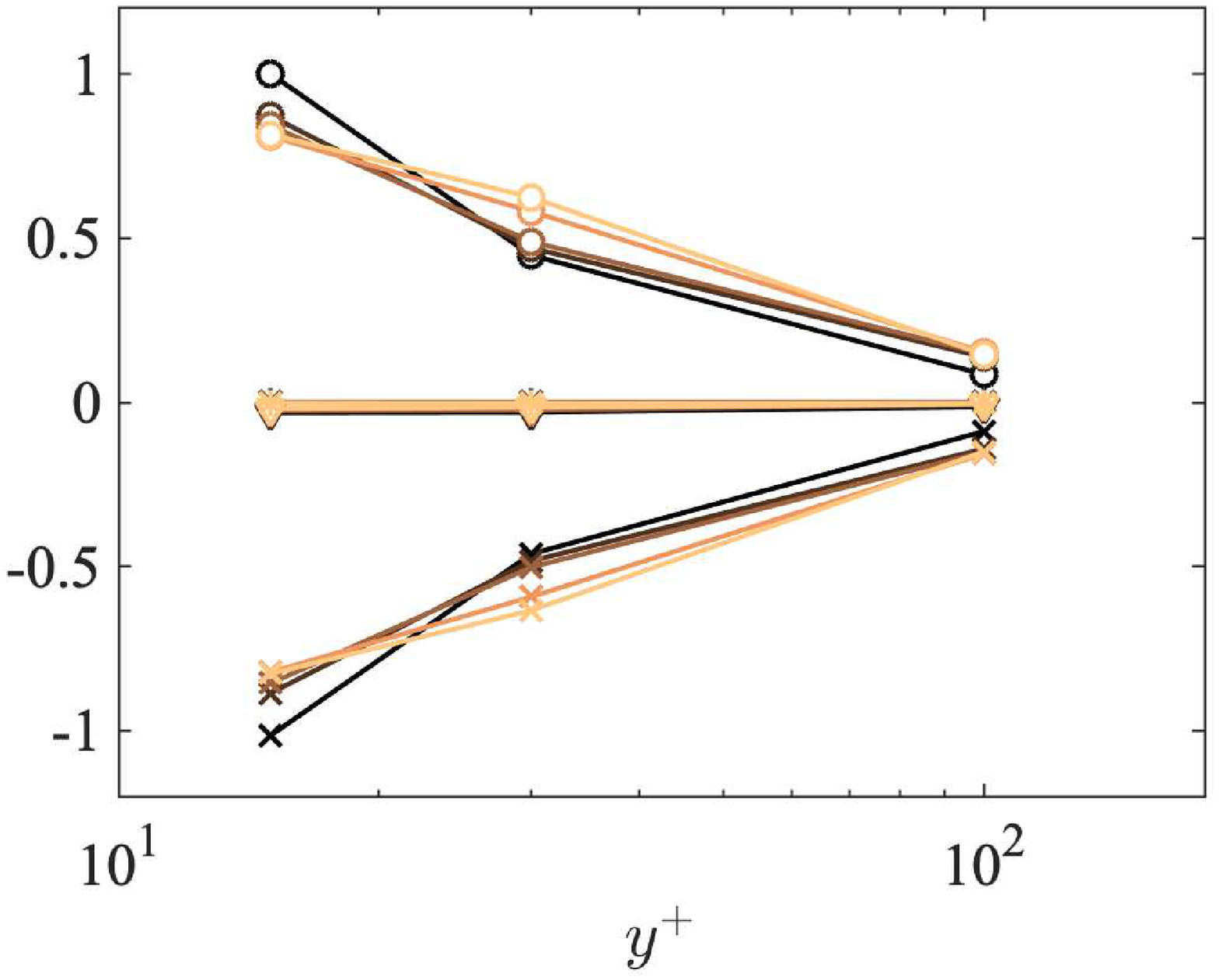}}
\hspace{0.1cm}
\subfloat[][]{\includegraphics[width=0.42\textwidth]{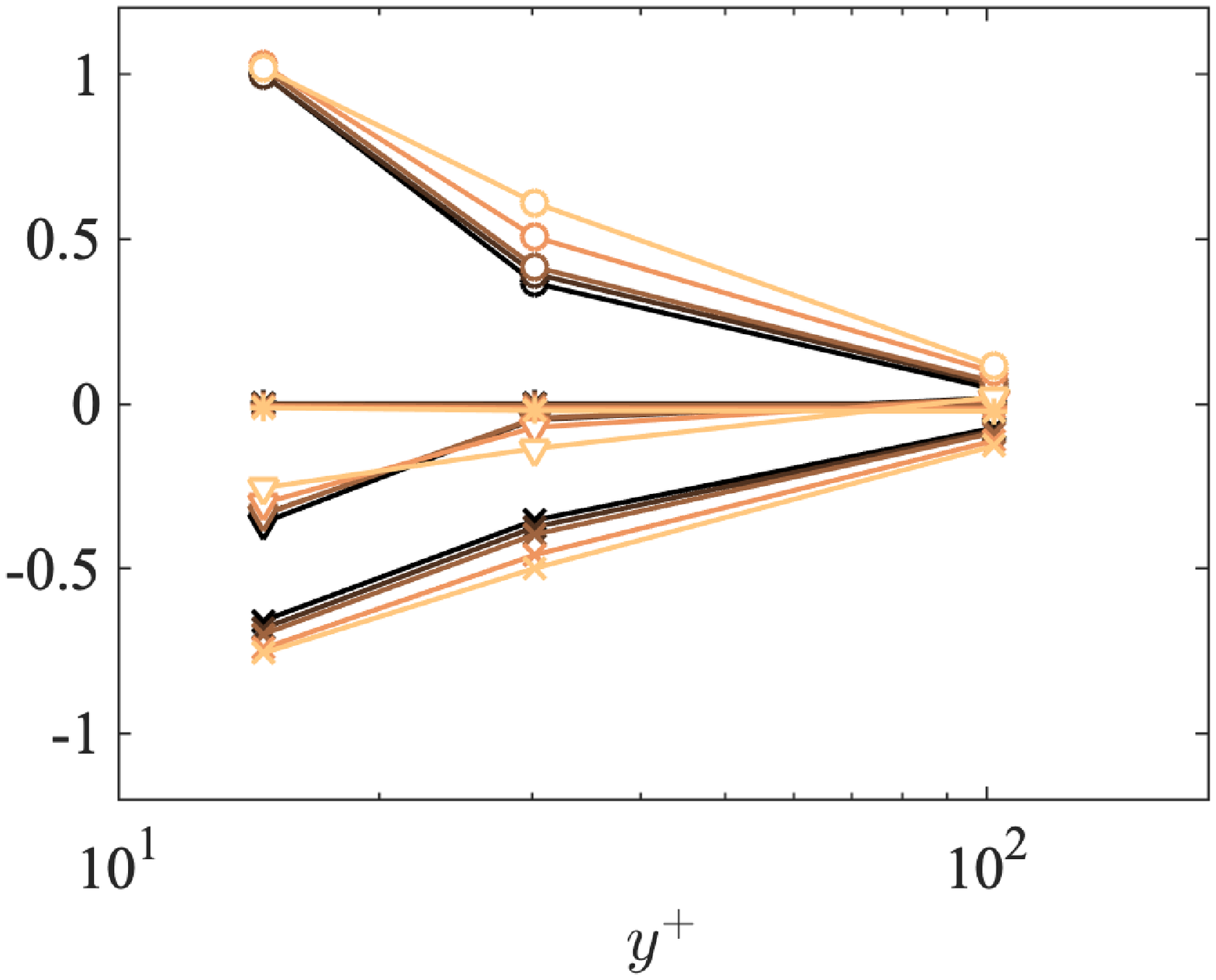}}
\caption{(a) Resolvent energy budget terms for wavenumber combinations E1, E2 and E3 evaluated at $y^+=15,30,100$, respectively, for $\Ritau=0,10,18,60,100$ (darker to lighter) normalized with $\mathcal{P}(y^+=15)$ for $\Ritau = 0$ and wavenumber combination E1. (b) DNS energy budget terms at $y^+=15,30,100$ for $\Ritau=0,10,18,60,100$ (darker to lighter) normalized with $\mathcal{P}_\text{DNS}(y^+=15)$ for $\Ritau = 0$.  Symbols are $\mathcal{P}$, circles; $\mathcal{T}$, triangles; $\mathcal{V}$, crosses; and $\mathcal{B}$, asterisks.}
\label{fig:budget_ratio}
\end{figure}

These results can be better quantified by plotting the values at each wall-normal location normalized by the peak production at $y^+=15$ for each case, as shown in Fig. \ref{fig:budget_ratio}. This shows that the overall trend of the budget terms are well captured by the resolvent budget terms, with the exception of the transport term close to the wall. This discrepancy may be attenuated by integrating over more wavespeeds.  

Note that the results are not expected to match that of DNS for all scales as the the energy captured in the wall-parallel resolvent modes are known to be overpredicted and the energy captured in the Reynolds stress and wall-normal resolvent modes underpredicted. This is a known issue for the resolvent analysis in the primitive variables due to the competing mechanisms of the Squire modes with the Orr-Sommerfeld modes \cite{Moarref2014,Rosenberg18}. Additionally, the underprediction of energy captured in the Reynolds stress and wall-normal resolvent modes could explain the underprediction of the transport term close to the wall.  Crucially, though, the most energetic scale can reproduce the integrated effect of all scales, which enables a quick predictive model of stratified boundary layers.   

\section{Conclusions}\label{sec:conclusions}

The resolvent framework for the Navier-Stokes equations with the Boussinesq approximation was applied to a stratified turbulent boundary layer. Computation of the leading resolvent modes is more cost-effective than performing a full-scale simulation or experiment, while being able to provide information on the flow. This quick model can provide meaningful insight into stratified flows with only information about the mean profile and prior knowledge of energetic scales of motion in the neutrally-buoyant boundary layers.

The results show that despite using only a very limited range of representative scales, the resolvent model was able to reproduce the relative magnitude of turbulence intensities and the balance of the energy budget as well as provide meaningful analysis of structures in the flow. We studied the amplitude of the resolvent response modes and their two-dimensional mode shapes of the rank-one approximation, which were then compared to the turbulence intensities and the two-dimensional auto-correlation of the velocity and density fields of the DNS, respectively. The resolvent response modes were able to predict the relative variation in turbulence intensities as a function of wall-normal distance and Richardson number for the $\Ritau$ under consideration in this study. The two-dimensional mode shapes also provided  insight into how the auto-correlation coefficient might shift as a function of $\Ritau$. Finally, the energy budget terms for the turbulent kinetic energy of the system were computed both using the rank-one approximation of the resolvent analysis and the DNS data. Again, the resolvent energy budget predicts well the relative distribution of energy between production, dissipation, transport, and buoyancy flux as a function of wall-normal distance and Richardson number. 

In the current study, the resolvent model was closed using mean velocity and density profiles obtained from DNS. The computational cost of calculating the forcing and response modes at certain scales was on the order of seconds on a laptop. Therefore, by obtaining only mean velocity and scalar profiles we could generate the salient modal structure for a given stratified wall-bounded flow. The next steps involve using in-situ data to generate modes that are representative of flow phenomena observed in nature. 

\section*{Acknowledgements}
The support of a Vannevar Bush Faculty Fellowship administered under
the U.S. Office of Naval Research, grant \#N00014-17-1-3022, is
gratefully acknowledged.  Additionally, the authors would like to
thank Dr.\ Angeliki Laskari for insightful discussions.

\bibliography{stratified_resolvent}

\begin{thebibliography}{73}%
\makeatletter
\providecommand \@ifxundefined [1]{%
 \@ifx{#1\undefined}
}%
\providecommand \@ifnum [1]{%
 \ifnum #1\expandafter \@firstoftwo
 \else \expandafter \@secondoftwo
 \fi
}%
\providecommand \@ifx [1]{%
 \ifx #1\expandafter \@firstoftwo
 \else \expandafter \@secondoftwo
 \fi
}%
\providecommand \natexlab [1]{#1}%
\providecommand \enquote  [1]{``#1''}%
\providecommand \bibnamefont  [1]{#1}%
\providecommand \bibfnamefont [1]{#1}%
\providecommand \citenamefont [1]{#1}%
\providecommand \href@noop [0]{\@secondoftwo}%
\providecommand \href [0]{\begingroup \@sanitize@url \@href}%
\providecommand \@href[1]{\@@startlink{#1}\@@href}%
\providecommand \@@href[1]{\endgroup#1\@@endlink}%
\providecommand \@sanitize@url [0]{\catcode `\\12\catcode `\$12\catcode
  `\&12\catcode `\#12\catcode `\^12\catcode `\_12\catcode `\%12\relax}%
\providecommand \@@startlink[1]{}%
\providecommand \@@endlink[0]{}%
\providecommand \url  [0]{\begingroup\@sanitize@url \@url }%
\providecommand \@url [1]{\endgroup\@href {#1}{\urlprefix }}%
\providecommand \urlprefix  [0]{URL }%
\providecommand \Eprint [0]{\href }%
\providecommand \doibase [0]{https://doi.org/}%
\providecommand \selectlanguage [0]{\@gobble}%
\providecommand \bibinfo  [0]{\@secondoftwo}%
\providecommand \bibfield  [0]{\@secondoftwo}%
\providecommand \translation [1]{[#1]}%
\providecommand \BibitemOpen [0]{}%
\providecommand \bibitemStop [0]{}%
\providecommand \bibitemNoStop [0]{.\EOS\space}%
\providecommand \EOS [0]{\spacefactor3000\relax}%
\providecommand \BibitemShut  [1]{\csname bibitem#1\endcsname}%
\let\auto@bib@innerbib\@empty
\bibitem [{\citenamefont {McKeon}\ and\ \citenamefont
  {Sharma}(2010)}]{McKeon2010}%
  \BibitemOpen
  \bibfield  {author} {\bibinfo {author} {\bibfnamefont {B.~J.}\ \bibnamefont
  {McKeon}}\ and\ \bibinfo {author} {\bibfnamefont {A.~S.}\ \bibnamefont
  {Sharma}},\ }\bibfield  {title} {\bibinfo {title} {A critical-layer framework
  for turbulent pipe flow},\ }\href@noop {} {\bibfield  {journal} {\bibinfo
  {journal} {J.~Fluid Mech.}\ }\textbf {\bibinfo {volume} {658}},\ \bibinfo
  {pages} {336} (\bibinfo {year} {2010})}\BibitemShut {NoStop}%
\bibitem [{\citenamefont {Dawson}\ \emph {et~al.}(2018)\citenamefont {Dawson},
  \citenamefont {Saxton-Fox},\ and\ \citenamefont {McKeon}}]{Dawson2018}%
  \BibitemOpen
  \bibfield  {author} {\bibinfo {author} {\bibfnamefont {S.~T.}\ \bibnamefont
  {Dawson}}, \bibinfo {author} {\bibfnamefont {T.}~\bibnamefont {Saxton-Fox}},\
  and\ \bibinfo {author} {\bibfnamefont {B.~J.}\ \bibnamefont {McKeon}},\
  }\bibfield  {title} {\bibinfo {title} {Modeling passive scalar dynamics in
  wall-bounded turbulence using resolvent analysis},\ }in\ \href@noop {} {\emph
  {\bibinfo {booktitle} {2018 AIAA Fluid Dynamics Conference}}}\ (\bibinfo
  {year} {2018})\ p.\ \bibinfo {pages} {4042}\BibitemShut {NoStop}%
\bibitem [{\citenamefont {Nieuwstadt}(1984)}]{Nieuwstadt1984}%
  \BibitemOpen
  \bibfield  {author} {\bibinfo {author} {\bibfnamefont {F.~T.~M.}\
  \bibnamefont {Nieuwstadt}},\ }\bibfield  {title} {\bibinfo {title} {The
  turbulent structure of the stable, nocturnal boundary layer},\ }\href@noop {}
  {\bibfield  {journal} {\bibinfo  {journal} {J.~Atm. Sci.}\ }\textbf {\bibinfo
  {volume} {41}},\ \bibinfo {pages} {2202} (\bibinfo {year}
  {1984})}\BibitemShut {NoStop}%
\bibitem [{\citenamefont {Stull}(2000)}]{Stull2000}%
  \BibitemOpen
  \bibfield  {author} {\bibinfo {author} {\bibfnamefont {R.~B.}\ \bibnamefont
  {Stull}},\ }\href@noop {} {\emph {\bibinfo {title} {Meteorology for
  scientists and engineers}}}\ (\bibinfo  {publisher} {Brooks/Cole},\ \bibinfo
  {year} {2000})\BibitemShut {NoStop}%
\bibitem [{\citenamefont {Wunsch}\ and\ \citenamefont
  {Ferrari}(2004)}]{Wunsch2004}%
  \BibitemOpen
  \bibfield  {author} {\bibinfo {author} {\bibfnamefont {C.}~\bibnamefont
  {Wunsch}}\ and\ \bibinfo {author} {\bibfnamefont {R.}~\bibnamefont
  {Ferrari}},\ }\bibfield  {title} {\bibinfo {title} {Vertical mixing, energy,
  and the general circulation of the oceans},\ }\href@noop {} {\bibfield
  {journal} {\bibinfo  {journal} {Annu. Rev. Fluid Mech.}\ }\textbf {\bibinfo
  {volume} {36}},\ \bibinfo {pages} {281} (\bibinfo {year} {2004})}\BibitemShut
  {NoStop}%
\bibitem [{\citenamefont {Thorpe}(2005)}]{Thorpe2005}%
  \BibitemOpen
  \bibfield  {author} {\bibinfo {author} {\bibfnamefont {S.~A.}\ \bibnamefont
  {Thorpe}},\ }\href@noop {} {\emph {\bibinfo {title} {The Turbulent Ocean}}}\
  (\bibinfo  {publisher} {Cambridge University Press},\ \bibinfo {year}
  {2005})\BibitemShut {NoStop}%
\bibitem [{\citenamefont {Panofsky}\ and\ \citenamefont
  {Dutton}(1984)}]{Panofsky1984}%
  \BibitemOpen
  \bibfield  {author} {\bibinfo {author} {\bibfnamefont {H.~A.}\ \bibnamefont
  {Panofsky}}\ and\ \bibinfo {author} {\bibfnamefont {J.~A.}\ \bibnamefont
  {Dutton}},\ }\href@noop {} {\emph {\bibinfo {title} {Atmospheric Turbulence:
  Models and methods for engineering applications}}}\ (\bibinfo  {publisher}
  {Wiley},\ \bibinfo {year} {1984})\BibitemShut {NoStop}%
\bibitem [{\citenamefont {Sorbjan}(1989)}]{Sorbjan1989}%
  \BibitemOpen
  \bibfield  {author} {\bibinfo {author} {\bibfnamefont {Z.}~\bibnamefont
  {Sorbjan}},\ }\href@noop {} {\emph {\bibinfo {title} {Structure of the
  atmospheric boundary layer}}}\ (\bibinfo  {publisher} {Prentice Hal},\
  \bibinfo {year} {1989})\BibitemShut {NoStop}%
\bibitem [{\citenamefont {Stull}(1988)}]{Stull1988}%
  \BibitemOpen
  \bibfield  {author} {\bibinfo {author} {\bibfnamefont {R.~B.}\ \bibnamefont
  {Stull}},\ }\href@noop {} {\emph {\bibinfo {title} {An Introduction to
  Boundary Layer Meteorology}}},\ Vol.~\bibinfo {volume} {13}\ (\bibinfo
  {publisher} {Springer Science \& Business Media},\ \bibinfo {year}
  {1988})\BibitemShut {NoStop}%
\bibitem [{\citenamefont {Wyngaard}(2010)}]{Wyngaard2010}%
  \BibitemOpen
  \bibfield  {author} {\bibinfo {author} {\bibfnamefont {J.~C.}\ \bibnamefont
  {Wyngaard}},\ }\href@noop {} {\emph {\bibinfo {title} {Turbulence in the
  Atmosphere}}}\ (\bibinfo  {publisher} {Cambridge University Press},\ \bibinfo
  {year} {2010})\BibitemShut {NoStop}%
\bibitem [{\citenamefont {Garratt}(1994)}]{Garratt1994}%
  \BibitemOpen
  \bibfield  {author} {\bibinfo {author} {\bibfnamefont {J.~R.}\ \bibnamefont
  {Garratt}},\ }\bibfield  {title} {\bibinfo {title} {The atmospheric boundary
  layer},\ }\href@noop {} {\bibfield  {journal} {\bibinfo  {journal}
  {Earth-Sci. Rev.}\ }\textbf {\bibinfo {volume} {37}},\ \bibinfo {pages} {89}
  (\bibinfo {year} {1994})}\BibitemShut {NoStop}%
\bibitem [{\citenamefont {Ivey}\ \emph {et~al.}(2008)\citenamefont {Ivey},
  \citenamefont {Winters},\ and\ \citenamefont {Koseff}}]{Ivey2008}%
  \BibitemOpen
  \bibfield  {author} {\bibinfo {author} {\bibfnamefont {G.~N.}\ \bibnamefont
  {Ivey}}, \bibinfo {author} {\bibfnamefont {K.~B.}\ \bibnamefont {Winters}},\
  and\ \bibinfo {author} {\bibfnamefont {J.~R.}\ \bibnamefont {Koseff}},\
  }\bibfield  {title} {\bibinfo {title} {Density stratification, turbulence,
  but how much mixing?},\ }\href@noop {} {\bibfield  {journal} {\bibinfo
  {journal} {Annu. Rev. Fluid Mech.}\ }\textbf {\bibinfo {volume} {40}},\
  \bibinfo {pages} {169} (\bibinfo {year} {2008})}\BibitemShut {NoStop}%
\bibitem [{\citenamefont {Mahrt}(2014)}]{Mahrt2014}%
  \BibitemOpen
  \bibfield  {author} {\bibinfo {author} {\bibfnamefont {L.}~\bibnamefont
  {Mahrt}},\ }\bibfield  {title} {\bibinfo {title} {Stably stratified
  atmospheric boundary layers},\ }\href@noop {} {\bibfield  {journal} {\bibinfo
   {journal} {Annu. Rev. Fluid Mech.}\ }\textbf {\bibinfo {volume} {46}},\
  \bibinfo {pages} {23} (\bibinfo {year} {2014})}\BibitemShut {NoStop}%
\bibitem [{\citenamefont {Mahrt}(1999)}]{Mahrt1999}%
  \BibitemOpen
  \bibfield  {author} {\bibinfo {author} {\bibfnamefont {L.}~\bibnamefont
  {Mahrt}},\ }\bibfield  {title} {\bibinfo {title} {Stratified atmospheric
  boundary layers},\ }\href@noop {} {\bibfield  {journal} {\bibinfo  {journal}
  {Bound.-Layer Meteorol.}\ }\textbf {\bibinfo {volume} {90}},\ \bibinfo
  {pages} {375} (\bibinfo {year} {1999})}\BibitemShut {NoStop}%
\bibitem [{\citenamefont {Williams}\ \emph {et~al.}(2017)\citenamefont
  {Williams}, \citenamefont {Hohman}, \citenamefont {Van~Buren}, \citenamefont
  {Bou-Zeid},\ and\ \citenamefont {Smits}}]{Williams2017}%
  \BibitemOpen
  \bibfield  {author} {\bibinfo {author} {\bibfnamefont {O.}~\bibnamefont
  {Williams}}, \bibinfo {author} {\bibfnamefont {T.}~\bibnamefont {Hohman}},
  \bibinfo {author} {\bibfnamefont {T.}~\bibnamefont {Van~Buren}}, \bibinfo
  {author} {\bibfnamefont {E.}~\bibnamefont {Bou-Zeid}},\ and\ \bibinfo
  {author} {\bibfnamefont {A.~J.}\ \bibnamefont {Smits}},\ }\bibfield  {title}
  {\bibinfo {title} {The effect of stable thermal stratification on turbulent
  boundary layer statistics},\ }\href@noop {} {\bibfield  {journal} {\bibinfo
  {journal} {J.~Fluid Mech.}\ }\textbf {\bibinfo {volume} {812}},\ \bibinfo
  {pages} {1039} (\bibinfo {year} {2017})}\BibitemShut {NoStop}%
\bibitem [{\citenamefont {Chauhan}\ \emph {et~al.}(2013)\citenamefont
  {Chauhan}, \citenamefont {Hutchins}, \citenamefont {Monty},\ and\
  \citenamefont {Marusic}}]{Chauhan2013}%
  \BibitemOpen
  \bibfield  {author} {\bibinfo {author} {\bibfnamefont {K.}~\bibnamefont
  {Chauhan}}, \bibinfo {author} {\bibfnamefont {N.}~\bibnamefont {Hutchins}},
  \bibinfo {author} {\bibfnamefont {J.}~\bibnamefont {Monty}},\ and\ \bibinfo
  {author} {\bibfnamefont {I.}~\bibnamefont {Marusic}},\ }\bibfield  {title}
  {\bibinfo {title} {Structure inclination angles in the convective atmospheric
  surface layer},\ }\href@noop {} {\bibfield  {journal} {\bibinfo  {journal}
  {Bound.-Layer Meteorol.}\ }\textbf {\bibinfo {volume} {147}},\ \bibinfo
  {pages} {41} (\bibinfo {year} {2013})}\BibitemShut {NoStop}%
\bibitem [{\citenamefont {Salesky}\ and\ \citenamefont
  {Anderson}(2018)}]{Salesky2018}%
  \BibitemOpen
  \bibfield  {author} {\bibinfo {author} {\bibfnamefont {S.~T.}\ \bibnamefont
  {Salesky}}\ and\ \bibinfo {author} {\bibfnamefont {W.}~\bibnamefont
  {Anderson}},\ }\bibfield  {title} {\bibinfo {title} {Buoyancy effects on
  large-scale motions in convective atmospheric boundary layers: implications
  for modulation of near-wall processes},\ }\href@noop {} {\bibfield  {journal}
  {\bibinfo  {journal} {J.~Fluid Mech.}\ }\textbf {\bibinfo {volume} {856}},\
  \bibinfo {pages} {135} (\bibinfo {year} {2018})}\BibitemShut {NoStop}%
\bibitem [{\citenamefont {Salesky}\ and\ \citenamefont
  {Anderson}(2020)}]{Salesky2020}%
  \BibitemOpen
  \bibfield  {author} {\bibinfo {author} {\bibfnamefont {S.~T.}\ \bibnamefont
  {Salesky}}\ and\ \bibinfo {author} {\bibfnamefont {W.}~\bibnamefont
  {Anderson}},\ }\bibfield  {title} {\bibinfo {title} {Revisiting inclination
  of large-scale motions in unstably stratified channel flow},\ }\href@noop {}
  {\bibfield  {journal} {\bibinfo  {journal} {J.~Fluid Mech.}\ }\textbf
  {\bibinfo {volume} {884}} (\bibinfo {year} {2020})}\BibitemShut {NoStop}%
\bibitem [{\citenamefont {Wyngaard}\ and\ \citenamefont
  {Cot{\'e}}(1971)}]{Wyngaard1971}%
  \BibitemOpen
  \bibfield  {author} {\bibinfo {author} {\bibfnamefont {J.~C.}\ \bibnamefont
  {Wyngaard}}\ and\ \bibinfo {author} {\bibfnamefont {O.~R.}\ \bibnamefont
  {Cot{\'e}}},\ }\bibfield  {title} {\bibinfo {title} {The budgets of turbulent
  kinetic energy and temperature variance in the atmospheric surface layer},\
  }\href@noop {} {\bibfield  {journal} {\bibinfo  {journal} {J.~Atm. Sci.}\
  }\textbf {\bibinfo {volume} {28}},\ \bibinfo {pages} {190} (\bibinfo {year}
  {1971})}\BibitemShut {NoStop}%
\bibitem [{\citenamefont {Kondo}\ \emph {et~al.}(1978)\citenamefont {Kondo},
  \citenamefont {Kanechika},\ and\ \citenamefont {Yasuda}}]{Kondo1978}%
  \BibitemOpen
  \bibfield  {author} {\bibinfo {author} {\bibfnamefont {J.}~\bibnamefont
  {Kondo}}, \bibinfo {author} {\bibfnamefont {O.}~\bibnamefont {Kanechika}},\
  and\ \bibinfo {author} {\bibfnamefont {N.}~\bibnamefont {Yasuda}},\
  }\bibfield  {title} {\bibinfo {title} {Heat and momentum transfers under
  strong stability in the atmospheric surface layer},\ }\href@noop {}
  {\bibfield  {journal} {\bibinfo  {journal} {J.~Atm. Sci.}\ }\textbf {\bibinfo
  {volume} {35}},\ \bibinfo {pages} {1012} (\bibinfo {year}
  {1978})}\BibitemShut {NoStop}%
\bibitem [{\citenamefont {Mahrt}(1998)}]{Mahrt1998}%
  \BibitemOpen
  \bibfield  {author} {\bibinfo {author} {\bibfnamefont {L.}~\bibnamefont
  {Mahrt}},\ }\bibfield  {title} {\bibinfo {title} {Nocturnal boundary-layer
  regimes},\ }\href@noop {} {\bibfield  {journal} {\bibinfo  {journal}
  {Bound.-Layer Meteorol.}\ }\textbf {\bibinfo {volume} {88}},\ \bibinfo
  {pages} {255} (\bibinfo {year} {1998})}\BibitemShut {NoStop}%
\bibitem [{\citenamefont {Fernando}\ and\ \citenamefont
  {Weil}(2010)}]{Fernando2010}%
  \BibitemOpen
  \bibfield  {author} {\bibinfo {author} {\bibfnamefont {H.~J.~S.}\
  \bibnamefont {Fernando}}\ and\ \bibinfo {author} {\bibfnamefont {J.~C.}\
  \bibnamefont {Weil}},\ }\bibfield  {title} {\bibinfo {title} {Whither the
  stable boundary layer? {A} shift in the research agenda},\ }\href@noop {}
  {\bibfield  {journal} {\bibinfo  {journal} {Bull. Am. Meteorol. Soc.}\
  }\textbf {\bibinfo {volume} {91}},\ \bibinfo {pages} {1475} (\bibinfo {year}
  {2010})}\BibitemShut {NoStop}%
\bibitem [{\citenamefont {Smedman}\ \emph {et~al.}(1994)\citenamefont
  {Smedman}, \citenamefont {Tjernstr{\"o}m},\ and\ \citenamefont
  {H{\"o}gstr{\"o}m}}]{Smedman1994}%
  \BibitemOpen
  \bibfield  {author} {\bibinfo {author} {\bibfnamefont {A.-S.}\ \bibnamefont
  {Smedman}}, \bibinfo {author} {\bibfnamefont {M.}~\bibnamefont
  {Tjernstr{\"o}m}},\ and\ \bibinfo {author} {\bibfnamefont {U.}~\bibnamefont
  {H{\"o}gstr{\"o}m}},\ }\bibfield  {title} {\bibinfo {title} {The near-neutral
  marine atmospheric boundary layer with no surface shearing stress: A case
  study},\ }\href@noop {} {\bibfield  {journal} {\bibinfo  {journal} {J. Atm.
  Sci.}\ }\textbf {\bibinfo {volume} {51}},\ \bibinfo {pages} {3399} (\bibinfo
  {year} {1994})}\BibitemShut {NoStop}%
\bibitem [{\citenamefont {Stacey}\ \emph {et~al.}(1999)\citenamefont {Stacey},
  \citenamefont {Monismith},\ and\ \citenamefont {Burau}}]{Stacey1999}%
  \BibitemOpen
  \bibfield  {author} {\bibinfo {author} {\bibfnamefont {M.~T.}\ \bibnamefont
  {Stacey}}, \bibinfo {author} {\bibfnamefont {S.~G.}\ \bibnamefont
  {Monismith}},\ and\ \bibinfo {author} {\bibfnamefont {J.~R.}\ \bibnamefont
  {Burau}},\ }\bibfield  {title} {\bibinfo {title} {Observations of turbulence
  in a partially stratified estuary},\ }\href@noop {} {\bibfield  {journal}
  {\bibinfo  {journal} {J.~Phys. Oceanogr.}\ }\textbf {\bibinfo {volume}
  {29}},\ \bibinfo {pages} {1950} (\bibinfo {year} {1999})}\BibitemShut
  {NoStop}%
\bibitem [{\citenamefont {Lu}\ \emph {et~al.}(2000)\citenamefont {Lu},
  \citenamefont {Lueck},\ and\ \citenamefont {Huang}}]{Lu2000}%
  \BibitemOpen
  \bibfield  {author} {\bibinfo {author} {\bibfnamefont {Y.}~\bibnamefont
  {Lu}}, \bibinfo {author} {\bibfnamefont {R.~G.}\ \bibnamefont {Lueck}},\ and\
  \bibinfo {author} {\bibfnamefont {D.}~\bibnamefont {Huang}},\ }\bibfield
  {title} {\bibinfo {title} {Turbulence characteristics in a tidal channel},\
  }\href@noop {} {\bibfield  {journal} {\bibinfo  {journal} {J.~Phys.
  Oceanogr.}\ }\textbf {\bibinfo {volume} {30}},\ \bibinfo {pages} {855}
  (\bibinfo {year} {2000})}\BibitemShut {NoStop}%
\bibitem [{\citenamefont {Large}\ \emph {et~al.}(1994)\citenamefont {Large},
  \citenamefont {McWilliams},\ and\ \citenamefont {Doney}}]{Large1994}%
  \BibitemOpen
  \bibfield  {author} {\bibinfo {author} {\bibfnamefont {W.~G.}\ \bibnamefont
  {Large}}, \bibinfo {author} {\bibfnamefont {J.~C.}\ \bibnamefont
  {McWilliams}},\ and\ \bibinfo {author} {\bibfnamefont {S.~C.}\ \bibnamefont
  {Doney}},\ }\bibfield  {title} {\bibinfo {title} {Oceanic vertical mixing:
  {A} review and a model with a nonlocal boundary layer parameterization},\
  }\href@noop {} {\bibfield  {journal} {\bibinfo  {journal} {Rev. Geophys.}\
  }\textbf {\bibinfo {volume} {32}},\ \bibinfo {pages} {363} (\bibinfo {year}
  {1994})}\BibitemShut {NoStop}%
\bibitem [{\citenamefont {Arya}(1975)}]{Arya1975}%
  \BibitemOpen
  \bibfield  {author} {\bibinfo {author} {\bibfnamefont {S.~P.~S.}\
  \bibnamefont {Arya}},\ }\bibfield  {title} {\bibinfo {title} {Buoyancy
  effects in a horizontal flat-plate boundary layer},\ }\href@noop {}
  {\bibfield  {journal} {\bibinfo  {journal} {J.~Fluid Mech.}\ }\textbf
  {\bibinfo {volume} {68}},\ \bibinfo {pages} {321} (\bibinfo {year}
  {1975})}\BibitemShut {NoStop}%
\bibitem [{\citenamefont {Britter}(1974)}]{Britter1974}%
  \BibitemOpen
  \bibfield  {author} {\bibinfo {author} {\bibfnamefont {R.~E.}\ \bibnamefont
  {Britter}},\ }\emph {\bibinfo {title} {An experiment on turbulence in a
  density stratified fluid}},\ \href@noop {} {Ph.D. thesis},\ \bibinfo
  {school} {Monash University, Australia} (\bibinfo {year} {1974})\BibitemShut
  {NoStop}%
\bibitem [{\citenamefont {Piat}\ and\ \citenamefont
  {Hopfinger}(1981)}]{Piat1981}%
  \BibitemOpen
  \bibfield  {author} {\bibinfo {author} {\bibfnamefont {J.~F.}\ \bibnamefont
  {Piat}}\ and\ \bibinfo {author} {\bibfnamefont {E.~J.}\ \bibnamefont
  {Hopfinger}},\ }\bibfield  {title} {\bibinfo {title} {A boundary layer topped
  by a density interface},\ }\href@noop {} {\bibfield  {journal} {\bibinfo
  {journal} {J.~Fluid Mech.}\ }\textbf {\bibinfo {volume} {113}},\ \bibinfo
  {pages} {411} (\bibinfo {year} {1981})}\BibitemShut {NoStop}%
\bibitem [{\citenamefont {Komori}\ \emph {et~al.}(1983)\citenamefont {Komori},
  \citenamefont {Ueda}, \citenamefont {Ogino},\ and\ \citenamefont
  {Mizushina}}]{Komori1983}%
  \BibitemOpen
  \bibfield  {author} {\bibinfo {author} {\bibfnamefont {S.}~\bibnamefont
  {Komori}}, \bibinfo {author} {\bibfnamefont {H.}~\bibnamefont {Ueda}},
  \bibinfo {author} {\bibfnamefont {F.}~\bibnamefont {Ogino}},\ and\ \bibinfo
  {author} {\bibfnamefont {T.}~\bibnamefont {Mizushina}},\ }\bibfield  {title}
  {\bibinfo {title} {Turbulence structure in stably stratified open-channel
  flow},\ }\href@noop {} {\bibfield  {journal} {\bibinfo  {journal} {J.~Fluid
  Mech.}\ }\textbf {\bibinfo {volume} {130}},\ \bibinfo {pages} {13} (\bibinfo
  {year} {1983})}\BibitemShut {NoStop}%
\bibitem [{\citenamefont {Fukui}\ \emph {et~al.}(1983)\citenamefont {Fukui},
  \citenamefont {Nakajima},\ and\ \citenamefont {Ueda}}]{Fukui1983}%
  \BibitemOpen
  \bibfield  {author} {\bibinfo {author} {\bibfnamefont {K.}~\bibnamefont
  {Fukui}}, \bibinfo {author} {\bibfnamefont {M.}~\bibnamefont {Nakajima}},\
  and\ \bibinfo {author} {\bibfnamefont {H.}~\bibnamefont {Ueda}},\ }\bibfield
  {title} {\bibinfo {title} {A laboratory experiment on momentum and heat
  transfer in the stratified surface layer},\ }\href@noop {} {\bibfield
  {journal} {\bibinfo  {journal} {Q. J. R. Meteorol. Soc.}\ }\textbf {\bibinfo
  {volume} {109}},\ \bibinfo {pages} {661} (\bibinfo {year}
  {1983})}\BibitemShut {NoStop}%
\bibitem [{\citenamefont {Miles}(1961)}]{Miles1961}%
  \BibitemOpen
  \bibfield  {author} {\bibinfo {author} {\bibfnamefont {J.~W.}\ \bibnamefont
  {Miles}},\ }\bibfield  {title} {\bibinfo {title} {On the stability of
  heterogeneous shear flows},\ }\href@noop {} {\bibfield  {journal} {\bibinfo
  {journal} {J.~Fluid Mech.}\ }\textbf {\bibinfo {volume} {10}},\ \bibinfo
  {pages} {496} (\bibinfo {year} {1961})}\BibitemShut {NoStop}%
\bibitem [{\citenamefont {Garg}\ \emph {et~al.}(2000)\citenamefont {Garg},
  \citenamefont {Ferziger}, \citenamefont {Monismith},\ and\ \citenamefont
  {Koseff}}]{Garg2000}%
  \BibitemOpen
  \bibfield  {author} {\bibinfo {author} {\bibfnamefont {R.~P.}\ \bibnamefont
  {Garg}}, \bibinfo {author} {\bibfnamefont {J.~H.}\ \bibnamefont {Ferziger}},
  \bibinfo {author} {\bibfnamefont {S.~G.}\ \bibnamefont {Monismith}},\ and\
  \bibinfo {author} {\bibfnamefont {J.~R.}\ \bibnamefont {Koseff}},\ }\bibfield
   {title} {\bibinfo {title} {Stably stratified turbulent channel flows. {I}.
  {S}tratification regimes and turbulence suppression mechanism},\ }\href@noop
  {} {\bibfield  {journal} {\bibinfo  {journal} {Phys. Fluids}\ }\textbf
  {\bibinfo {volume} {12}},\ \bibinfo {pages} {2569} (\bibinfo {year}
  {2000})}\BibitemShut {NoStop}%
\bibitem [{\citenamefont {Armenio}\ and\ \citenamefont
  {Sarkar}(2002)}]{Armenio2002}%
  \BibitemOpen
  \bibfield  {author} {\bibinfo {author} {\bibfnamefont {V.}~\bibnamefont
  {Armenio}}\ and\ \bibinfo {author} {\bibfnamefont {S.}~\bibnamefont
  {Sarkar}},\ }\bibfield  {title} {\bibinfo {title} {An investigation of stably
  stratified turbulent channel flow using large-eddy simulation},\ }\href@noop
  {} {\bibfield  {journal} {\bibinfo  {journal} {J.~Fluid Mech.}\ }\textbf
  {\bibinfo {volume} {459}},\ \bibinfo {pages} {1} (\bibinfo {year}
  {2002})}\BibitemShut {NoStop}%
\bibitem [{\citenamefont {Basu}\ and\ \citenamefont
  {Port{\'e}-Agel}(2006)}]{Basu2006}%
  \BibitemOpen
  \bibfield  {author} {\bibinfo {author} {\bibfnamefont {S.}~\bibnamefont
  {Basu}}\ and\ \bibinfo {author} {\bibfnamefont {F.}~\bibnamefont
  {Port{\'e}-Agel}},\ }\bibfield  {title} {\bibinfo {title} {Large-eddy
  simulation of stably stratified atmospheric boundary layer turbulence: a
  scale-dependent dynamic modeling approach},\ }\href@noop {} {\bibfield
  {journal} {\bibinfo  {journal} {J. Atmos. Sci.}\ }\textbf {\bibinfo {volume}
  {63}},\ \bibinfo {pages} {2074} (\bibinfo {year} {2006})}\BibitemShut
  {NoStop}%
\bibitem [{\citenamefont {Stoll}\ and\ \citenamefont
  {Port{\'e}-Agel}(2008)}]{Stoll2008}%
  \BibitemOpen
  \bibfield  {author} {\bibinfo {author} {\bibfnamefont {R.}~\bibnamefont
  {Stoll}}\ and\ \bibinfo {author} {\bibfnamefont {F.}~\bibnamefont
  {Port{\'e}-Agel}},\ }\bibfield  {title} {\bibinfo {title} {Large-eddy
  simulation of the stable atmospheric boundary layer using dynamic models with
  different averaging schemes},\ }\href@noop {} {\bibfield  {journal} {\bibinfo
   {journal} {Bound.-Layer Meteorol.}\ }\textbf {\bibinfo {volume} {126}},\
  \bibinfo {pages} {1} (\bibinfo {year} {2008})}\BibitemShut {NoStop}%
\bibitem [{\citenamefont {Iida}\ \emph {et~al.}(2002)\citenamefont {Iida},
  \citenamefont {Kasagi},\ and\ \citenamefont {Nagano}}]{Iida2002}%
  \BibitemOpen
  \bibfield  {author} {\bibinfo {author} {\bibfnamefont {O.}~\bibnamefont
  {Iida}}, \bibinfo {author} {\bibfnamefont {N.}~\bibnamefont {Kasagi}},\ and\
  \bibinfo {author} {\bibfnamefont {Y.}~\bibnamefont {Nagano}},\ }\bibfield
  {title} {\bibinfo {title} {Direct numerical simulation of turbulent channel
  flow under stable density stratification},\ }\href@noop {} {\bibfield
  {journal} {\bibinfo  {journal} {Int. J. Heat Mass Transf.}\ }\textbf
  {\bibinfo {volume} {45}},\ \bibinfo {pages} {1693} (\bibinfo {year}
  {2002})}\BibitemShut {NoStop}%
\bibitem [{\citenamefont {Nieuwstadt}(2005)}]{Nieuwstadt2005}%
  \BibitemOpen
  \bibfield  {author} {\bibinfo {author} {\bibfnamefont {F.~T.~M.}\
  \bibnamefont {Nieuwstadt}},\ }\bibfield  {title} {\bibinfo {title} {Direct
  numerical simulation of stable channel flow at large stability},\ }\href@noop
  {} {\bibfield  {journal} {\bibinfo  {journal} {Bound.-Layer Meteorol.}\
  }\textbf {\bibinfo {volume} {116}},\ \bibinfo {pages} {277} (\bibinfo {year}
  {2005})}\BibitemShut {NoStop}%
\bibitem [{\citenamefont {Brethouwer}\ \emph {et~al.}(2007)\citenamefont
  {Brethouwer}, \citenamefont {Billant}, \citenamefont {Lindborg},\ and\
  \citenamefont {Chomaz}}]{Brethouwer2007}%
  \BibitemOpen
  \bibfield  {author} {\bibinfo {author} {\bibfnamefont {G.}~\bibnamefont
  {Brethouwer}}, \bibinfo {author} {\bibfnamefont {P.}~\bibnamefont {Billant}},
  \bibinfo {author} {\bibfnamefont {E.}~\bibnamefont {Lindborg}},\ and\
  \bibinfo {author} {\bibfnamefont {J.-M.}\ \bibnamefont {Chomaz}},\ }\bibfield
   {title} {\bibinfo {title} {Scaling analysis and simulation of strongly
  stratified turbulent flows},\ }\href
  {https://doi.org/10.1017/S0022112007006854} {\bibfield  {journal} {\bibinfo
  {journal} {J. Fluid Mech.}\ }\textbf {\bibinfo {volume} {585}},\ \bibinfo
  {pages} {343–368} (\bibinfo {year} {2007})}\BibitemShut {NoStop}%
\bibitem [{\citenamefont {Flores}\ and\ \citenamefont
  {Riley}(2011)}]{Flores2011}%
  \BibitemOpen
  \bibfield  {author} {\bibinfo {author} {\bibfnamefont {O.}~\bibnamefont
  {Flores}}\ and\ \bibinfo {author} {\bibfnamefont {J.~J.}\ \bibnamefont
  {Riley}},\ }\bibfield  {title} {\bibinfo {title} {Analysis of turbulence
  collapse in the stably stratified surface layer using direct numerical
  simulation},\ }\href@noop {} {\bibfield  {journal} {\bibinfo  {journal}
  {Bound.-Layer Meteorol.}\ }\textbf {\bibinfo {volume} {139}},\ \bibinfo
  {pages} {241} (\bibinfo {year} {2011})}\BibitemShut {NoStop}%
\bibitem [{\citenamefont {Garc\'ia-Villalba}\ and\ \citenamefont
  {Del~Alamo}(2011)}]{Garcia2011}%
  \BibitemOpen
  \bibfield  {author} {\bibinfo {author} {\bibfnamefont {M.}~\bibnamefont
  {Garc\'ia-Villalba}}\ and\ \bibinfo {author} {\bibfnamefont {J.~C.}\
  \bibnamefont {Del~Alamo}},\ }\bibfield  {title} {\bibinfo {title} {Turbulence
  modification by stable stratification in channel flow},\ }\href@noop {}
  {\bibfield  {journal} {\bibinfo  {journal} {Phys. Fluids}\ }\textbf {\bibinfo
  {volume} {23}},\ \bibinfo {pages} {045104} (\bibinfo {year}
  {2011})}\BibitemShut {NoStop}%
\bibitem [{\citenamefont {Yeh}\ and\ \citenamefont {Taira}(2018)}]{Yeh18}%
  \BibitemOpen
  \bibfield  {author} {\bibinfo {author} {\bibfnamefont {C.-A.}\ \bibnamefont
  {Yeh}}\ and\ \bibinfo {author} {\bibfnamefont {K.}~\bibnamefont {Taira}},\
  }\bibfield  {title} {\bibinfo {title} {Resolvent-analysis-based design of
  airfoil separation control},\ }\href@noop {} {\bibfield  {journal} {\bibinfo
  {journal} {J. Fluid Mech.}\ }\textbf {\bibinfo {volume} {867}},\ \bibinfo
  {pages} {572} (\bibinfo {year} {2018})}\BibitemShut {NoStop}%
\bibitem [{\citenamefont {Towne}\ \emph {et~al.}(2018)\citenamefont {Towne},
  \citenamefont {Schmidt},\ and\ \citenamefont {Colonius}}]{Towne18}%
  \BibitemOpen
  \bibfield  {author} {\bibinfo {author} {\bibfnamefont {A.}~\bibnamefont
  {Towne}}, \bibinfo {author} {\bibfnamefont {O.~T.}\ \bibnamefont {Schmidt}},\
  and\ \bibinfo {author} {\bibfnamefont {T.}~\bibnamefont {Colonius}},\
  }\bibfield  {title} {\bibinfo {title} {Spectral proper orthogonal
  decomposition and its relationship to dynamic mode decomposition and
  resolvent analysis},\ }\href@noop {} {\bibfield  {journal} {\bibinfo
  {journal} {J. Fluid Mech.}\ }\textbf {\bibinfo {volume} {847}},\ \bibinfo
  {pages} {821} (\bibinfo {year} {2018})}\BibitemShut {NoStop}%
\bibitem [{\citenamefont {Bae}\ \emph {et~al.}(2020)\citenamefont {Bae},
  \citenamefont {Dawson},\ and\ \citenamefont {McKeon}}]{Bae2020}%
  \BibitemOpen
  \bibfield  {author} {\bibinfo {author} {\bibfnamefont {H.~J.}\ \bibnamefont
  {Bae}}, \bibinfo {author} {\bibfnamefont {S.~T.}\ \bibnamefont {Dawson}},\
  and\ \bibinfo {author} {\bibfnamefont {B.~J.}\ \bibnamefont {McKeon}},\
  }\bibfield  {title} {\bibinfo {title} {Resolvent-based study of
  compressibility effects on supersonic turbulent boundary layers},\
  }\href@noop {} {\bibfield  {journal} {\bibinfo  {journal} {J.~Fluid Mech.}\
  }\textbf {\bibinfo {volume} {883}} (\bibinfo {year} {2020})}\BibitemShut
  {NoStop}%
\bibitem [{\citenamefont {McMullen}\ \emph {et~al.}(2020)\citenamefont
  {McMullen}, \citenamefont {Rosenberg},\ and\ \citenamefont
  {McKeon}}]{McMullen2020}%
  \BibitemOpen
  \bibfield  {author} {\bibinfo {author} {\bibfnamefont {R.~M.}\ \bibnamefont
  {McMullen}}, \bibinfo {author} {\bibfnamefont {K.}~\bibnamefont
  {Rosenberg}},\ and\ \bibinfo {author} {\bibfnamefont {B.~J.}\ \bibnamefont
  {McKeon}},\ }\bibfield  {title} {\bibinfo {title} {Interaction of forced
  {O}rr-{S}ommerfeld and {S}quire modes in a low-order representation of
  turbulent channel flow},\ }\href
  {https://doi.org/10.1103/PhysRevFluids.5.084607} {\bibfield  {journal}
  {\bibinfo  {journal} {Phys. Rev. Fluids}\ }\textbf {\bibinfo {volume} {5}},\
  \bibinfo {pages} {084607} (\bibinfo {year} {2020})}\BibitemShut {NoStop}%
\bibitem [{\citenamefont {Nogueira}\ \emph {et~al.}(2021)\citenamefont
  {Nogueira}, \citenamefont {Morra}, \citenamefont {Martini}, \citenamefont
  {Cavalieri},\ and\ \citenamefont {Henningson}}]{Nogueira2021}%
  \BibitemOpen
  \bibfield  {author} {\bibinfo {author} {\bibfnamefont {P.~A.~S.}\
  \bibnamefont {Nogueira}}, \bibinfo {author} {\bibfnamefont {P.}~\bibnamefont
  {Morra}}, \bibinfo {author} {\bibfnamefont {E.}~\bibnamefont {Martini}},
  \bibinfo {author} {\bibfnamefont {A.~V.~G.}\ \bibnamefont {Cavalieri}},\ and\
  \bibinfo {author} {\bibfnamefont {D.~S.}\ \bibnamefont {Henningson}},\
  }\bibfield  {title} {\bibinfo {title} {Forcing statistics in resolvent
  analysis: application in minimal turbulent couette flow},\ }\href
  {https://doi.org/10.1017/jfm.2020.918} {\bibfield  {journal} {\bibinfo
  {journal} {J. Fluid Mech.}\ }\textbf {\bibinfo {volume} {908}},\ \bibinfo
  {pages} {A32} (\bibinfo {year} {2021})}\BibitemShut {NoStop}%
\bibitem [{\citenamefont {McKeon}(2017)}]{McKeon2017}%
  \BibitemOpen
  \bibfield  {author} {\bibinfo {author} {\bibfnamefont {B.~J.}\ \bibnamefont
  {McKeon}},\ }\bibfield  {title} {\bibinfo {title} {The engine behind (wall)
  turbulence: perspectives on scale interactions},\ }\href@noop {} {\bibfield
  {journal} {\bibinfo  {journal} {J. Fluid Mech.}\ }\textbf {\bibinfo {volume}
  {817}} (\bibinfo {year} {2017})}\BibitemShut {NoStop}%
\bibitem [{\citenamefont {Jovanovi{\'c}}(2020)}]{Jovanovic2021}%
  \BibitemOpen
  \bibfield  {author} {\bibinfo {author} {\bibfnamefont {M.~R.}\ \bibnamefont
  {Jovanovi{\'c}}},\ }\bibfield  {title} {\bibinfo {title} {From bypass
  transition to flow control and data-driven turbulence modeling: An
  input-output viewpoint},\ }\href
  {https://doi.org/10.1146/annurev-fluid-010719-060244} {\bibfield  {journal}
  {\bibinfo  {journal} {Annu. Rev. Fluid Mech.}\ }\textbf {\bibinfo {volume}
  {53}},\ \bibinfo {pages} {null} (\bibinfo {year} {2020})}\BibitemShut
  {NoStop}%
\bibitem [{\citenamefont {Moarref}\ \emph {et~al.}(2013)\citenamefont
  {Moarref}, \citenamefont {Sharma}, \citenamefont {Tropp},\ and\ \citenamefont
  {McKeon}}]{Moarref2013}%
  \BibitemOpen
  \bibfield  {author} {\bibinfo {author} {\bibfnamefont {R.}~\bibnamefont
  {Moarref}}, \bibinfo {author} {\bibfnamefont {A.~S.}\ \bibnamefont {Sharma}},
  \bibinfo {author} {\bibfnamefont {J.~A.}\ \bibnamefont {Tropp}},\ and\
  \bibinfo {author} {\bibfnamefont {B.~J.}\ \bibnamefont {McKeon}},\ }\bibfield
   {title} {\bibinfo {title} {Model-based scaling of the streamwise energy
  density in high-{R}eynolds-number turbulent channels},\ }\href@noop {}
  {\bibfield  {journal} {\bibinfo  {journal} {J.~Fluid Mech.}\ }\textbf
  {\bibinfo {volume} {734}},\ \bibinfo {pages} {275} (\bibinfo {year}
  {2013})}\BibitemShut {NoStop}%
\bibitem [{\citenamefont {Madhusudanan}(2020)}]{madhusudanan2020coherent}%
  \BibitemOpen
  \bibfield  {author} {\bibinfo {author} {\bibfnamefont {A.}~\bibnamefont
  {Madhusudanan}},\ }\emph {\bibinfo {title} {Coherent structures from the
  linearized Navier-Stokes equations for wall-bounded turbulent flows}},\
  \href@noop {} {Ph.D. thesis},\ \bibinfo  {school} {University of Melbourne}
  (\bibinfo {year} {2020})\BibitemShut {NoStop}%
\bibitem [{\citenamefont {Lorenz}(1955)}]{Lorenz1955}%
  \BibitemOpen
  \bibfield  {author} {\bibinfo {author} {\bibfnamefont {E.~N.}\ \bibnamefont
  {Lorenz}},\ }\bibfield  {title} {\bibinfo {title} {Available potential energy
  and the maintenance of the general circulation},\ }\href@noop {} {\bibfield
  {journal} {\bibinfo  {journal} {Tellus}\ }\textbf {\bibinfo {volume} {7}},\
  \bibinfo {pages} {157} (\bibinfo {year} {1955})}\BibitemShut {NoStop}%
\bibitem [{\citenamefont {Turner}(1979)}]{Turner1979}%
  \BibitemOpen
  \bibfield  {author} {\bibinfo {author} {\bibfnamefont {J.~S.}\ \bibnamefont
  {Turner}},\ }\href@noop {} {\emph {\bibinfo {title} {Buoyancy Effects in
  Fluids}}}\ (\bibinfo  {publisher} {Cambridge University Press},\ \bibinfo
  {year} {1979})\BibitemShut {NoStop}%
\bibitem [{\citenamefont {Orlandi}(2000)}]{Orlandi2000}%
  \BibitemOpen
  \bibfield  {author} {\bibinfo {author} {\bibfnamefont {P.}~\bibnamefont
  {Orlandi}},\ }\href@noop {} {\emph {\bibinfo {title} {Fluid Flow Phenomena: A
  Numerical Toolkit}}}\ (\bibinfo  {publisher} {Springer},\ \bibinfo {year}
  {2000})\BibitemShut {NoStop}%
\bibitem [{\citenamefont {Wray}(1990)}]{Wray1990}%
  \BibitemOpen
  \bibfield  {author} {\bibinfo {author} {\bibfnamefont {A.~A.}\ \bibnamefont
  {Wray}},\ }\href@noop {} {\emph {\bibinfo {title} {{Minimal-storage time
  advancement schemes for spectral methods}}}},\ \bibinfo {type} {Tech. Rep.}\
  (\bibinfo  {institution} {NASA Ames Research Center},\ \bibinfo {year}
  {1990})\BibitemShut {NoStop}%
\bibitem [{\citenamefont {Chorin}(1968)}]{Chorin1968}%
  \BibitemOpen
  \bibfield  {author} {\bibinfo {author} {\bibfnamefont {A.~J.}\ \bibnamefont
  {Chorin}},\ }\bibfield  {title} {\bibinfo {title} {Numerical solution of the
  {Navier-Stokes} equations},\ }\href@noop {} {\bibfield  {journal} {\bibinfo
  {journal} {Math. Comput.}\ }\textbf {\bibinfo {volume} {22}},\ \bibinfo
  {pages} {745} (\bibinfo {year} {1968})}\BibitemShut {NoStop}%
\bibitem [{\citenamefont {Bae}\ \emph {et~al.}(2019)\citenamefont {Bae},
  \citenamefont {Lozano-Dur{\'a}n}, \citenamefont {Bose},\ and\ \citenamefont
  {Moin}}]{Bae2019}%
  \BibitemOpen
  \bibfield  {author} {\bibinfo {author} {\bibfnamefont {H.~J.}\ \bibnamefont
  {Bae}}, \bibinfo {author} {\bibfnamefont {A.}~\bibnamefont
  {Lozano-Dur{\'a}n}}, \bibinfo {author} {\bibfnamefont {S.~T.}\ \bibnamefont
  {Bose}},\ and\ \bibinfo {author} {\bibfnamefont {P.}~\bibnamefont {Moin}},\
  }\bibfield  {title} {\bibinfo {title} {Dynamic slip wall model for large-eddy
  simulation},\ }\href@noop {} {\bibfield  {journal} {\bibinfo  {journal} {J.
  Fluid Mech.}\ }\textbf {\bibinfo {volume} {859}},\ \bibinfo {pages} {400}
  (\bibinfo {year} {2019})}\BibitemShut {NoStop}%
\bibitem [{\citenamefont {Lozano-Dur{\'a}n}\ and\ \citenamefont
  {Bae}(2019)}]{Lozano-Duran2019}%
  \BibitemOpen
  \bibfield  {author} {\bibinfo {author} {\bibfnamefont {A.}~\bibnamefont
  {Lozano-Dur{\'a}n}}\ and\ \bibinfo {author} {\bibfnamefont {H.~J.}\
  \bibnamefont {Bae}},\ }\bibfield  {title} {\bibinfo {title} {Characteristic
  scales of {T}ownsend’s wall-attached eddies},\ }\href@noop {} {\bibfield
  {journal} {\bibinfo  {journal} {J. Fluid Mech.}\ }\textbf {\bibinfo {volume}
  {868}},\ \bibinfo {pages} {698} (\bibinfo {year} {2019})}\BibitemShut
  {NoStop}%
\bibitem [{\citenamefont {Ribeiro}\ \emph {et~al.}(2020)\citenamefont
  {Ribeiro}, \citenamefont {Yeh},\ and\ \citenamefont {Taira}}]{Ribeiro20}%
  \BibitemOpen
  \bibfield  {author} {\bibinfo {author} {\bibfnamefont {J.~H.~M.}\
  \bibnamefont {Ribeiro}}, \bibinfo {author} {\bibfnamefont {C.-A.}\
  \bibnamefont {Yeh}},\ and\ \bibinfo {author} {\bibfnamefont {K.}~\bibnamefont
  {Taira}},\ }\bibfield  {title} {\bibinfo {title} {Randomized resolvent
  analysis},\ }\href@noop {} {\bibfield  {journal} {\bibinfo  {journal} {Phys.
  Rev. Fluids}\ }\textbf {\bibinfo {volume} {5}},\ \bibinfo {pages} {033902}
  (\bibinfo {year} {2020})}\BibitemShut {NoStop}%
\bibitem [{\citenamefont {Symon}\ \emph {et~al.}(2018)\citenamefont {Symon},
  \citenamefont {Rosenberg}, \citenamefont {Dawson},\ and\ \citenamefont
  {McKeon}}]{Symon2018}%
  \BibitemOpen
  \bibfield  {author} {\bibinfo {author} {\bibfnamefont {S.}~\bibnamefont
  {Symon}}, \bibinfo {author} {\bibfnamefont {K.}~\bibnamefont {Rosenberg}},
  \bibinfo {author} {\bibfnamefont {S.~T.~M.}\ \bibnamefont {Dawson}},\ and\
  \bibinfo {author} {\bibfnamefont {B.~J.}\ \bibnamefont {McKeon}},\ }\bibfield
   {title} {\bibinfo {title} {Non-normality and classification of amplification
  mechanisms in stability and resolvent analysis},\ }\href@noop {} {\bibfield
  {journal} {\bibinfo  {journal} {Phys. Rev. Fluids}\ }\textbf {\bibinfo
  {volume} {3}},\ \bibinfo {pages} {053902} (\bibinfo {year}
  {2018})}\BibitemShut {NoStop}%
\bibitem [{\citenamefont {del Alamo}\ \emph {et~al.}(2004)\citenamefont {del
  Alamo}, \citenamefont {Jim{\'e}nez}, \citenamefont {Zandonade},\ and\
  \citenamefont {Moser}}]{delAlamo2004}%
  \BibitemOpen
  \bibfield  {author} {\bibinfo {author} {\bibfnamefont {J.~C.}\ \bibnamefont
  {del Alamo}}, \bibinfo {author} {\bibfnamefont {J.}~\bibnamefont
  {Jim{\'e}nez}}, \bibinfo {author} {\bibfnamefont {P.}~\bibnamefont
  {Zandonade}},\ and\ \bibinfo {author} {\bibfnamefont {R.~D.}\ \bibnamefont
  {Moser}},\ }\bibfield  {title} {\bibinfo {title} {Scaling of the energy
  spectra of turbulent channels},\ }\href@noop {} {\bibfield  {journal}
  {\bibinfo  {journal} {J. Fluid Mech.}\ }\textbf {\bibinfo {volume} {500}},\
  \bibinfo {pages} {135} (\bibinfo {year} {2004})}\BibitemShut {NoStop}%
\bibitem [{\citenamefont {Hopfinger}(1987)}]{Hopfinger1987}%
  \BibitemOpen
  \bibfield  {author} {\bibinfo {author} {\bibfnamefont {E.~J.}\ \bibnamefont
  {Hopfinger}},\ }\bibfield  {title} {\bibinfo {title} {Turbulence in
  stratified fluids: A review},\ }\href@noop {} {\bibfield  {journal} {\bibinfo
   {journal} {J.~Geophys. Research}\ }\textbf {\bibinfo {volume} {92}},\
  \bibinfo {pages} {5287} (\bibinfo {year} {1987})}\BibitemShut {NoStop}%
\bibitem [{\citenamefont {Tritton}(1967)}]{Tritton1967}%
  \BibitemOpen
  \bibfield  {author} {\bibinfo {author} {\bibfnamefont {D.~J.}\ \bibnamefont
  {Tritton}},\ }\bibfield  {title} {\bibinfo {title} {Some new correlation
  measurements in a turbulent boundary layer},\ }\href@noop {} {\bibfield
  {journal} {\bibinfo  {journal} {J. Fluid Mech.}\ }\textbf {\bibinfo {volume}
  {28}},\ \bibinfo {pages} {439} (\bibinfo {year} {1967})}\BibitemShut
  {NoStop}%
\bibitem [{\citenamefont {Liu}\ \emph {et~al.}(2001)\citenamefont {Liu},
  \citenamefont {Adrian},\ and\ \citenamefont {Hanratty}}]{Liu2001}%
  \BibitemOpen
  \bibfield  {author} {\bibinfo {author} {\bibfnamefont {Z.}~\bibnamefont
  {Liu}}, \bibinfo {author} {\bibfnamefont {R.~J.}\ \bibnamefont {Adrian}},\
  and\ \bibinfo {author} {\bibfnamefont {T.~J.}\ \bibnamefont {Hanratty}},\
  }\bibfield  {title} {\bibinfo {title} {Large-scale modes of turbulent channel
  flow: transport and structure},\ }\href@noop {} {\bibfield  {journal}
  {\bibinfo  {journal} {J. Fluid Mech.}\ }\textbf {\bibinfo {volume} {448}},\
  \bibinfo {pages} {53} (\bibinfo {year} {2001})}\BibitemShut {NoStop}%
\bibitem [{\citenamefont {Ganapathisubramani}\ \emph
  {et~al.}(2005)\citenamefont {Ganapathisubramani}, \citenamefont {Hutchins},
  \citenamefont {Hambleton}, \citenamefont {Longmire},\ and\ \citenamefont
  {Marusic}}]{Ganapathisubramani2005}%
  \BibitemOpen
  \bibfield  {author} {\bibinfo {author} {\bibfnamefont {B.}~\bibnamefont
  {Ganapathisubramani}}, \bibinfo {author} {\bibfnamefont {N.}~\bibnamefont
  {Hutchins}}, \bibinfo {author} {\bibfnamefont {W.~T.}\ \bibnamefont
  {Hambleton}}, \bibinfo {author} {\bibfnamefont {E.~K.}\ \bibnamefont
  {Longmire}},\ and\ \bibinfo {author} {\bibfnamefont {I.}~\bibnamefont
  {Marusic}},\ }\bibfield  {title} {\bibinfo {title} {Investigation of
  large-scale coherence in a turbulent boundary layer using two-point
  correlations},\ }\href@noop {} {\bibfield  {journal} {\bibinfo  {journal} {J.
  Fluid Mech.}\ }\textbf {\bibinfo {volume} {524}},\ \bibinfo {pages} {57}
  (\bibinfo {year} {2005})}\BibitemShut {NoStop}%
\bibitem [{\citenamefont {Lee}\ and\ \citenamefont {Sung}(2011)}]{Lee2011}%
  \BibitemOpen
  \bibfield  {author} {\bibinfo {author} {\bibfnamefont {J.~H.}\ \bibnamefont
  {Lee}}\ and\ \bibinfo {author} {\bibfnamefont {H.~J.}\ \bibnamefont {Sung}},\
  }\bibfield  {title} {\bibinfo {title} {Very-large-scale motions in a
  turbulent boundary layer},\ }\href@noop {} {\bibfield  {journal} {\bibinfo
  {journal} {J. Fluid Mech.}\ }\textbf {\bibinfo {volume} {673}},\ \bibinfo
  {pages} {80} (\bibinfo {year} {2011})}\BibitemShut {NoStop}%
\bibitem [{\citenamefont {Pirozzoli}\ and\ \citenamefont
  {Bernardini}(2011)}]{Pirozzoli2011}%
  \BibitemOpen
  \bibfield  {author} {\bibinfo {author} {\bibfnamefont {S.}~\bibnamefont
  {Pirozzoli}}\ and\ \bibinfo {author} {\bibfnamefont {M.}~\bibnamefont
  {Bernardini}},\ }\bibfield  {title} {\bibinfo {title} {Turbulence in
  supersonic boundary layers at moderate {R}eynolds number},\ }\href@noop {}
  {\bibfield  {journal} {\bibinfo  {journal} {J. Fluid Mech.}\ }\textbf
  {\bibinfo {volume} {688}},\ \bibinfo {pages} {120} (\bibinfo {year}
  {2011})}\BibitemShut {NoStop}%
\bibitem [{\citenamefont {Sillero}\ \emph {et~al.}(2014)\citenamefont
  {Sillero}, \citenamefont {Jim{\'e}nez},\ and\ \citenamefont
  {Moser}}]{Sillero2014}%
  \BibitemOpen
  \bibfield  {author} {\bibinfo {author} {\bibfnamefont {J.~A.}\ \bibnamefont
  {Sillero}}, \bibinfo {author} {\bibfnamefont {J.}~\bibnamefont
  {Jim{\'e}nez}},\ and\ \bibinfo {author} {\bibfnamefont {R.~D.}\ \bibnamefont
  {Moser}},\ }\bibfield  {title} {\bibinfo {title} {Two-point statistics for
  turbulent boundary layers and channels at {R}eynolds numbers up to
  $\delta^+\approx 2000$},\ }\href@noop {} {\bibfield  {journal} {\bibinfo
  {journal} {Phys. Fluids}\ }\textbf {\bibinfo {volume} {26}},\ \bibinfo
  {pages} {105109} (\bibinfo {year} {2014})}\BibitemShut {NoStop}%
\bibitem [{\citenamefont {Kline}\ \emph {et~al.}(1967)\citenamefont {Kline},
  \citenamefont {Reynolds}, \citenamefont {Schraub},\ and\ \citenamefont
  {Runstadler}}]{Kline1967}%
  \BibitemOpen
  \bibfield  {author} {\bibinfo {author} {\bibfnamefont {S.~J.}\ \bibnamefont
  {Kline}}, \bibinfo {author} {\bibfnamefont {W.~C.}\ \bibnamefont {Reynolds}},
  \bibinfo {author} {\bibfnamefont {F.~A.}\ \bibnamefont {Schraub}},\ and\
  \bibinfo {author} {\bibfnamefont {P.~W.}\ \bibnamefont {Runstadler}},\
  }\bibfield  {title} {\bibinfo {title} {The structure of turbulent boundary
  layers},\ }\href@noop {} {\bibfield  {journal} {\bibinfo  {journal} {J.~Fluid
  Mech.}\ }\textbf {\bibinfo {volume} {30}},\ \bibinfo {pages} {741} (\bibinfo
  {year} {1967})}\BibitemShut {NoStop}%
\bibitem [{\citenamefont {Smith}\ and\ \citenamefont
  {Metzler}(1983)}]{Smith1983}%
  \BibitemOpen
  \bibfield  {author} {\bibinfo {author} {\bibfnamefont {C.~R.}\ \bibnamefont
  {Smith}}\ and\ \bibinfo {author} {\bibfnamefont {S.~P.}\ \bibnamefont
  {Metzler}},\ }\bibfield  {title} {\bibinfo {title} {The characteristics of
  low-speed streaks in the near-wall region of a turbulent boundary layer},\
  }\href@noop {} {\bibfield  {journal} {\bibinfo  {journal} {J.~Fluid Mech.}\
  }\textbf {\bibinfo {volume} {129}},\ \bibinfo {pages} {27} (\bibinfo {year}
  {1983})}\BibitemShut {NoStop}%
\bibitem [{\citenamefont {Symon}\ \emph {et~al.}(2020)\citenamefont {Symon},
  \citenamefont {Illingworth},\ and\ \citenamefont {Marusic}}]{Symon21}%
  \BibitemOpen
  \bibfield  {author} {\bibinfo {author} {\bibfnamefont {S.}~\bibnamefont
  {Symon}}, \bibinfo {author} {\bibfnamefont {S.~J.}\ \bibnamefont
  {Illingworth}},\ and\ \bibinfo {author} {\bibfnamefont {I.}~\bibnamefont
  {Marusic}},\ }\bibfield  {title} {\bibinfo {title} {Energy transfer in
  turbulent channel flows and implications for resolvent modelling},\
  }\href@noop {} {\bibfield  {journal} {\bibinfo  {journal} {J. Fluid Mech. (to
  appear), arXiv:2004.13266}\ } (\bibinfo {year} {2020})}\BibitemShut {NoStop}%
\bibitem [{\citenamefont {Mizuno}(2016)}]{Mizuno2016}%
  \BibitemOpen
  \bibfield  {author} {\bibinfo {author} {\bibfnamefont {Y.}~\bibnamefont
  {Mizuno}},\ }\bibfield  {title} {\bibinfo {title} {Spectra of energy
  transport in turbulent channel flows for moderate {R}eynolds numbers},\
  }\href@noop {} {\bibfield  {journal} {\bibinfo  {journal} {J. Fluid Mech.}\
  }\textbf {\bibinfo {volume} {805}},\ \bibinfo {pages} {171} (\bibinfo {year}
  {2016})}\BibitemShut {NoStop}%
\bibitem [{\citenamefont {Moarref}\ \emph {et~al.}(2014)\citenamefont
  {Moarref}, \citenamefont {Jovanovi{\'c}}, \citenamefont {Tropp},
  \citenamefont {Sharma},\ and\ \citenamefont {McKeon}}]{Moarref2014}%
  \BibitemOpen
  \bibfield  {author} {\bibinfo {author} {\bibfnamefont {R.}~\bibnamefont
  {Moarref}}, \bibinfo {author} {\bibfnamefont {M.~R.}\ \bibnamefont
  {Jovanovi{\'c}}}, \bibinfo {author} {\bibfnamefont {J.~A.}\ \bibnamefont
  {Tropp}}, \bibinfo {author} {\bibfnamefont {A.~S.}\ \bibnamefont {Sharma}},\
  and\ \bibinfo {author} {\bibfnamefont {B.~J.}\ \bibnamefont {McKeon}},\
  }\bibfield  {title} {\bibinfo {title} {A low-order decomposition of turbulent
  channel flow via resolvent analysis and convex optimization},\ }\href@noop {}
  {\bibfield  {journal} {\bibinfo  {journal} {Phys. Fluids}\ }\textbf {\bibinfo
  {volume} {26}},\ \bibinfo {pages} {051701} (\bibinfo {year}
  {2014})}\BibitemShut {NoStop}%
\bibitem [{\citenamefont {Rosenberg}\ and\ \citenamefont
  {McKeon}(2019)}]{Rosenberg18}%
  \BibitemOpen
  \bibfield  {author} {\bibinfo {author} {\bibfnamefont {K.}~\bibnamefont
  {Rosenberg}}\ and\ \bibinfo {author} {\bibfnamefont {B.~J.}\ \bibnamefont
  {McKeon}},\ }\bibfield  {title} {\bibinfo {title} {Efficient representation
  of exact coherent states of the {N}avier-{S}tokes equations using resolvent
  analysis},\ }\href@noop {} {\bibfield  {journal} {\bibinfo  {journal} {Fluid
  Dyn. Res.}\ }\textbf {\bibinfo {volume} {51}} (\bibinfo {year}
  {2019})}\BibitemShut {NoStop}%
\end{thebibliography}%
\end{document}